\documentclass[10pt]{article} % onecolumn (standard format)

\usepackage[utf8]{inputenc}
\usepackage{amsmath}
\usepackage{amsfonts}
\usepackage{amssymb}
\usepackage{mathtools}
\usepackage{listings}
\usepackage{mathrsfs}
\usepackage{times}
\usepackage{float}
\usepackage{color}
\usepackage{setspace}
\usepackage{color,soul}

\usepackage{amsmath,amsfonts,amsthm,eucal}
\usepackage{enumerate}
\usepackage[letterpaper,margin=3cm]{geometry}
\usepackage{float}
\usepackage[title]{appendix}
\usepackage{color}
\usepackage{tabularx}
\usepackage{siunitx}
\usepackage[tableposition=below]{caption}
\usepackage[
pdfstartview=XYZ,
bookmarks=true,
colorlinks=true,
linkcolor=blue,
urlcolor=blue,
citecolor=blue,
pdftex,
bookmarks=true,
linktocpage=true,   % makes the page number as hyperlink in table of content
hyperindex=true
]{hyperref}

\usepackage{lipsum}
\usepackage[FIGBOTCAP,TABTOPCAP,bf,tight]{subfigure}

\usepackage{natbib}
\usepackage[pdftex]{graphicx}
\usepackage{epstopdf}
\usepackage{booktabs} 
\usepackage{tikz,mathpazo}
\usetikzlibrary{shapes.geometric, arrows}
\usepackage{caption}

\newcommand{\tensor}[1]{\ensuremath{\boldsymbol{#1}}}
\newcommand{\set}[1]{\ensuremath{\mathbb{#1}}}
\newcommand{\tuple}[1]{\ensuremath{\mathbb{#1}}}

\theoremstyle{remark}

\renewcommand{\vec}[1]{\ensuremath{\boldsymbol{#1}}}

\usepackage{algorithm}
\usepackage{algorithmicx}
\usepackage[noend]{algpseudocode}

\theoremstyle{definition}

\usepackage{longtable}
\usepackage{adjustbox}
\usepackage{framed}

\title{A non-cooperative meta-modeling game for automated third-party calibrating, validating, and falsifying constitutive laws with parallelized adversarial attacks} 

\begin{document}

\author{
Kun Wang\thanks{Fluid Dynamics and Solid Mechanics Group, Theoretical Division, Los Alamos National Laboratory, Los Alamos, NM 87545. kunw@lanl.gov}       
\and
WaiChing Sun\thanks{Department of Civil Engineering and Engineering Mechanics, Columbia University, New York, NY 10027. wsun@columbia.edu (corresponding author)}
\and
Qiang Du\thanks{Department of Applied Physics and Applied Mathematics, and Data Science Institute, Columbia University, New York, NY 10027. qd2125@columbia.edu}
}

\maketitle

\begin{abstract}
The evaluation of constitutive models, especially for high-risk and high-regret engineering applications, requires efficient and rigorous third-party calibration, validation and falsification. 
While there are numerous efforts to develop paradigms and standard procedures to validate models, difficulties may arise due to the sequential, manual and often biased nature of the commonly adopted calibration and validation processes, thus slowing down data collections, hampering the progress towards discovering new physics, increasing expenses and possibly leading to misinterpretations of the credibility and application ranges of proposed models. 
This work attempts to introduce concepts from game theory and machine learning techniques to overcome many of these existing difficulties. 
We introduce an automated meta-modeling game where two competing AI agents systematically generate experimental data to calibrate a given constitutive model and to explore its weakness, in order to improve experiment design and model robustness through competition. 
The two agents automatically search for the Nash equilibrium of the meta-modeling game in an adversarial reinforcement learning framework without human intervention. 
In particular, a protagonist agent seeks to find the more effective ways to generate data for model calibrations, while an adversary agent tries to find the most devastating test scenarios that expose the weaknesses of the constitutive model calibrated by the protagonist. 
By capturing all possible design options of the laboratory experiments into a single decision tree, we recast the design of experiments as a game of combinatorial moves that can be resolved through deep reinforcement learning by the two competing players. 
Our adversarial framework emulates idealized scientific collaborations and competitions among researchers to achieve better understanding of the application range of the learned material laws and prevent misinterpretations caused by conventional AI-based third-party validation. 
Numerical examples are given to demonstrate the wide applicability of the proposed meta-modeling game with adversarial attacks on both human-crafted constitutive models and machine learning models.
\end{abstract}

\section{Introduction and background}
\label{intro}
As constitutive models that predict material responses become increasingly sophisticated and complex,  the demands and difficulties for accurately calibrating and validating those constitutive laws also increase \citep{dafalias1984modelling, thacker2004concepts, borja2013plasticity, liu2016determining, de2017multiscale, bryant2018mixed, na2019configurational, na2019multi}. 
Engineering applications, particularly those involve high-risk and high-regret decision-makings, require models to maintain robustness and accuracy in unforeseen scenarios using as little amount of necessary calibration data as possible. 
Quantifying the reliability of a new constitutive law, however, is nontrivial. 
As many constitutive models are calibrated against limited amount and types of experimental data,  identifying the reliable application range of these constitutive laws beyond the loading paths used in the calibration could be challenging. 
While an apparent good match between predictions and calibration data can be easily achieved  by increasing the dimensionality of the parametric space of a given model, over-fitting may also jeopardize the application of the learned models in the case where the predicted loading path bears little resemblance  to the calibration data \citep{wang2016identifying, gupta2019open, heider2020so}. 
Using such constitutive models therefore bears more risks for a third-party user who is unaware of the sensitivity of material parameters on the stress predictions. 
Furthermore, the culture and the ecosystem of the scientific communities  often place a more significant focus on reporting the success of the material models on limited cases. Yet, precise and thorough investigations on the weakness and shortcomings of material models are important and often necessary, but they are less
reported or documented in the literature due to the lack of incentive \citep{pack2014sandia, boyce2016second}.

Model calibration issues are critical not only for hand-crafted models but for  many machine learning models and data-driven framework that either directly use experimental data to replace constitutive laws \cite{kirchdoerfer2016data, kirchdoerfer2017data, he2019physics} or generate optimal response surfaces via optimization problems \citep{bessa2017framework, yang2019adversarial, zhu2019physics}. 
%Due to the inherent lack of interpretability of predictions generated by neural network, reliability of the deep neural network constitutive laws is often ensured by uncertainty quantification 
%This uncertainty quantification can be conducted via different procedures, including Bayesian statistics, 
%polynomial chaos expansion and Monte Carlo sampling where one seeks to understand how probability distribution of the input material parameters affect the outcomes of the predictions, which is often a stress measure or performance metrics for solid mechanics applications. 
%While uncertainty quantification is crucial step to ensure the readiness of a constitutive law for engineering applications,
%a common challenge is to detect the rare event where a catastrophic drop in prediction accuracy may occur in an otherwise 
%highly accurate constitutive laws. 
The recent trend of using black-box deep neural network (DNN) to generate constitutive laws has made the reliability analysis even more crucial. 
At present, due to the lack of interpretability of predictions generated by neural network, reliability of DNN generated constitutive laws is often assessed through uncertainty quantification (UQ) \citep{yang2019adversarial, zhu2019physics}. UQ  can be conducted via different procedures, including Bayesian statistics, polynomial chaos expansion and Monte Carlo sampling where one seeks to understand how probability distributions of the  input material parameters affect the outcomes of predictions, as often represented by some stress measures or performance metrics for solid mechanics applications. 
While UQ is a crucial step to ensure the readiness of constitutive laws for engineering applications, a common challenge is to detect  rare events where a catastrophic loss of the prediction accuracy may occur in otherwise highly accurate constitutive laws. 

The machine learning research community has been proposing methods  to improve the interpolation and generalization capabilities,  hence improving the predictive capability with exogenous data as well as reducing  the epistemic uncertainties of trained neural network models. 
For instance,  the active learning approaches (e.g. \citep{settles2009active}),  which is sometimes also referred as "optimal experimental design" \citep{olsson2009literature}, introduce query strategies to choose  what data to be generated to reduce generalization errors, balance exploration and  exploitation and quantify uncertainties. 
These approaches have repeatedly outperformed traditional "passive learning" methods which involve randomly gathering a large amount of training data. 
Active learning is widely investigated using different deep learning algorithms like CNNs and LSTMS \citep{sener2017active, shen2017deep}. There is also research on implementing Generative Adversarial Networks (GANs) into the active learning framework \citep{zhu2017generative}. With the increasing interest in deep reinforcement learning, researchers are trying to re-frame active learning as a reinforcement learning problem \citep{fang2017learning}. 
Another recent study focuses on the "semi-supervised learning" approaches \citep{zhu2005semi, verma2019interpolation, berthelot2019mixmatch}, which take advantage of the structures of unlabeled input data to enhance the "interpolation consistency", in addition to labeled training data. 
These recently developed techniques have shown some degrees of  successes for image recognition, natural language processing, and therefore could potentially be helpful for mechanics problems. 

%The calibration, validation and falsification of material models have similar issues to be improved. 
%Experimental data are expensive to get in both time and cost. 
%Hence, experimentalists need to provide least amount of data that can calibrate a constitutive model with highest reliability and also identify the limitations of a constitutive model. 
The calibration, validation and falsification of material models have issues similar to those discussed above.
Moreover, experimental data are often expensive to get in both time and cost. 
Hence, experimentalists would like to generate the least amount of data that can calibrate a constitutive model with the highest reliability and can also identify its limitations. 
Traditionally the decisions on which experiments to conduct are based on the  human knowledge and experiences. 
We make an effort here to use AI to assist the decision-makings of experimentalists, which will be the  first of its kind specifically targeting the  automated design of  data generation  that can efficiently calibrate and falsify a constitutive model. 
%Our method,  different from existing machine learning techniques, is a novel formulation of the experimentalist-model-critic environment as Markov games via decision-trees. 
%In the game environment, the generation of calibration data is handled by a protagonist agent, and once the model is calibrated, the testing data are generated by an adversary agent to evaluate the forward prediction accuracy. 
%The goal of the adversary is to identify potentially all application scenarios that the model would fail according to a user-defined objective function (falsification). 

The major contribution of this paper is the introduction of a non-cooperative game that leads to new optimized experimental designs that both improves the accuracy and robustness  of the predictions on unseen data, while at the same time exposing any potential weakness  and shortcoming of a constitutive law.

We create a non-cooperative game in which a pair of agents are trained to emulate a form of artificial   intelligence capable of improving their performance through trial and errors. 
%\alert{which component of the game playing is related to long-term planning? also, the term higher form is not explained here, but in later paragraphs, maybe we should refrain using this word until later}
The two agents play against each other in a turn-based strategy game, according to  their own agendas and purposes  respectively that serve to achieve  opposite objectives. 
This setup constitutes a zero-sum game in which each agent is  rewarded by competing against the opponent. 
While the protagonist agent  learns to validate models by designing the experiments than enhance the model predictions,  the adversary agent learns how to undermine the protagonist agent by designing experiments that expose the weakness of the models. 
The optimal game strategies for both players  are explored by searching for Nash equilibrium \citep{nash1950equilibrium} of the games using deep reinforcement learning (DRL). 

With recent rapid development,  DRL techniques have  found unprecedented success in the last decades on achieving superhuman intelligence and performance in  playing increasingly complex games: Atari \citep{mnih2013playing}, board games \citep{silver2017mastering, silver2017masteringchess}, Starcraft \citep{vinyals2019grandmaster}. 
AlphaZero \citep{silver2017mastering} is also capable of learning the game strategies of our game without human knowledge. 
By emulating the learning process of human learners through trial-and-error and competition, the DRL process enables both AI agents to learn from their own successes and failures but also through their competitions to  master the tasks of calibrating and falsifying a constitutive law. 
The knowledge gained from the competitions will help us  understanding the relative rewards of different experimental setup for validation and falsification mathematically represented by the decision trees corresponding to  the protagonist and adversary agents. 

The rest of the paper is organized as follows. We first describe the meta-modeling non-cooperative game, including the method to recast the design of experiments into decision trees (Section \ref{sec:decision_tree_graph}). 
Following this, we will introduce the detailed design of the calibration-falsification game for modeling the competition between the AI experimental agent and the AI adversarial agent (Section \ref{sec:metamodelinggame}). 
In Section \ref{sec:gamedrl}, we present the multi-agent reinforcement learning algorithms that enable us to find the optimal decision for calibrating and falsifying constitutive laws. 
For numerical examples in Section \ref{sec:numericalExp}, we report the performances of the non-cooperative game on two classical elasto-plasticity models proposed by human experts for bulk granular materials, and one neural networks model of traction-separation law on granular interface.

As for notations and symbols, bold-faced letters denote tensors (including vectors which are rank-one tensors); 
%\textcolor{red}{OK here if we include vectors as special rank-one tensors? since the notation $\vec{a}$ has a single index below} 
the symbol '$\cdot$' denotes a single contraction of adjacent indices of two tensors (e.g. $\vec{a} \cdot \vec{b} = a_{i}b_{i}$ or $\tensor{c} \cdot \vec{d} = c_{ij}d_{jk}$ ); the symbol `:' denotes a double contraction of adjacent indices of tensor of rank two or higher ( e.g. $\tensor{C} : \vec{\epsilon^{e}}$ = $C_{ijkl} \epsilon_{kl}^{e}$ ); the symbol `$\otimes$' denotes a juxtaposition of two vectors (e.g. $\vec{a} \otimes \vec{b} = a_{i}b_{j}$) or two symmetric second order tensors (e.g. $(\vec{\alpha} \otimes \vec{\beta})_{ijkl} = \alpha_{ij}\beta_{kl}$). Moreover, $(\tensor{\alpha}\oplus\tensor{\beta})_{ijkl} = \alpha_{jl} \beta_{ik}$ and $(\tensor{\alpha}\ominus\tensor{\beta})_{ijkl} = \alpha_{il} \beta_{jk}$. We also define identity tensors $(\tensor{I})_{ij} = \delta_{ij}$, $(\tensor{I}^4)_{ijkl} = \delta_{ik}\delta_{jl}$, and $(\tensor{I}^4_{\text{sym}})_{ijkl} = \frac{1}{2} (\delta_{ik}\delta_{jl} + \delta_{il}\delta_{kj})$, where $\delta_{ij}$ is the Kronecker delta. 

\section{AI-designed experiments: selecting paths in arborescences of decisions}
\label{sec:decision_tree_graph}
Traditionally the decisions on which experiments to conduct are based a combination of intuition, knowledge and experience from human. 
We make the first effort to use AI to assist the decision-makings of experimentalists on how to get data that can efficiently calibrate and falsify a constitutive model. 
Our method differs from the existing machine learning techniques that we formulate the experimentalist-model-critic environment as Markov games via decision-trees. 
In this game, the generation of calibration data is handled by a protagonist agent, once the model is calibrated, the testing data are generated by an adversary agent to evaluate the forward prediction accuracy. 
The goal of the adversary is to identify all application scenarios that the model will fail according to a user-defined objective function (falsification). 
Hence the validation will be simultaneously achieved: the model is valid within the calibration scenarios picked by protagonist and the testing scenarios that the adversary has not picked. 
Practically, the model is safe to use unless the adversary "warns" that the model is at high risk. 
The formalization of decisions (or actions) as decision-trees, along with the communication mechanism designed in the game, enable AI agents to play this game competitively instead of human players. 

Here we idealize the process of designing or planning an experiment as a sequence of decision making among different available options. 
All the available options and choices in the design process considered by the AI experimentalists (protagonist and adversary) are modeled by "arborescences" in graph theory with labeled vertices and edges. 
An arborescence is a rooted polytree in which, for a single root vertex $u$ and any other vertex $v$, there exists one unique directed path from $u$ to $v$. 
A polytree (or directed tree) is a directed graph whose underlying graph is a singly connected acyclic graph. 
A brief review of the essential terminologies are given in \citet{wang2019cooperative},  and their detailed definitions can be found in, for instance, \citet{graham1989concrete, west2001introduction, bang2008digraphs}.
Mathematically, the arborescence for decision making (referred to as "decision tree" hereafter) can be expressed as an 8-tuple $\tuple{G}=(\set{L_{V}}, \set{L_{E}}, \set{V},\set{E},\vec{s},\vec{t}, \vec{n_{V}}, \vec{n_{E}})$ where
$\set{V}$ and $\set{E}$ are the sets of vertices and edges, $\set{L_{V}}$ and $\set{L_{E}}$ are the sets  of labels for the vertices and edges, $\vec{s} : \set{E} \rightarrow \set{V}$ and $\vec{t} : \set{E} \rightarrow \set{V}$ are the mappings that map each edge to its source vertex and its target vertex, 
$\vec{n_{V}} : \set{V} \rightarrow \set{L_{V}}$ and $\vec{n_{E}} : \set{E} \rightarrow \set{L_{E}}$ are the mappings that give the vertices and edges their corresponding labels (names) in $\set{L_{V}}$ and $\set{L_{E}}$.

The decision trees are constructed based on a hierarchical series of test conditions (e.g., category of test, pressure level, target strain level) that an experimentalist needs to decide in order to design an experiment on a material. 
Assuming that an experiment can be completely and uniquely defined by an ordered list of selected test conditions $tc = [tc_1, tc_2, tc_3, ..., tc_n]$, where $N_{\set{TC}}$ is the total number of test conditions. 
Each $tc_i$ is selected from a finite set of choices $\set{TC}_i = \{tc^1_i, tc^2_i, tc^3_i, ..., tc^{m_i}_i\}$, where $m_i$ is the number of choices for the $i$th test condition. 
For test conditions with inherently continuous design variables, $\set{TC}_i$ can include preset discrete values. 
For example, the target strain for a loading can be chosen from discrete values of $1\%$, $2\%$, $3\%$, etc. 
All design choices available to experimentalists are represented by an ordered list of sets $\set{TC} = [\set{TC}_1, \set{TC}_2, \set{TC}_3, ..., \set{TC}_n]$ with a hierarchical relationship such that, if $i<j$, $tc_i \in \set{TC}_i$ must be selected prior to the selection of $tc_j \in \set{TC}_j$. 

After the construction of $\set{TC}$ for experimentalists, a decision tree is built top-down from a root node representing the 'Null' state that no test condition is decided. 
The root node is split into $m_1$ subnodes according to the first level of decisions $\set{TC}_1$. 
Each subnode is further split into $m_2$ subnodes according to the second level of decisions $\set{TC}_2$. 
The splitting process on the subnodes is carried out recursively for all the $N_{\set{TC}}$ levels of decisions in $\set{TC}$. 
Finally, the down-most leaf nodes represent all possible combinations of test conditions. 
The maximum number of possible configurations of experiments is $N^{max}_{test} = \prod_{i=1}^{N_{\set{TC}}} m_i$, when all decisions across $\set{TC}_i$ are independent. 
The number of possible experiments is reduced ($N_{test} < N^{max}_{test}$) when restrictions are specified for the selections of test conditions. 
E.g., the selection of $tc_i \in \set{TC}_i$ may prohibit the selections of certain choices $tc_j$ in subsequent test conditions $\set{TC}_j, j>i$. 
The experimentalists can choose multiple experiments by taking multiple paths in the decision tree from the root node to the leaf nodes. 
The total number of possible combination of paths, if the maximum allowed number of simultaneously chosen paths is $N^{max}_{path}$, is $\sum_{k=1}^{N^{max}_{path}} C_{N_{test}}^{k}$, where $C_{N_{test}}^{k} = \frac{N_{test} !}{k! (N_{test}-k)!}$ is the combination number.

\begin{proof}[Example for hierarchical test conditions and experimental decision tree]
	Consider a simple design of mechanical experiments for geomaterials, for which all choices are listed in
	\begin{equation}
	\set{TC} = [\text{'Sample'},\ \text{'Type'},\ \text{'Target'}].
	\label{eq:TC_example}
	\end{equation}
	The first decision is to pick the initial geomaterial sample to test. Assuming that a sample is fully characterized by its initial pressure $p_0$, a simple set of discrete sample choices is given as
	\begin{equation}
	\set{TC}_1 = \text{'Sample'} = \{\text{'300kPa'},\ \text{'400kPa'}\}.
	\label{eq:TC1_example}
	\end{equation}
	The second test condition is the type of the experiment. The experiment can be either drained triaxial compression test
	('DTC') or drained triaxial extension test ('DTE'). Then
	\begin{equation}
	\set{TC}_2 = \text{'Type'} = \{\text{'DTC'},\ \text{'DTE'}\}.
	\label{eq:TC2_example}
	\end{equation}
	The third test condition to decide is the target strain magnitude for the loading. For example,  
	\begin{equation}
	\set{TC}_3 = \text{'Target'} = \{\text{'1\%'},\ \text{'3\%'}\}.
	\label{eq:TC3_example}
	\end{equation}
	After all three decisions are sequentially made (taking a path in the decision tree), the experiment is completely determined by an ordered list, e.g., $tc = [\text{'300kPa'},\ \text{'DTE'},\ \text{'3\%'}]$. 
	It indicates that the AI experimentalist decides to perform a monotonic drained triaxial extension test on a sample with $p_0=300kPa$ until the axial strain reaches $3\%$.
	
	The decision tree $\tuple{G}$ for the hierarchical design of geomaterial experiments specified by Equations (\ref{eq:TC_example}), (\ref{eq:TC1_example}), (\ref{eq:TC2_example}), (\ref{eq:TC3_example}) is shown in Fig. \ref{fig:decision_tree_example_tree}. 
	The vertex sets and edge sets of the graph are
	\begin{equation}
	\begin{aligned}
	\set{V} = & \{ \text{'Null'},\ \text{'300kPa'},\ \text{'400kPa'},\ \text{'300kPa\_DTC'},\ \text{'300kPa\_DTE'},\ \text{'400kPa\_DTC'},\ \text{'400kPa\_DTE'}, \\
	& \text{'300kPa\_DTC\_1\%'},\ \text{'300kPa\_DTC\_3\%'},\ \text{'300kPa\_DTE\_1\%'},\ \text{'300kPa\_DTE\_3\%'},\\
	& \text{'400kPa\_DTC\_1\%'},\ \text{'400kPa\_DTC\_3\%'},\ \text{'400kPa\_DTE\_1\%'},\ \text{'400kPa\_DTE\_3\%'} \}, \\
	\set{E} = & \{ \text{'Null'} \rightarrow \text{'300kPa'},\ \text{'Null'} \rightarrow \text{'400kPa'},\ \text{'300kPa'} \rightarrow \text{'300kPa\_DTC'}, \\
	& \text{'300kPa'} \rightarrow \text{'300kPa\_DTE'},\ \text{'400kPa'} \rightarrow \text{'400kPa\_DTC'},\ \text{'400kPa'} \rightarrow \text{'400kPa\_DTE'},\\ 
	& \text{'300kPa\_DTC'} \rightarrow \text{'300kPa\_DTC\_1\%'},\ \text{'300kPa\_DTC'} \rightarrow \text{'300kPa\_DTC\_3\%'},\\
	& \text{'300kPa\_DTE'} \rightarrow \text{'300kPa\_DTE\_1\%'},\ \text{'300kPa\_DTE'} \rightarrow \text{'300kPa\_DTE\_3\%'},\\
	& \text{'400kPa\_DTC'} \rightarrow \text{'400kPa\_DTC\_1\%'},\ \text{'400kPa\_DTC'} \rightarrow \text{'400kPa\_DTC\_3\%'},\\ 
	& \text{'400kPa\_DTE'} \rightarrow \text{'400kPa\_DTE\_1\%'},\ \text{'400kPa\_DTE'} \rightarrow \text{'400kPa\_DTE\_3\%'} \},\\
	\set{L_{V}} = & \set{V},\\
	\set{L_{E}} = & \{ \text{'300kPa'},\ \text{'400kPa'},\ \text{'DTC'},\ \text{'DTE'},\ \text{'1\%'},\ \text{'3\%'} \}.
	\end{aligned}
	\end{equation}
	In this example, $N_{test} = N^{max}_{test} = 2*2*2 = 8$. 
	If an experimentalist only collects data from one or two experiments, i.e., $N^{max}_{path}=2$, the total number of possible combinations is $C_{8}^1 + C_{8}^2 = 36$. 
	Fig. \ref{fig:decision_tree_example_paths} presents two example paths with edge labels illustrating the hierarchical decisions on the test conditions in order to arrive at the final experimental designs '300kPa\_DTE\_1\%' and '400kPa\_DTC\_3\%'. 
	\begin{figure}[h!]\center
	\subfigure[Decision tree with labels of vertices and edges]{
		\includegraphics[width=0.48\textwidth]{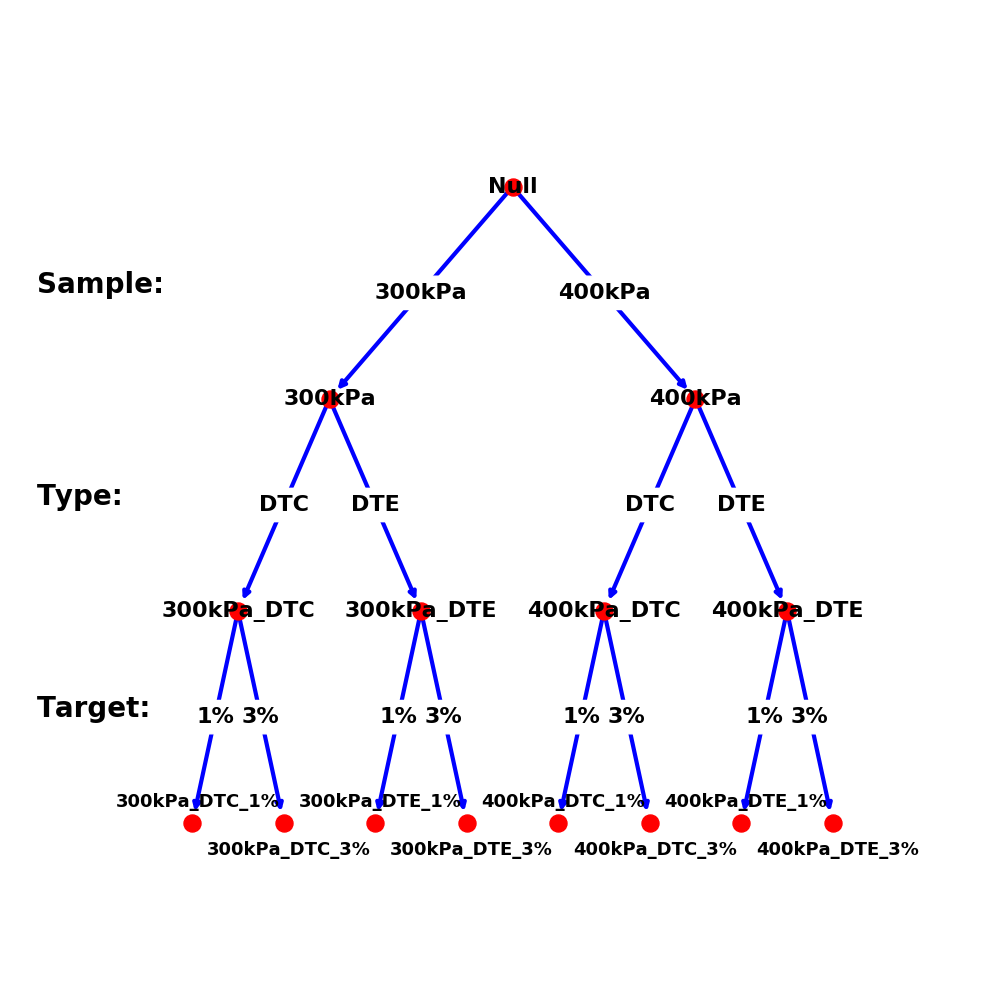}
	\label{fig:decision_tree_example_tree}
	}
	\subfigure[Example of paths in the decision tree, the selected tests are '300kPa\_DTE\_1\%' and '400kPa\_DTC\_3\%']{
		\includegraphics[width=0.48\textwidth]{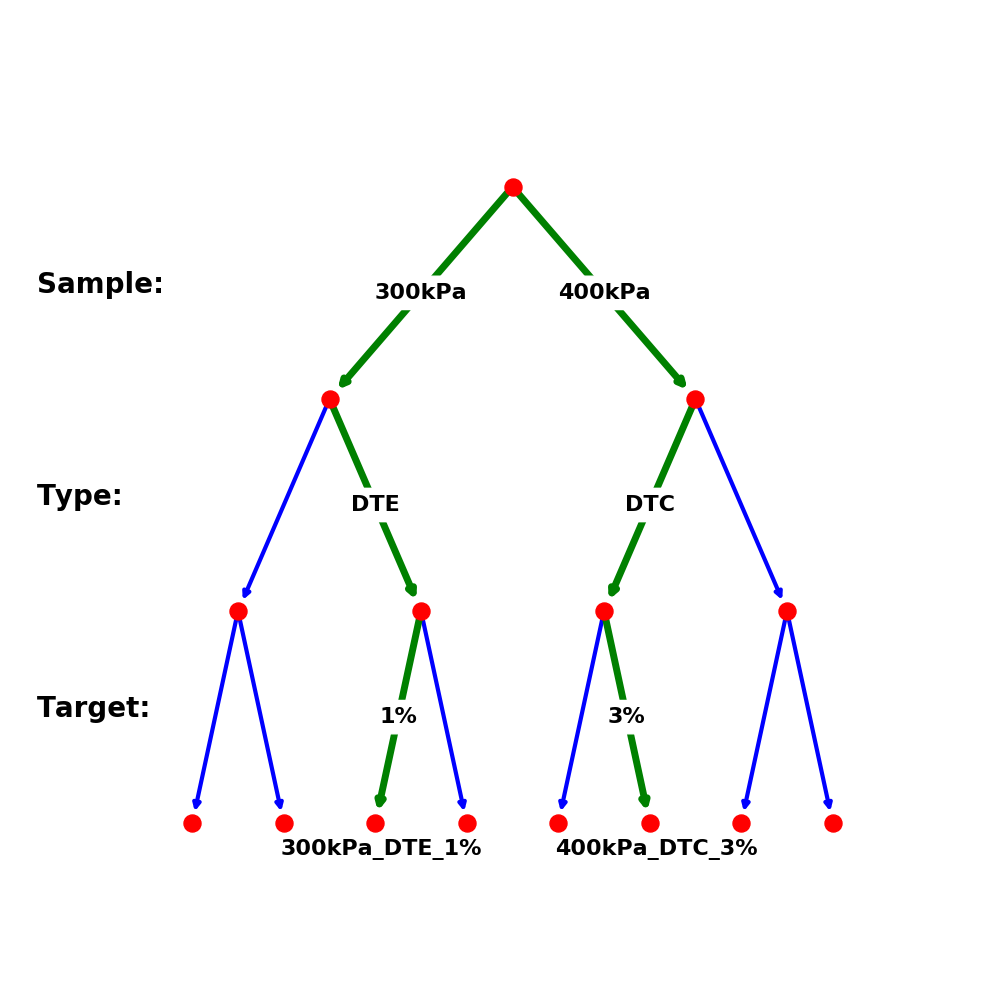}
	\label{fig:decision_tree_example_paths}
	}
	\caption{Decision tree for a simple experimental design for geomaterials (Eq. (\ref{eq:TC_example}), (\ref{eq:TC1_example}), (\ref{eq:TC2_example}), (\ref{eq:TC3_example})).}
	\label{fig:decision_tree_example}
	\end{figure}
\end{proof}

In this section, we present two decision trees for the design of geomechanical experiments, one for the bulk mechanical behavior of granular materials, another for the traction-separation behaviour of granular interfaces. 
We later study the intelligence of the reinforcement-learning-based experimentalists (protagonist and adversary) on these decision trees in Section \ref{sec:numericalExp}.

\subsection{Decision tree for AI-guided experimentation on bulk granular materials}
\label{subsec:decision_tree_bulk}
This section defines a representative decision tree for the AI-guided experimentation on bulk geomaterials. 
The hierarchical series of test conditions includes six elements, $\set{TC}$ = [$\set{TC}_1$, $\set{TC}_2$, $\set{TC}_3$, $\set{TC}_4$, $\set{TC}_5$, $\set{TC}_6$], such that the AI experimentalists can choose isotropic granular samples of different initial pressure $p_0$ and initial void ratio $e_0$, perform different drained triaxial tests, and design different loading-unloading-reloading paths. 

The choices for each test conditions are shown in Table \ref{tab:test_cond_bulk}, represented by decision labels. The decision labels for the test types $\set{TC}_3$ are defined as follows, 
\begin{enumerate}
	\item 'DTC': drained conventional triaxial compression test ($\dot{\epsilon}_{11} < 0$, $\dot{\sigma}_{22}=\dot{\sigma}_{33}=\dot{\sigma}_{12}=\dot{\sigma}_{23}=\dot{\sigma}_{13}=0$),
	\item 'DTE': drained conventional triaxial extension test ($\dot{\epsilon}_{11} > 0$, $\dot{\sigma}_{22}=\dot{\sigma}_{33}=\dot{\sigma}_{12}=\dot{\sigma}_{23}=\dot{\sigma}_{13}=0$),
	\item 'TTC': drained true triaxial test with $b=0.5$ ($\dot{\epsilon}_{11} < 0$, $b = \frac{\sigma_{22}-\sigma_{33}}{\sigma_{11}-\sigma_{33}} = const$, $\dot{\sigma}_{33}=\dot{\sigma}_{12}=\dot{\sigma}_{23}=\dot{\sigma}_{13}=0$),
\end{enumerate}
with the loading conditions represented by constraints on the components of the stress rate and strain rate tensors
\begin{equation}
\dot{\tensor{\epsilon}} =
\begin{bmatrix}
\dot{\epsilon}_{11} & \dot{\epsilon}_{12} & \dot{\epsilon}_{13}\\
& \dot{\epsilon}_{22} & \dot{\epsilon}_{23}\\
\text{sym} &  & \dot{\epsilon}_{33}\\
\end{bmatrix},\ 
\dot{\tensor{\sigma}} =
\begin{bmatrix}
\dot{\sigma}_{11} & \dot{\sigma}_{12} & \dot{\sigma}_{13}\\
& \dot{\sigma}_{22} & \dot{\sigma}_{23}\\
\text{sym} &  & \dot{\sigma}_{33}\\
\end{bmatrix}.
\end{equation}
Since 'DTC' and 'DTE' are special cases of true triaxial tests, the choices $\{\text{'DTC'},\ \text{'DTE'},\ \text{'TTC'}\}$ for $\set{TC}_3$ are equivalent to choosing the value of $b = \frac{\sigma_{22}-\sigma_{33}}{\sigma_{11}-\sigma_{33}}$ from $\{\text{'0.0'},\ \text{'1.0'},\ \text{'0.5'}\}$, respectively \citep{rodriguez2013true}. 

The decision labels $\text{'NaN'}$ in $\set{TC}_5$ and $\set{TC}_6$ indicate that the unloading or reloading is not activated. 
This design enables the freedom of generating monotonic loading paths (e.g., '5\%\_NaN\_NaN'), loading-unloading paths (e.g., '5\%\_0\%\_NaN') and loading-unloading-reloading paths (e.g., '5\%\_0\%\_3\%'). 
There are restrictions in choosing the strain targets. 
The experimentalist picks the loading target in $\set{TC}_4$ first and the unloading target in $\set{TC}_5$ must be, if not 'NaN' (stop the experiment), smaller than the loading strain. Then the reloading target in $\set{TC}_6$ must be, if not 'NaN', larger than the unloading strain. 

\begin{table}[h!]
	\centering
	\begin{tabular}{| p{5cm} | p{5cm} |}
		\hline
		$\set{TC}$ Test Conditions & Choices\\ \hline
		$\set{TC}_1=$ 'Sample $p_0$' & $\{\text{'300kPa'},\ \text{'400kPa'},\ \text{'500kPa'}\}$\\ \hline
		$\set{TC}_2=$ 'Sample $e_0$' & $\{\text{'0.60'},\ \text{'0.55'} \}$\\ \hline
		$\set{TC}_3=$ 'Type' & $\{\text{'DTC'},\ \text{'DTE'},\ \text{'TTC'}\}$\\ \hline
		$\set{TC}_4=$ 'Load Target' & $\{\text{'3\%'},\ \text{'5\%'}\}$ \\ \hline
		$\set{TC}_5=$ 'Unload Target' &  $\{\text{'NaN'},\ \text{'0\%'},\ \text{'3\%'}\}$ \\ \hline
		$\set{TC}_6=$ 'Reload Target' &  $\{\text{'NaN'},\ \text{'3\%'},\ \text{'5\%'}\}$ \\ \hline
	\end{tabular}
	\caption{Choices of test conditions for AI-guided experimentation on bulk granular materials.}
	\label{tab:test_cond_bulk}
\end{table} 

\begin{figure}[h!]\center
\includegraphics[width=0.95\textwidth]{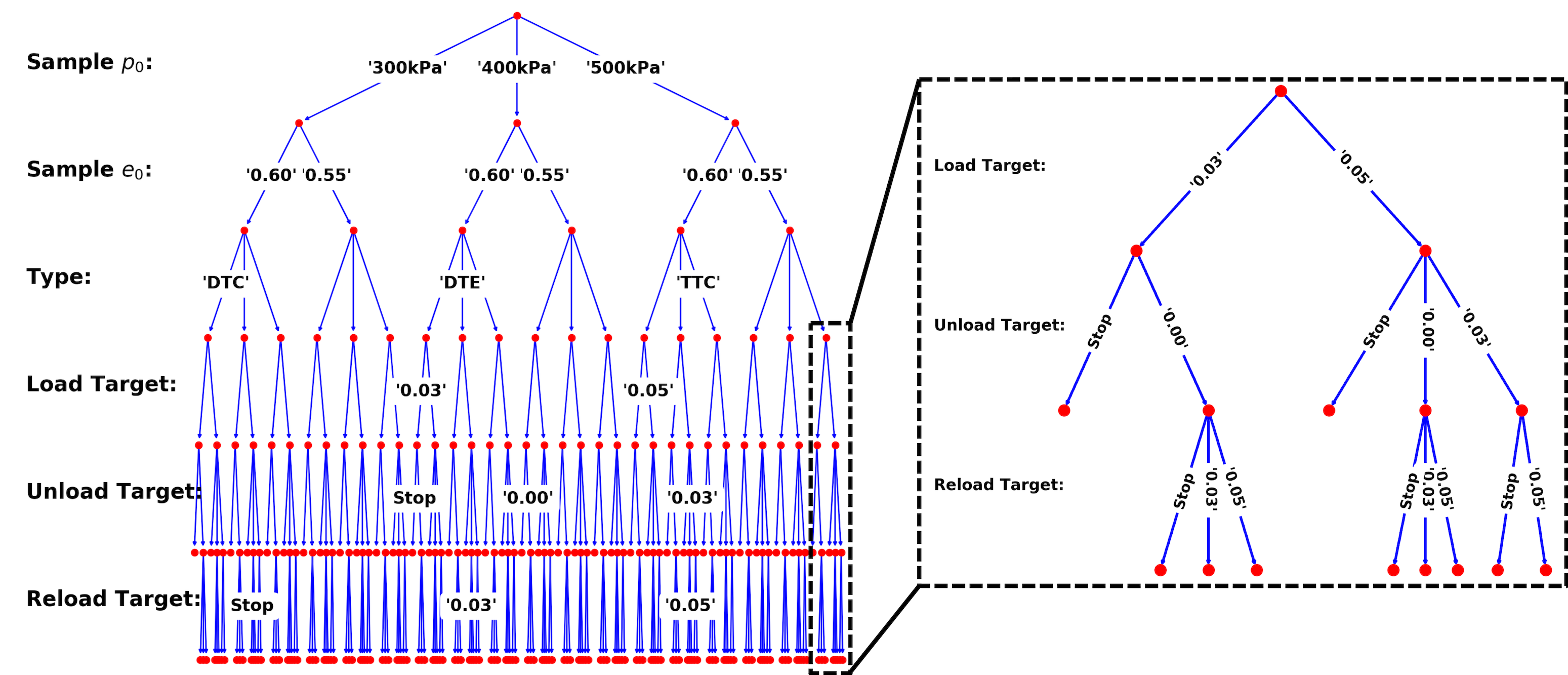}
\caption{Decision tree for AI-guided drained true triaxial tests on bulk granular materials. Due to the complexity of the graph, the vertex labels are omitted, and only a few edge labels are shown. See Fig. \ref{fig:decision_tree_example} for exhaustive vertex and edge labels in a simple decision tree example.}
\label{fig:decision_tree_experimentation_bulk}
\end{figure}

\begin{figure}[h!]\center
	\includegraphics[width=0.9\textwidth]{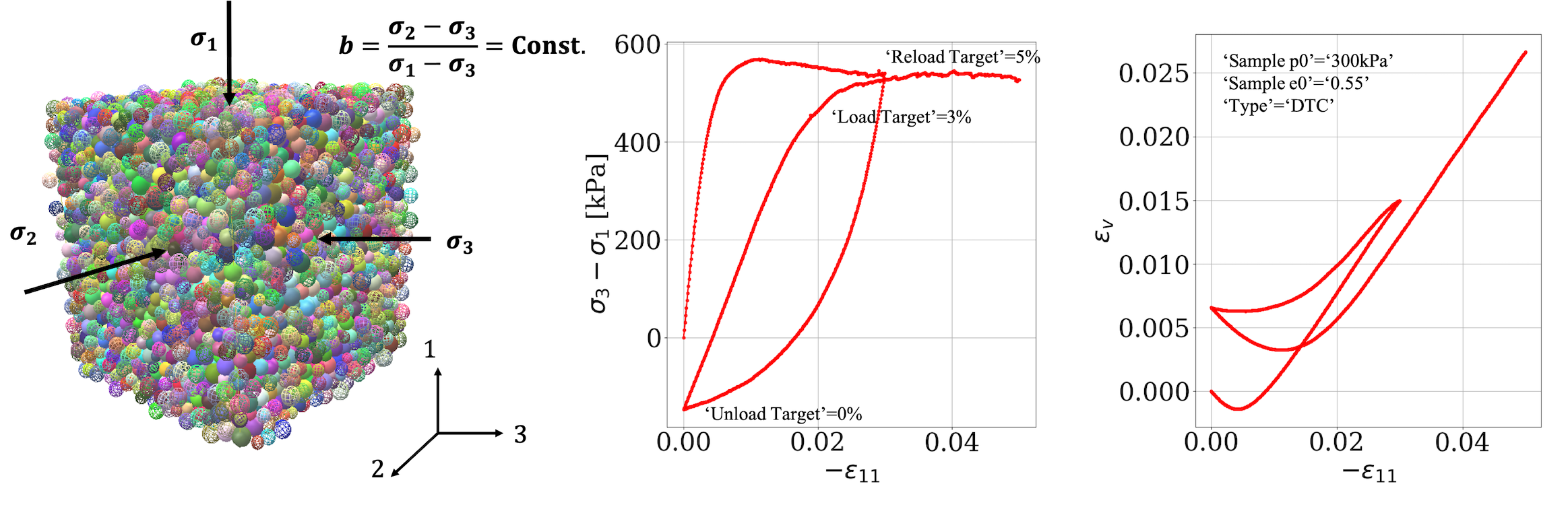}
	\caption{Experimental settings of drained true triaxial tests on numerical specimen of bulk granular materials using DEM (discrete element methods). The test conditions for the AI experimentalist are presented in Table \ref{tab:test_cond_bulk}. As an example, the differential stress data and volumetric strain data obtained from a test designed by the decision tree path '300kPa' $\rightarrow$ '0.55' $\rightarrow$ 'DTC' $\rightarrow$ '3\%' $\rightarrow$ '0\%' $\rightarrow$ '5\%' are presented.}
	\label{fig:TTC_Experiment_Illust}
\end{figure}

The corresponding decision tree is shown in Fig. \ref{fig:decision_tree_experimentation_bulk}. 
The subtree concerning the restricted decision-making in $\set{TC}_4$, $\set{TC}_5$ and $\set{TC}_6$ is also detailed in the figure. 
The total number of experimental designs (which equals to the number of leaf nodes in the tree) is $N_{test}=180$. 
Fig. \ref{fig:TTC_Experiment_Illust} provides the experimental settings on DEM (discrete element methods) numerical specimens and data from one example of the experiments. 
The total number of experimental data combinations increases significantly when the maximum allowed simultaneous paths $N^{max}_{path}$ increases. 
The combination number equals to $C_{180}^1 = 180$ when $N^{max}_{path}=1$, equals to $C_{180}^1 + C_{180}^2 = 16290$ when $N^{max}_{path}=2$, equals to $C_{180}^1 + C_{180}^2 + C_{180}^3 = 972150$ when $N^{max}_{path}=3$, etc. 

\subsection{Decision tree for AI-guided experimentation on granular interfaces}
\label{subsec:decision_tree_interface}
This section defines a representative decision tree for the AI-guided experimentation on granular interfaces. 
The hierarchical series of test conditions includes six elements, $\set{TC}$ = [$\set{TC}_1$, $\set{TC}_2$, $\set{TC}_3$, $\set{TC}_4$, $\set{TC}_5$, $\set{TC}_6$], such that the AI experimentalists can choose the direction of the prescribed displacement jump, the number of loading cycles, and different target displacement values to design complex loading paths.

The choices for each test conditions are shown in Table \ref{tab:test_cond_TS}, represented by decision labels. 
'NormTangAngle' represents the angle between the displacement jump vector and the tangential direction vector, the corresponding values in the Choices column are in units of degree. 'NumCycle' represents the number of loading-unloading cycles. 
The conditions 'Target1', 'Target2', 'Target3', 'Target4' represent the target displacement jump magnitudes along the loading-unloading cycles, the corresponding values in the Choices column are in units of millimeters. 
Regardless of the loading-unloading cycles, the final displacement jump reaches the magnitude of 0.4 mm. 
The decision label $\text{'NaN'}$ indicates that the unloading or reloading is not activated. 
For example, 'NumCycle'='0' means a monotonic loading to 0.4 mm, hence all the target conditions should adopt the values of 'NaN'; 'NumCycle'='1' means a loading-unloading-reloading path to 0.4 mm, hence 'Target1' (loading target) and 'Target2'(unloading target) can adopt values within $\text{'0.0'},\ \text{'0.1'},\ \text{'0.2'},\ \text{'0.3'}$, while 'Target3' and 'Target4' should be 'NaN's. 

\begin{table}[h!]
	\centering
	\begin{tabular}{| p{5cm} | p{8cm} |}
		\hline
		$\set{TC}$ Test Conditions & Choices\\ \hline
		$\set{TC}_1=$ 'NormTangAngle' & $\{\text{'0'},\ \text{'15'},\ \text{'30'},\ \text{'45'},\ \text{'60'},\ \text{'75'}\}$\\ \hline
		$\set{TC}_2=$ 'NumCycle' & $\{\text{'0'},\ \text{'1'},\ \text{'2'}\}$\\ \hline
		$\set{TC}_3=$ 'Target1' & $\{\text{'NaN'},\ \text{'0.1'},\ \text{'0.2'},\ \text{'0.3'}\}$\\ \hline
		$\set{TC}_4=$ 'Target2' & $\{\text{'NaN'},\ \text{'0.0'},\ \text{'0.1'},\ \text{'0.2'}\}$\\ \hline
		$\set{TC}_5=$ 'Target3' & $\{\text{'NaN'},\ \text{'0.1'},\ \text{'0.2'},\ \text{'0.3'}\}$\\ \hline
		$\set{TC}_6=$ 'Target4' & $\{\text{'NaN'},\ \text{'0.0'},\ \text{'0.1'},\ \text{'0.2'}\}$\\ \hline
	\end{tabular}
	\caption{Choices of test conditions for AI-guided experimentation on granular interfaces.}
	\label{tab:test_cond_TS}
\end{table}

\begin{figure}[h!]\center
\includegraphics[width=0.75\textwidth]{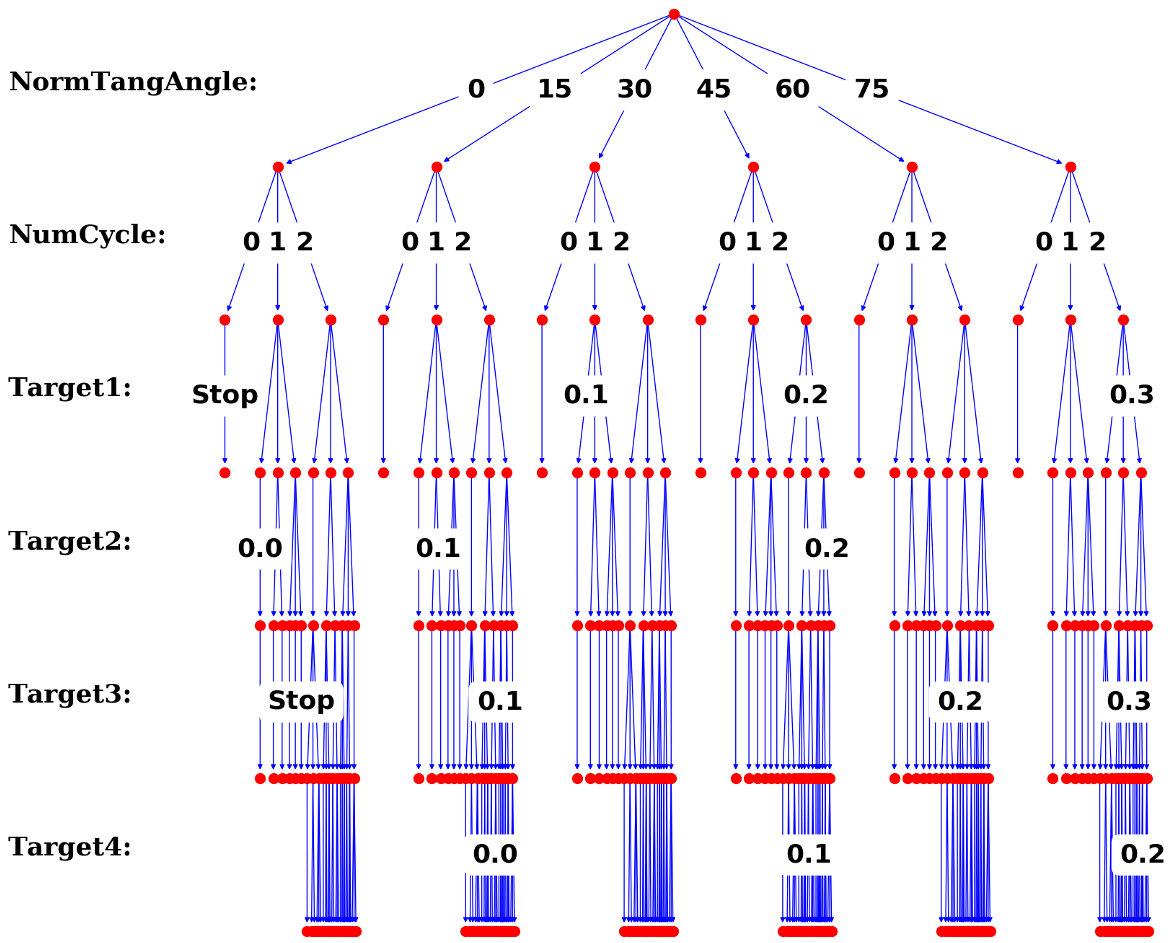}
\caption{Decision tree for AI-guided displacement-driven mixed-mode shear tests on granular interfaces. Due to the complexity of the graph, the vertex labels are omitted, and only a few edge labels are shown.}
\label{fig:Decision_Tree_TS}
\end{figure}

\begin{figure}[h!]\center
	\includegraphics[width=0.9\textwidth]{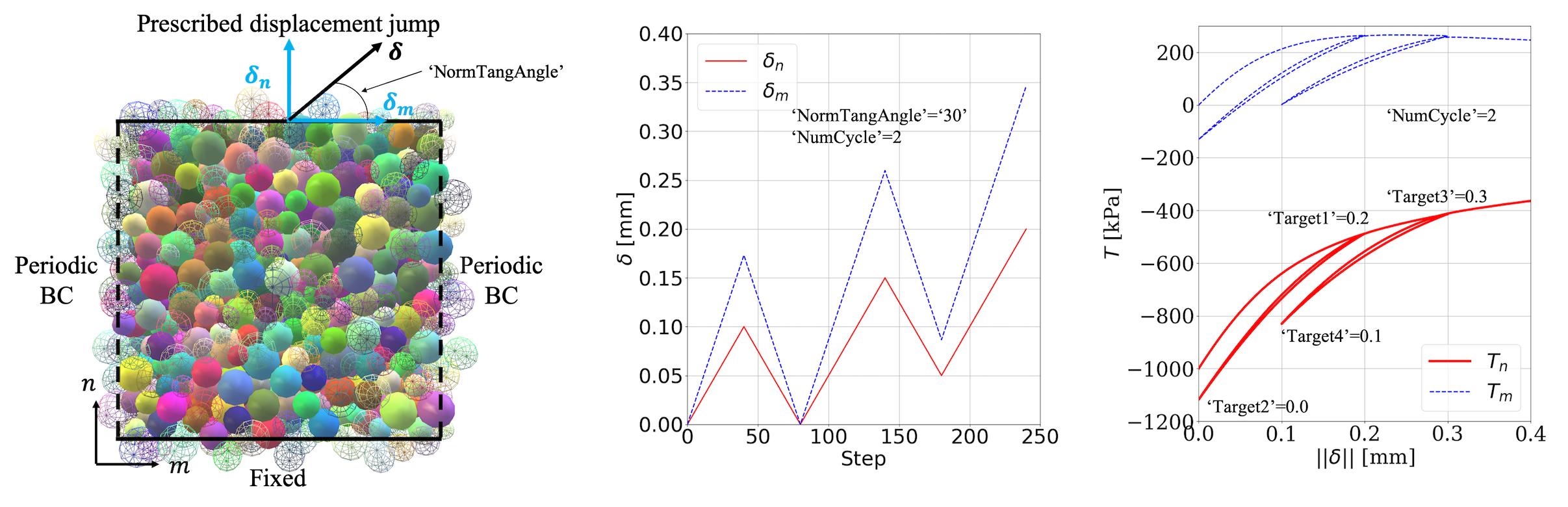}
	\caption{Experimental settings of displacement-driven mixed-mode shear tests on numerical specimen of granular interfaces using DEM (discrete element methods). The test conditions for the AI experimentalist are presented in Table \ref{tab:test_cond_TS}. As an example, the loading path and traction in normal and tangential directions obtained from a test designed by the decision tree path '30' $\rightarrow$ '2' $\rightarrow$ '0.2' $\rightarrow$ '0.0' $\rightarrow$ '0.3' $\rightarrow$ '0.1' are presented. 
	Regardless of the designed loading-unloading cycles, the final displacement jump reaches the magnitude of 0.4 mm.
	}
	\label{fig:TS_Experiment_Illust}
\end{figure}

The corresponding decision tree is shown in Fig. \ref{fig:Decision_Tree_TS}. 
The total number of experimental designs (which equals to the number of leaf nodes in the tree) is $N_{test}=228$. 
Fig. \ref{fig:TS_Experiment_Illust} provides the experimental settings on DEM (discrete element methods) numerical specimens and data from one example of the experiments. 
The total number of experimental data combinations, for example, equals to $C_{228}^1 + C_{228}^2 + C_{228}^3 \approx 1.97e6$ when $N^{max}_{path}=3$. 
Such number is already impractical for human to find the optimal data sets for calibration and falsification by trial and error. 
For high efficiency, the decisions in performing experiments should be guided by experienced experts or, in this paper, reinforcement-learning-based AI.

\section{Multi-agent non-cooperative game for model calibration/falsification with adversarial attacks}
\label{sec:metamodelinggame}
This section presents the design of a data acquisition game for both AI experimentalists (protagonist and adversary) to play, based on the decision trees defined in Section \ref{sec:decision_tree_graph} involving the common actions in testing the mechanical properties of geomaterials. 
The goal of this game is to enable  the protagonist agent to find the optimal design of experiments  that best calibrate a constitutive law, while having the adversary agent designs a counterpart set of experiments that expose the weakness of the models in the same decision tree that represents the application range. 
For simplicity, we assume that all experiments conducted by both agents are fully reproducible and free of noise. 
We will introduce a more comprehensive treatment for more general situations in which the bias and sensitivity of the data as well as the possibility of erroneous and even fabricated data are considered in future. 
Such a treatment is, nevertheless out of the scope of this work. 

Multi-agent multi-objective Markov games \citep{littman1994markov} have been widely studied and applied in robotics \citep{pinto2017robust}, traffic control \citep{wiering2000multi}, social dilemmas \citep{leibo2017multi}, etc. 
In our previous work, \citet{wang2019cooperative}, our focus was on designing agents that have different actions and states but share the same goal. 
In this work, our new innovation is on designing a zero-sum game in which the agents are competing against each other for a pair of opposite goals. 
While the reinforcement learning may lead to improved game play through repeated trial-and-error,  the non-cooperative nature of this new game will force the protagonist to act differently in response to the weakness exposed by the adversary. This treatment therefore may lead to a more robust and regularized model. 
In this work, the protagonist and the adversary are given the exact same action space mathematically characterized as a decision tree. 
While a non-cooperative game with non-symmetric action spaces can enjoy great performance  as demonstrated
in some of the OpenAI systems \citep{pinto2017robust}, such an extension is out of the scope of this study and will be considered in the future.

\subsection{Non-cooperative calibration/falsification game involving protagonist and adversary} 
We follow the general setup in \citet{pinto2017robust} to create a two-player Markov game with competing objectives to calibrate and falsify a constitutive model. 
Both calibration and falsification are idealized as  procedures that involves sequences of actions taken to maximize (in the case of calibration) and minimize (in the case of the falsification) a metric that assesses the prediction accuracy and robustness. 

Consider the Markov decision process (MDP) in this game expressed as a tuple 
$(\mathcal{S}, \mathcal{A}_{p}, \mathcal{A}_{a}, \mathcal{P}, r_p, r_a, s_{0})$ where $\mathcal{S}$ is the set of game states and $s_0$ is the initial state distribution.  $\mathcal{A}_{p}$ is the set of actions taken by the protagonist in charge of generating the experimental data to calibrate a given material model. 
$\mathcal{A}_{a}$ is the set of actions taken by the adversary in charge of falsifying the material model. 
$\mathcal{P} : \mathcal{S} \times \mathcal{A}_{p} \times \mathcal{A}_{a} \times \mathcal{S} \rightarrow \mathbb{R}$ is the transition probability density. 
$r_p : \mathcal{S} \times \mathcal{A}_{p} \times \mathcal{A}_{a} \rightarrow \mathbb{R}$ and $r_a : \mathcal{S} \times \mathcal{A}_{p} \times \mathcal{A}_{a} \rightarrow \mathbb{R}$ are the rewards of protagonist and adversary, respectively. 
If $r_p = r_a$, the game is fully cooperative. If $r_p = -r_a$, the game is zero-sum competitive. 
At the current state $s$ of the game, if the protagonist is taking action $a_p$ sampled from a stochastic policy $\mu_p$ and the adversary is taking action $a_a$ sampled from a stochastic policy $\mu_a$, the reward functions are $r_p^{\mu_p, \mu_a} = E_{a_p \sim \mu_p(\cdot|s), a_a \sim \mu_a(\cdot|s)}[r_p(s, a_p, a_a)]$ and $r_a^{\mu_p, \mu_a} = E_{a_p \sim \mu_p(\cdot|s), a_a \sim \mu_a(\cdot|s)}[r_a(s, a_p, a_a)]$.

In this work, all the possible actions of the protagonist and the adversary agent are mathematically represented by decision trees (Section \ref{sec:decision_tree_graph}). 
The protagonist first selects one or more paths in its own tree which provide the detailed experimental setups to generate calibration data for the material model, then the adversary selects one or more paths in its own tree (identical to the protagonist's tree) to generate test data for the calibrated model, aiming to find the worst prediction scenarios. 
The rewards are based on the prediction accuracy measures $\text{SCORE}$ of the constitutive model against data. 
This measure of 'win' or 'lose' is only available when the game is terminated, similar to Chess and Go \citep{silver2017mastering, silver2017masteringchess}, thus the final rewards are back-propagated to inform all intermediate rewards $r_p(s, a_p, a_a)$ and $r_a(s, a_p, a_a)$. 
$r_p$ is defined to encourage the increase of $\text{SCORE}$ of model calibrations, while $r_a$ is defined to favor the decrease of $\text{SCORE}$ of forward predictions. 
In this setting, the game is non-cooperative, and generally not zero-sum.

\subsection{Components of the game for the experimentalist agents}\label{sec:game}
\begin{figure}[h!]\center
	\includegraphics[width=0.9\textwidth]{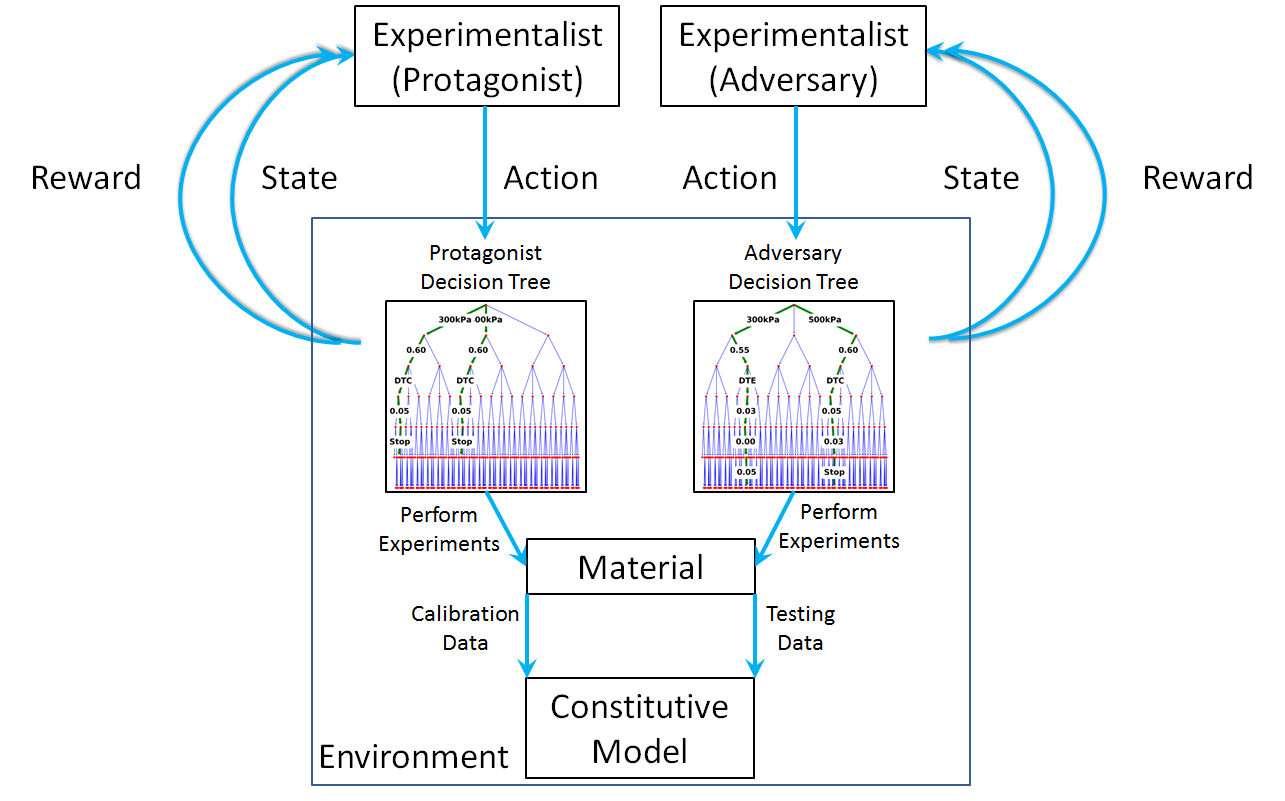}
	\caption{Key ingredients (environment, agents, states, actions, rules, and rewards) of the two-player non-cooperative agent-environment interactive system (game) for the experimentalist agents.}
	\label{fig:game_ingredients}
\end{figure}

The agent-environment interactive system (game) for the experimentalist agents consists of the game environment, game states, game actions, game rules, and game rewards \citep{bonabeau2002agent, wang2019meta} (Fig. \ref{fig:game_ingredients}). 
These key ingredients are detailed as follows. 

\textit{Game Environment} consists of the geomaterial samples, the constitutive model for performance evaluation, and the experimental decision trees. 
The samples in this game are representative volume elements (RVEs) of virtual granular assemblies modeled by the discrete element method (DEM) (e.g., Fig. \ref{fig:TTC_Experiment_Illust}, Fig. \ref{fig:TS_Experiment_Illust}). 
The preparation of such DEM RVEs are detailed in the numerical examples. 
The constitutive model can be given by the modeler agent in a meta-modeling game \citep{wang2019meta, wang2019cooperative}. 
In this paper, we focus on the interactive learning of data acquisition strategies for a certain constitutive model of interest. 
Three-agent protagonist-modeler-adversary reinforcement learning games is out of the scope of the current study. 
The protagonist and adversary agents determine the experiments on the RVEs in order to collect data for model parameter identification and testing the forward prediction accuracy of the constitutive model, respectively, via taking paths in their own decision trees (e.g., Fig. \ref{fig:decision_tree_example}, Fig. \ref{fig:decision_tree_experimentation_bulk}, Fig. \ref{fig:Decision_Tree_TS}). 

\textit{Game State} For the convenience of deep reinforcement learning using policy/value neural networks, we use a 2D array $s_{(2)}$ to concisely represent the paths that the protagonist or adversary has selected in the experimental decision tree. 
The mapping from the set of the 2D arrays to the set of path combinations in the decision tree is injective. 
The array has a row size of $N^{max}_{path}$ and a column size of $N_{\set{TC}}$. 
Each row represents one path in the decision tree from the root node to a leaf node, i.e., a complete design of one experiment. 
The number of allowed experiments is restricted by the row size $N^{max}_{path}$, which is defined by the user. 
Each array entry in the $N_{\set{TC}}$ columns represents the selected decision label of each test condition in $\set{TC}$. 
The entry $a$ (integer) in the $j$th row and $i$th column indicates that the $a$th decision label in the set $\set{TC}_i$ is selected for the $j$th experiment. 
Before the decision tree selections, the agent first decide a 1D array $s_{(1)}$ of size $N^{max}_{path}$, with its $k$th entry indicating whether the agent decides to take a new path in the decision tree (perform an another experiment) after the current $k$th experiment is done. A value of 1 indicates continuation and 2 indicates stop. 
The total state $s$ of the game combines $s_{(1)}$ and $s_{(2)}$, with $s_{(2)}$ flattened to a 1D array of size $N^{max}_{path}*N_{\set{TC}}$ and then input into the policy/value neural networks for policy evaluations. 
Initially, all entries in the arrays are 0, indicating no decisions has been made. 

\textit{Game Action} The AI agent works on the game state arrays by changing the initial zero entries into integers representing the decision labels. 
The agent firstly selects 1 for continuation or selects 2 for stop in $s_{(1)}$, in the left-to-right order. 
The agent then works on $s_{(2)}$ in the left-to-right then top-to-bottom order. 
Suppose that the first zero element of the current state array $s_{(2)}$ is in the $j$th row and $i$th column, the agent will select an integer $1 \leq a \leq m_i$ (number of choices) to choose a decision label in $\set{TC}_i$. 
The size of the action space is $N_{action} = \max_{i\in[1,N_{\set{TC}}]} m_i$. 

\begin{figure}[h!]\center
	\includegraphics[width=0.8\textwidth]{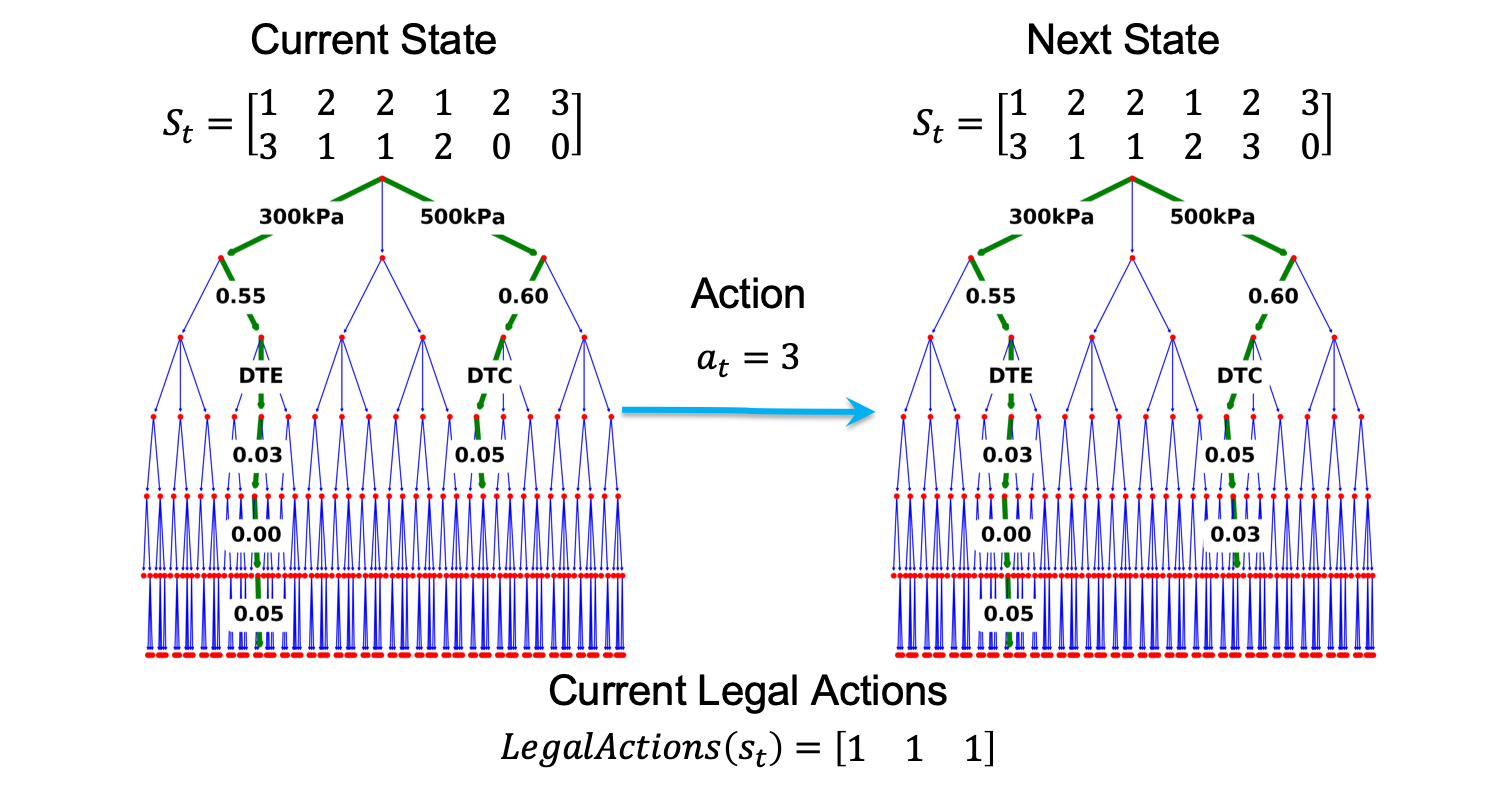}
	\caption{Example of the current $s_t$ and next $s_{t+1}$ game states describing the selected edges in the decision tree, action by the agent $a_t$ to "advance" in the decision tree, and the legal actions at the current state, with $N^{max}_{path}=2$.}
	\label{fig:game_state_action_example}
\end{figure}

\textit{Game Rule}
The AI agents are restricted to follow existing edges in the constructed decision tree, which has already incorporated decision limitations such as the choices of loading/unloading/reloading strain targets. 
The game rules are reflected by a list of $N_{action}$ binaries $LegalActions(s) = [ii_{1}, ii_{2}, ..., ii_{N_{action}}]$ at the current state $s$. 
If the $a$th decision is allowed, the $a$th entry is 1. 
Otherwise, the entry is 0. 
Figure \ref{fig:game_state_action_example} provides an example of the mathematical representations of the game states, actions and rules of the decision tree game. 

\textit{Game Reward}
The rewards from the game environment to the experimentalists should consider the performance of a given constitutive model on calibration data and testing data. 
After the decision of experiments by the protagonist, these experiments are performed on material samples to collect data. 
Then the constitutive model is calibrated with these data, and the accuracy is evaluated by a model score $\text{SCORE}_{\text{protagonist}}$. 
After the decision of experiments by the adversary, the calibrated constitutive model gives forward predictions on these testing data. 
The accuracy is evaluated by a model score $\text{SCORE}_{\text{adversary}}$. 
$+\text{SCORE}_{\text{protagonist}}$ is returned to the protagonist to inform its game reward, while $-\text{SCORE}_{\text{adversary}}$ is returned to the adversary. 
This adversary attack reward system is the key to ensure that the protagonist generates calibration data to maximize the prediction strength of the constitutive model, while the adversary tries to explore the weakness of the model.

\subsection{Evaluation of model scores and game rewards}
The accuracy of model calibrations and forward predictions are quantified by calculating the discrepancy between the vector of data points $[\overline{Y}^{\text{data}}_i]_{i=1}^{N_{data}}$ and the vector of predicted values $[\overline{Y}^{\text{model}}_i]_{i=1}^{N_{data}}$ under the same experimental conditions. 
For both data points and predictions, $\overline{Y}_i = \mathcal{S}_{j}(Y_i^j)$, where $Y_i^j$ is the data that falls into the $j$th category of output features (quantities of interest, such as deviatoric stress $q$ and void ratio $e$). 
$\mathcal{S}_{j}$ is the scaling operator (standardization, min-max scaling, ...) for the $j$th output feature. 

The predictions $[\overline{Y}^{\text{model}}_i]_{i=1}^{N_{data}}$ come from a given constitutive model that is calibrated with data generated by the protagonist. 
In this work, for elasto-plastic models, the nonlinear least-squares solver "NL2SOL" in Dakota software \citep{adams2014dakota} is used to find the optimal parameter values. The initial guess, upper and lower bounds of each parameter are given by domain experts' knowledge and preliminary estimations. 
For models of artificial neural networks, parameters are tuned by backpropagation algorithms using Tensorflow \citep{tensorflow2015-whitepaper}. 
In both cases, the optimal material parameters minimize the scaled mean squared error objective function
\begin{equation}
{\rm scaled\;MSE}\,=\,\frac{1}{N_{data}}\sum_{i=1}^{N_{data}} (\overline{Y}^{\text{model}}_i\,-\,\overline{Y}^{\text{data}}_i)^2\, . 
\label{eq:scaledMSE}
\end{equation} 

The model scores measuring the prediction accuracy are based on the modified Nash-Sutcliffe efficiency index \citep{nash1970river, krause2005comparison},
\begin{equation}
E^j_{NS} = 1-\frac{\sum_{i=1}^{N_{data}}|\overline{Y}^{\text{data}}_i-\overline{Y}^{\text{model}}_i|^j}{\sum_{i=1}^{N_{data}}|\overline{Y}^{\text{data}}_i-\text{mean}({\overline{Y}^{\text{data}}})|^j} %,\ E^j_{NS}
 \in (-\infty, 1.0].
\end{equation}
When $j=2$, it recovers the conventional Nash-Sutcliffe efficiency index. Here we adopt $j=1$, and 
\begin{equation}
\text{SCORE}_{\text{protagonist or adversary}} = 2 * \frac{\min (\max(E^1_{NS}, E^{min}_{NS}), E^{max}_{NS}) - 0.5*(E^{min}_{NS}+E^{max}_{NS})}{E^{max}_{NS}-E^{min}_{NS}},
\label{eq:score}
\end{equation}
where $E^{max}_{NS}$ and $E^{min}_{NS}$ are maximum and minimum cutoff values of the modified Nash-Sutcliffe efficiency index, $\text{SCORE} \in [-1.0, 1.0]$.

The game reward returned to the protagonist can consider both the calibration accuracy and the forward prediction accuracy, by including an exponential decay term: 
\begin{equation}
\text{Reward}_{\text{protagonist}} = -1 + (\text{SCORE}_{\text{protagonist}}+1)*\exp{[-\alpha_{\text{SCORE}} * \max (E^{min}_{NS}-\min (\{E^1_{NS}\}),0)]}.
\label{eq:reward_protagonist}
\end{equation}
where $\min (\{E^1_{NS}\})$ is the minimum N-S index observed in the gameplay history, $\alpha_{\text{SCORE}}$ is a user-defined decay coefficient. When $\min (\{E^1_{NS}\}) < E^{min}_{NS}$, the decay term starts to drop the reward of the protagonist, otherwise $\text{Reward}_{\text{protagonist}} = +\text{SCORE}_{\text{protagonist}}$.
On the other hand, the game reward returned to the adversary is
\begin{equation}
\text{Reward}_{\text{adversary}} = -\text{SCORE}_{\text{adversary}}.
\label{eq:reward_adversary}
\end{equation}
Since the adversary is rewarded at the expense of the protagonist's failure, it is progressively learning to create increasingly devastating experimental design to falsify the model, thus forcing the protagonist to calibrate material models that are robust to any disturbances created by the adversary. 
In this work, we refer to the move of  the protagonist as calibration or defense, while the move of the adversary as falsification or attack.

\section{Parallel reinforcement learning algorithm for the non-cooperative experimental/adversarial game}\label{sec:gamedrl}
In the language of game theory, the meta-modeling game defined in the previous section is categorized as non-cooperative, asymmetric (the payoff of a particular strategy depends on whether protagonist or adversary is playing), non-zero-sum, sequential (the adversary is aware of the protagonist's strategy in order to attack accordingly), imperfect information (the protagonist does not know how the adversary will attack). 
Let ($\mathcal{M}$, $\mathcal{R}$) be a representation of this two-player (denoted by subscripts $p$ and $a$) non-cooperative game, with $\mathcal{M} = \mathcal{M}_p \times \mathcal{M}_a$ the set of strategy profiles. $\mathcal{R}(\mu) = (\mathcal{R}_p(\mu), \mathcal{R}_a(\mu))$ is the payoff (final reward) function evaluated at a strategy profile $\mu = (\mu_p, \mu_a) \in \mathcal{M}$. 
A strategy profile $\mu^*$ is a Nash equilibrium if no unilateral change in $\mu^*$ by any player is more profitable for that player, i.e.,
\begin{equation}
\left\{
\begin{aligned}
\forall \mu_p \in \mathcal{M}_p,\ \mathcal{R}_p( (\mu_p^*, \mu_a^*) ) \geq \mathcal{R}_p( (\mu_p, \mu_a^*) )\\
\forall \mu_a \in \mathcal{M}_a,\ \mathcal{R}_a( (\mu_p^*, \mu_a^*) ) \geq \mathcal{R}_a( (\mu_p^*, \mu_a) )
\end{aligned}
\right. .
\end{equation}
The existence of at least one such equilibrium point is proven by \citet{nash1950equilibrium}.

\begin{figure}[h!]\center
	\includegraphics[width=0.95\textwidth]{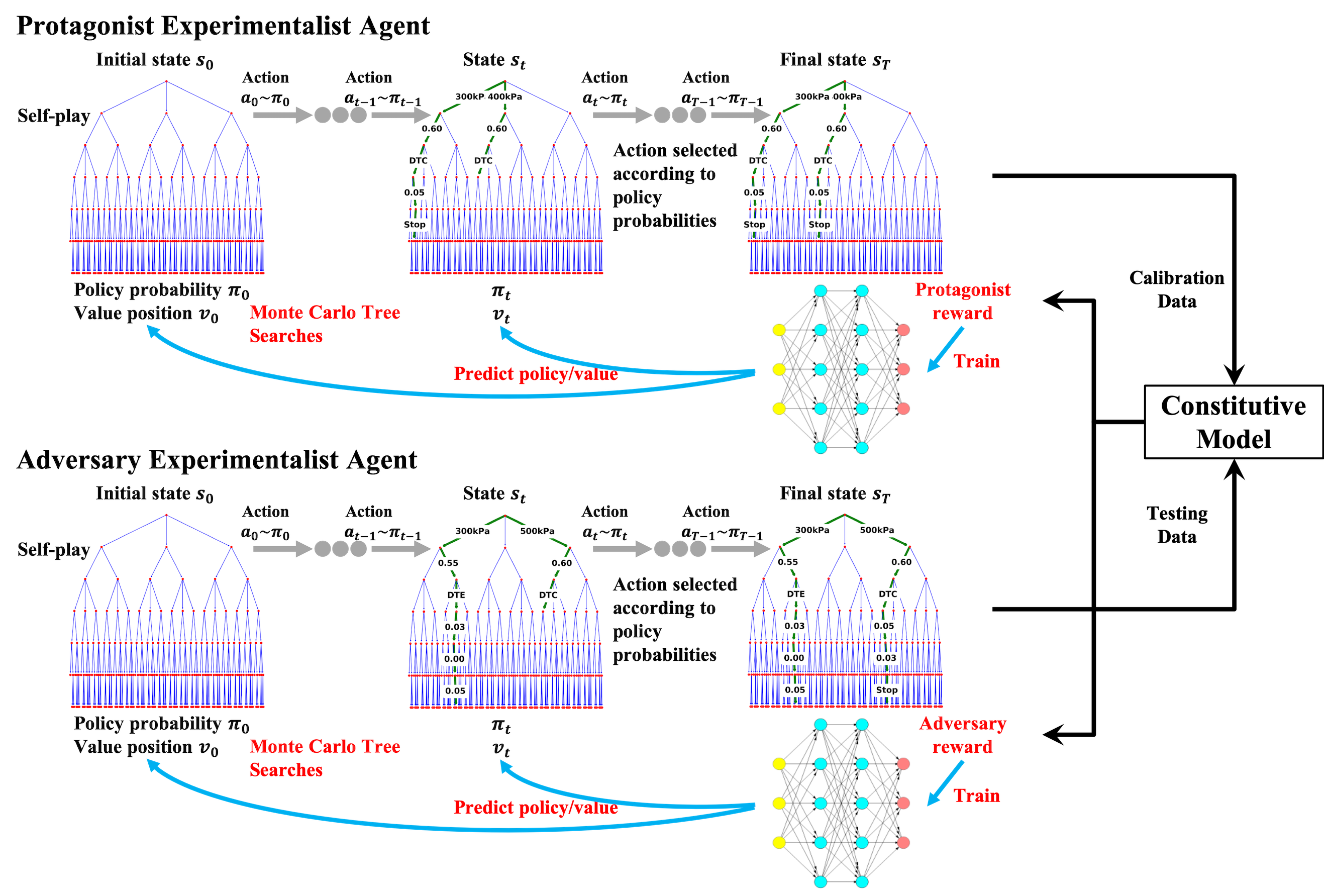}
	\caption{Two-player adversarial reinforcement learning for generating optimal strategies to automate the calibration and falsification of a constitutive model.}
	\label{fig:selfplay_learn}
\end{figure}

Solving the optimization problem directly to find the Nash equilibria strategies for this complex game is prohibitive \citep{pmlr-v37-perolat15}. 
Instead, deep reinforcement learning (DRL) algorithm is employed. In this technique the strategy of each player ($\mu_p$ or $\mu_a$) is parameterized by an artificial neural network $f_{\theta}$ that takes in the description of the current state $s$ of the game and outputs a policy vector $\vec{p}$ with each component representing the probability of taking actions from state $s$, as well as a scalar $v$ for estimating the expected reward of the game from state $s$, i.e., 
\begin{equation}
    (\vec{p}, v) = f_{\theta}(s).
\end{equation}
These policy/value networks provide guidance in learning optimal strategies of both protagonist and adversary in order to maximize the final game rewards. 
The learning is completely free of human interventions after the complete game settings. 
This tactic is considered one of the key ideas leading to the major breakthrough in AI playing the game of Go (AlphaGo Zero) \citep{silver2017mastering}, Chess and shogi (Alpha Zero) \citep{silver2017masteringchess} and many other games. 
In \citet{wang2019meta}, the key ingredients (policy/value network, upper confidence bound for Q-value, Monte Carlo Tree Search) of the DRL technique are detailed and applied to a meta-modeling game for modeler agent only, focusing on finding the optimal topology of physical relations from fixed training/testing datasets. 
Since DRL needs to figure out the optimal strategies for both agents, the algorithm is extended to multi-agent multi-objective DRL \citep{tan1993multi,foerster2016learning,tampuu2017multiagent}. 
The AI for protagonist and adversary are improved simultaneously during the self-plays of the entire meta-modeling game, according to the individual rewards they receive from the game environment and the communications between themselves (Figure \ref{fig:selfplay_learn}). 

The pseudocode of the reinforcement learning algorithm to play the non-cooperative game is presented in Algorithm \ref{mcts_algorithm}. 
This is an extension of the algorithm in \citep{wang2019meta}. 
As demonstrated in Algorithm \ref{mcts_algorithm},  each complete DRL procedure involves $numIters$ number of training iterations and one final iteration for generating the converged selected paths in decision trees.
Each iteration involves $numEpisodes$ number of game episodes that construct the training example set $trainExamples$ for the training of the policy/value networks $f_{\theta}^{\text{Protagonist}}$ and $f_{\theta}^{\text{Adversary}}$. 
For decision makings in each game episode, the action probabilities are estimated from $numMCTSSims$ runs of MCTS simulations. 

\begin{figure}[h!]\center
	\includegraphics[width=0.95\textwidth]{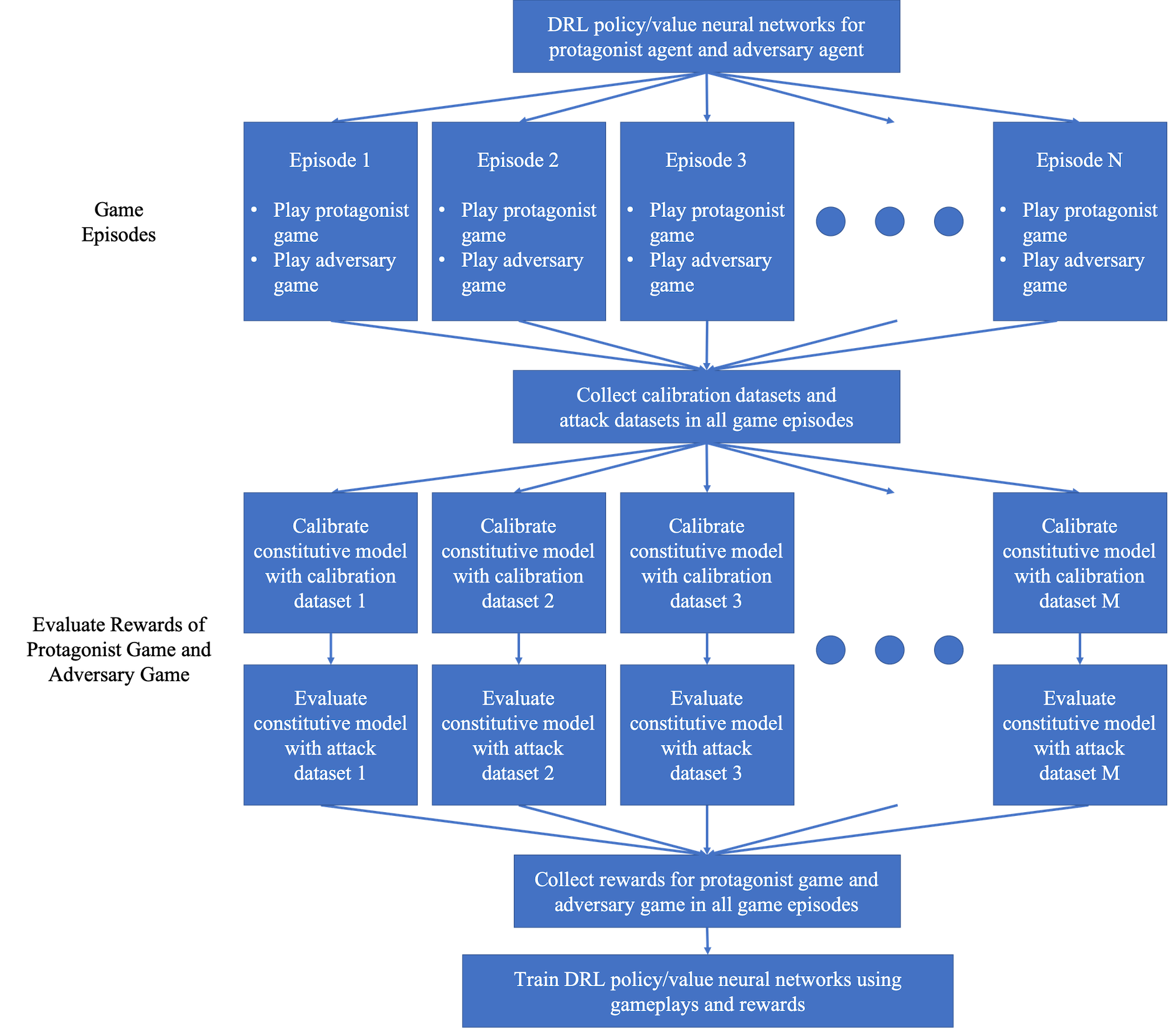}
	\caption{Workflow of parallel gameplays and reward evaluations in DRL.}
	\label{fig:parallel_DRL}
\end{figure}

The state values $v$ can be equal to the continuous reward functions (\ref{eq:reward_protagonist}), (\ref{eq:reward_adversary}) to train the policy/value neural networks. 
To further improve the convergence rate of the DRL algorithm, we propose an empirical method to train the networks with binary state values (1 or -1) post-processed from the reward values, which is similar to the concept of "win" (1) and "lose" (-1) in the game of Chess. 
Consider the set of rewards $\{ \text{Reward}_{\text{protagonist}} \}_{i}$ of the $numEpisodes$ games played in the $i$th DRL iteration. The maximum reward encountered from Iteration 0 to the current Iteration $k$, is $R^{max}_{p} = \max_{i \in [0,k]} ( \max(\{ \text{Reward}_{\text{protagonist}} \}_{i}))$. 
A minimum reward is chosen as $R^{min}_{p} = \max_{i \in [0,k]} ( \min(\{ \text{Reward}_{\text{protagonist}} \}_{i}))$. 
A reward range in Iteration $k$ is $R^{range}_{p} = R^{max}_{p} - R^{min}_{p}$. 
A strategy $\mu_p$ is considered as a "win" ($v=1$) when its reward $\text{Reward}_{\text{protagonist}} \geq R^{max}_{p} - R^{range}_{p}*\alpha_{range}$, while it is a "lose" ($v=-1$) when $\text{Reward}_{\text{protagonist}} < R^{max}_{p} - R^{range}_{p}*\alpha_{range}$. 
$\alpha_{range}$ is a user-defined coefficient which influences the degree of "exploration and exploitation" of the AI agents. 
Similarly, for the adversary agent, $R^{max}_{a_{\mu_p}} = \min_{i \in [0,k]} ( \max(\{ -\text{Reward}^{\mu_p}_{\text{adversary}} \}_{i}))$, $R^{min}_{a_{\mu_p}} = \min_{i \in [0,k]} ( \min(\{ -\text{Reward}^{\mu_p}_{\text{adversary}} \}_{i}))$ and $R^{range}_{a_{\mu_p}} = R^{max}_{a_{\mu_p}} - R^{min}_{a_{\mu_p}}$ are collected for each protagonist strategy $\mu_p$. 
Then an attack strategy $\mu_a$ corresponding to $\mu_p$ is considered as a "win" ($v=1$) when its reward $-\text{Reward}^{\mu_p}_{\text{adversary}} \leq R^{min}_{a_{\mu_p}} + R^{range}_{a_{\mu_p}}*\alpha_{range}$, while it is a "lose" ($v=-1$) when $-\text{Reward}^{\mu_p}_{\text{adversary}} > R^{min}_{a_{\mu_p}} + R^{range}_{a_{\mu_p}}*\alpha_{range}$. 
The training examples for the policy/value neural networks are limited to the gameplays in the DRL iterations $i \in [\max(k-i_{lookback}, 0), k]$, where $i_{lookback}$ is a user-defined hyperparameter controlling the degree of "forget" of the AI agents.

\begin{algorithm}
	\caption{Self-play reinforcement learning of the non-cooperative meta-modeling game}\label{mcts_algorithm}
	\begin{algorithmic}[1]
		\Require The definitions of the non-cooperative meta-modeling game: game environment, game states, game actions, game rules, game rewards (Sections \ref{sec:metamodelinggame}).
		\State Initialize the policy/value networks $f_{\theta}^{\text{Protagonist}}$ and $f_{\theta}^{\text{Adversary}}$. For fresh learning, the networks are randomly initialized. For transfer learning, load pre-trained networks instead.
		\State Initialize empty sets of the training examples for both protagonist and adversary $trainExamples^{\text{Protagonist}} \leftarrow []$, $trainExamples^{\text{adversary}} \leftarrow []$.
		\For{i in [0,..., $numIters-1$]}
		\For{j in [0,..., $numEpisodes-1$]}
		\State Initialize the starting game state $s$.
		\For{$player$ in [\text{Protagonist},\ \text{Adversary}]}
		\State Initialize empty tree of the Monte Carlo Tree search (MCTS), set the temperature parameter $\tau_{train}$ for "exploration and exploitation".
		\While{True}
		\State Check for all legal actions at current state $s$ according to the game rules.
		\State Get the action probabilities $\pi(s,\cdot)$ for all legal actions by performing $numMCTSSims$ times of MCTS simulations.
		\State Sample action $a$ from the probabilities $\pi(s,\cdot)$
		\State Modify the current game state to a new state $s$ by taking the action $a$.
		\If{$s$ is the end state of the game of $player$}
		\State Evaluate the score of the selected paths in the decision tree.
		\State Evaluate the reward $r$ of this gameplay according to the score.
		\State \textbf{Break}.
		\EndIf
		\EndWhile
		\State Append the gameplay history $[s,a,\pi(s,\cdot),r]$ to $trainExamples^{player}$.
		\EndFor
		\EndFor
		\State Train the policy/value networks $f_{\theta}^{\text{Protagonist}}$ and $f_{\theta}^{\text{Adversary}}$ with $trainExamples^{\text{Protagonist}}$ and $trainExamples^{\text{Adversary}}$.
		\EndFor
		
		\State Use the final trained networks $f_{\theta}^{\text{Protagonist}}$ and $f_{\theta}^{\text{Adversary}}$ in MCTS with temperature parameter $\tau_{test}$ for one more iteration of "competitive gameplays" to generate the final converged selected experiments.
		\State Exit
	\end{algorithmic}
\end{algorithm}

Another new contribution to the DRL framework is that we improve the computational efficiency of DRL by executing the mutually independent gameplays and reward evaluations in a parallel manner, instead of serial executions as in previous works \citep{wang2018multiscale, wang2019meta, wang2019cooperative}. We use the parallel python library "Ray" \citep{moritz2018ray} for its simplicity and speed in building and running distributed applications. 
The new workflow of parallel playing of game episodes in each training iteration for DRL is illustrated in Figure \ref{fig:parallel_DRL}.

\section{Automated calibration and falsification experiments} \label{sec:numericalExp}
We demonstrate the applications of the non-cooperative game for automated calibration and falsification on three types of constitutive models. 
The material samples are representative volume elements (RVEs) of densely-packed spherical DEM particles. 
The decision-tree-based experiments are performed via numerical simulations on these samples. 
The preparation and experiments of the samples are detailed in Appendix \ref{sec:dem_samples}. 
The three constitutive models studied in this paper are Drucker-Prager model \citep{tu2009return}, SANISAND model \citep{dafalias2004simple}, and data-driven traction-separation model \citep{wang2018multiscale}. 
Their formulations are detailed in Appendix \ref{sec:example_models}. 
The results shown in this section are representatives of the AI agents' performances, since the policy/value networks are randomly initialized and the MCTS simulations involve samplings from action probabilities. 
The gameplays during DRL iterations may vary, but similar convergence performances are expected for different executions of the algorithm. 
Furthermore, the material calibration procedures, such as the initial guesses in Dakota and the hyperparameters in training of the neural networks, may affect the game scores and the converged Nash equilibrium points. 
Finally, since simplifications and assumptions are involved in the DEM samples, their mechanical properties differ from real-world geo-materials. 
The conclusions of the three investigated constitutive models are only on these artificial and numerical samples. 
However, the same DRL algorithm is also applicable for real materials, when the actions of the AI experimentalists can be programmed in laboratory instruments.

The policy/value networks $f_{\theta}$ are deel neural network in charge of updating the Q table that determines the optimal strategies. The design of the policy/value networks are identical for both agents in this paper. 
Both of them consist of one input layer of the game state $s$, two densely connected hidden layers, and two output layers for the action probabilities $\vec{p}$ and the state value $v$, respectively. Each hidden layer contains 256 artificial neurons, followed by Batch Normalization, ReLU activation and Dropout. The dropout layer is a popular regularization mechanism designed to reduce overfitting and improve generalization errors in deep neural network (cf. \citet{srivastava2014dropout}). 
The dropout rate is 0.5 for the protagonist and 0.25 for the adversary. 
These different dropout rates are used such that the higher dropout rate for the protagonist will motivate the protagonist to calibrate the Drucker-Prager model with less generalization errors, while the smaller  dropout rate will help the adversary to find the hidden catastrophic failures in response to a large amount of protagonist's strategies. In addition, the non-cooperative game requires the hyperparameters listed in Table \ref{tab:hyperparamter} to configure the game. 

\begin{table}[h!]
\centering
\begin{tabular}{|c|c|c|}
\hline
Hyperparameters & Definition & Usage \\ \hline
$N^{max}_{path}$ & \shortstack{Maximum number of decision tree \\ paths chosen by the agents} & \shortstack{Define the dimension of \\ the game states} \\ \hline
$E^{max}_{NS}$ & \shortstack{Maximum cutoff value of \\ the modified Nash-Sutcliffe efficiency index} & See Eq. (\ref{eq:score}) \\ \hline
$E^{min}_{NS}$ & \shortstack{Minimum cutoff value of \\ the modified Nash-Sutcliffe efficiency index} & See Eq. (\ref{eq:score}) \\ \hline
$numIters$ & Number of training iterations & \shortstack{Define DRL iterations for \\ training policy/value networks} \\ \hline
$numEpisodes$ & \shortstack{Number of gameplay episodes \\ in each training iterations} & \shortstack{Define the amount of \\ collected gameplay evidences} \\ \hline
$numMCTSSims$ & \shortstack{Number of Monte Carlo Tree Search \\ simulations in each gameplay step} & \shortstack{Control the agents' estimations \\ of action probabilities} \\ \hline
$\alpha_{\text{SCORE}}$ & Decay coefficient for the protagonist's reward & See Eq. (\ref{eq:reward_protagonist}) \\ \hline
$\alpha_{range}$                     & \shortstack{Coefficient for determining \\ "win" or "lose" of a game episode} & \shortstack{Set the agents' balance between \\ "exploration and exploitation"}\\ \hline
$i_{lookback}$ & \shortstack{Number of gameplay iterations \\ for training of the policy/value networks} & \shortstack{Control the agents'  \\ "memory depth"}\\ \hline
$\tau_{train}$ & \shortstack{Temperature parameter for \\ training iterations} & \shortstack{Set the agents' balance between \\ "exploration and exploitation"}\\ \hline
$\tau_{test}$ & \shortstack{Temperature parameter for \\ competitative gameplays} & \shortstack{Set the agents' balance between\\ "exploration and exploitation"}\\
\hline 
\end{tabular}
\caption{Hyperparameters required to setup the non-coorperative game.}
\label{tab:hyperparamter}
\end{table}

\subsection{Experiment 1: Drucker-Prager model}
\label{subsec:dpmodel_example}
The two-player non-cooperative game is played by DRL-based AI experimentalists for Drucker-Prager model. 
The formulations of the model are detailed by Eq. (\ref{eq:dp_eq_1}), (\ref{eq:dp_eq_2}), (\ref{eq:dp_eq_3}). 
The initial guesses, upper and lower bounds of the material parameters for Dakota calibration are presented in Table \ref{tab:dp_parameters_guess}. 
The game settings are $N^{max}_{path}=5$ for both the protagonist and the adversary, $E^{max}_{NS}=1.0$, $E^{min}_{NS}=-1.0$. 
Hence the combination number of the selected experimental decision tree paths in this example is  $180!/(5! (180-5)!) \approx 1.5e9$ where 180 is the total number of leaves in the decision tree and 5 is the maximum number of paths chosen by either agent. 
The hyperparameters for the DRL algorithm used in this game are $numIters=10$, $numEpisodes=50$, $numMCTSSims=50$, $\alpha_{\text{SCORE}}=0.0$, $\alpha_{range} = 0.2$, $i_{lookback}=4$, $\tau_{train} = 1.0$, $\tau_{test} = 0.1$. 

\begin{figure}[h!]\center
	\subfigure[Protagonist]{
		\includegraphics[width=0.45\textwidth]{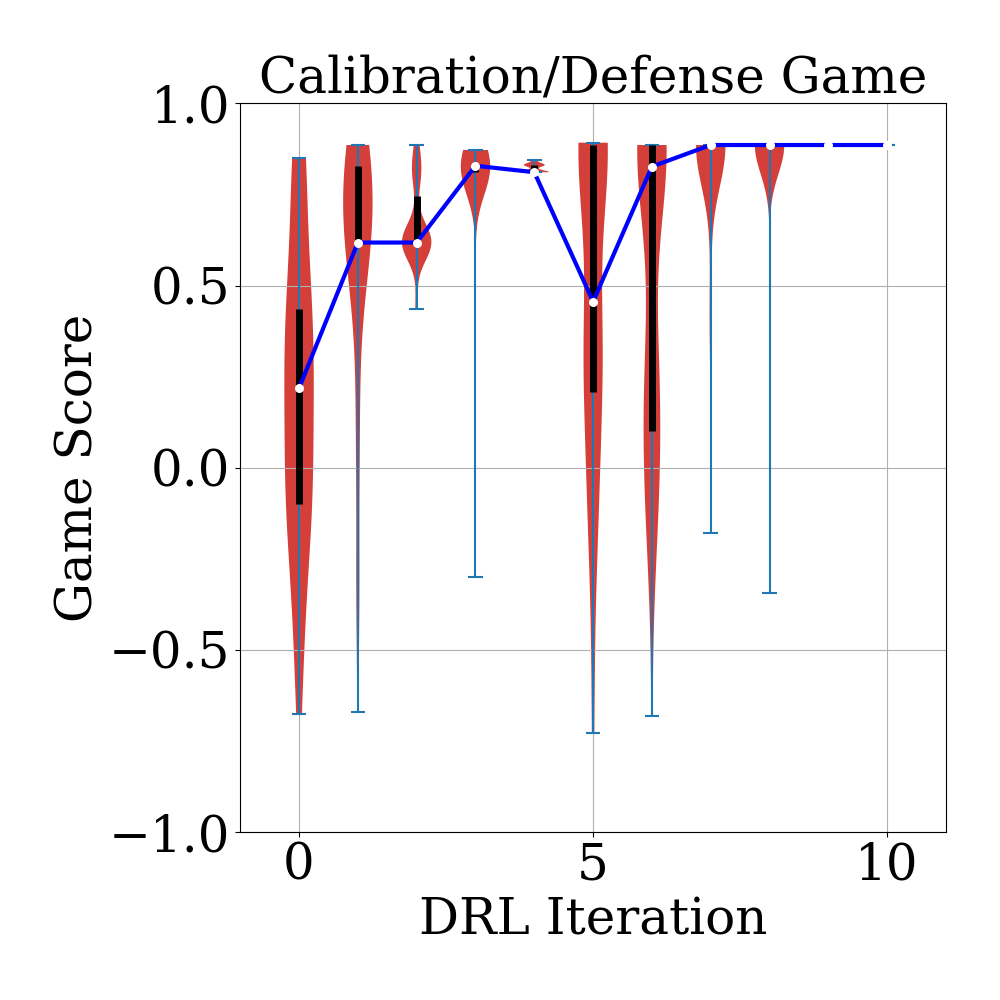}
	}
	\subfigure[Adversary]{
		\includegraphics[width=0.45\textwidth]{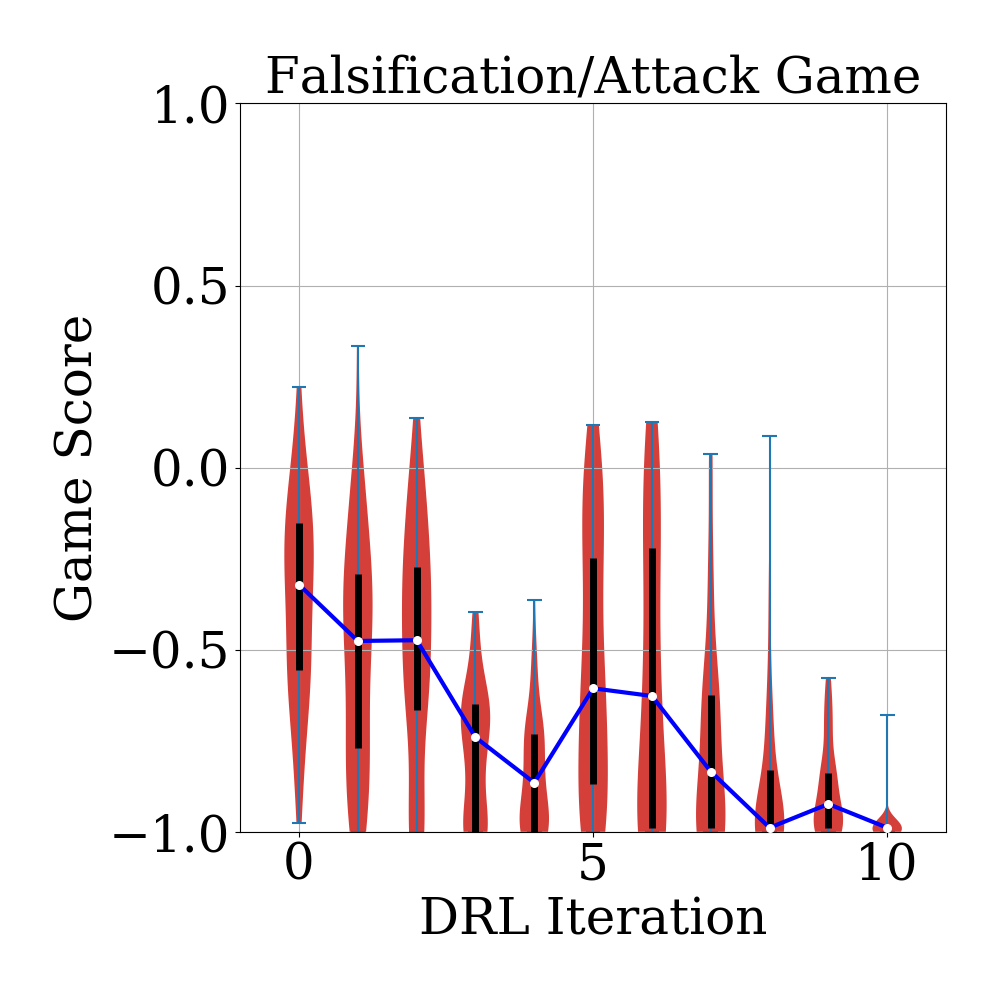}
	}
	\caption{Violin plots of the density distributions of game scores in each DRL iteration in Drucker-Prager model. The shaded area represents the density distribution of scores. The white point represents the median. The thick black bar represents the inter-quantile range between 25\% quantile and 75\% quantile. The maximum and minimum scores played in each iteration are marked.}
	\label{fig:DP_learn_violinplot}
\end{figure}

The statistics of the game scores played for the "Calibration/Defense" by the protagonist and the "Falsification/Attack" by the adversary during the DRL iterations are shown in Fig. \ref{fig:DP_learn_violinplot}. 
The AI agents only know the experimental decision tree and the rules of the two-player game without any prior knowledge on the strengths and weaknesses of the Drucker-Prager model. 
At the first DRL iteration, the agents play the game through trial and error guided by randomly initialized policy/value networks and MCTS. This lack of knowledge on proper gameplay strategies can be seen from the widely spread density distribution of game scores and the large inter-quantile range between 25\% and 75\% in both "Calibration/Defense" and "Falsification/Attack".  
In the subsequent iterations, the agents progressively understand the "winning strategies" via reinforcement learning on the gameplay histories and the associated game rewards, hence intend to play games with better outcomes. 
This is shown in the increase of the median of game scores by the protagonist and the decrease of the median by the adversary, and also the narrowing of inter-quantile ranges. 
In these intermediate training iterations, games can sometimes be played badly, since the agents are allowed to explore various game policies in order to avoid convergence to local extremum. The strengths of the AI agents after the 0th to 9th training iterations are tested by suppressing the "exploration plays”, and the ultimate game scores show outstanding performances.

\begin{figure}[h!]\center
	\subfigure[Iteration 0, Episode 10, \newline \hspace{\linewidth} Defense Game Score: 0.262]{
		\includegraphics[width=0.235\textwidth]{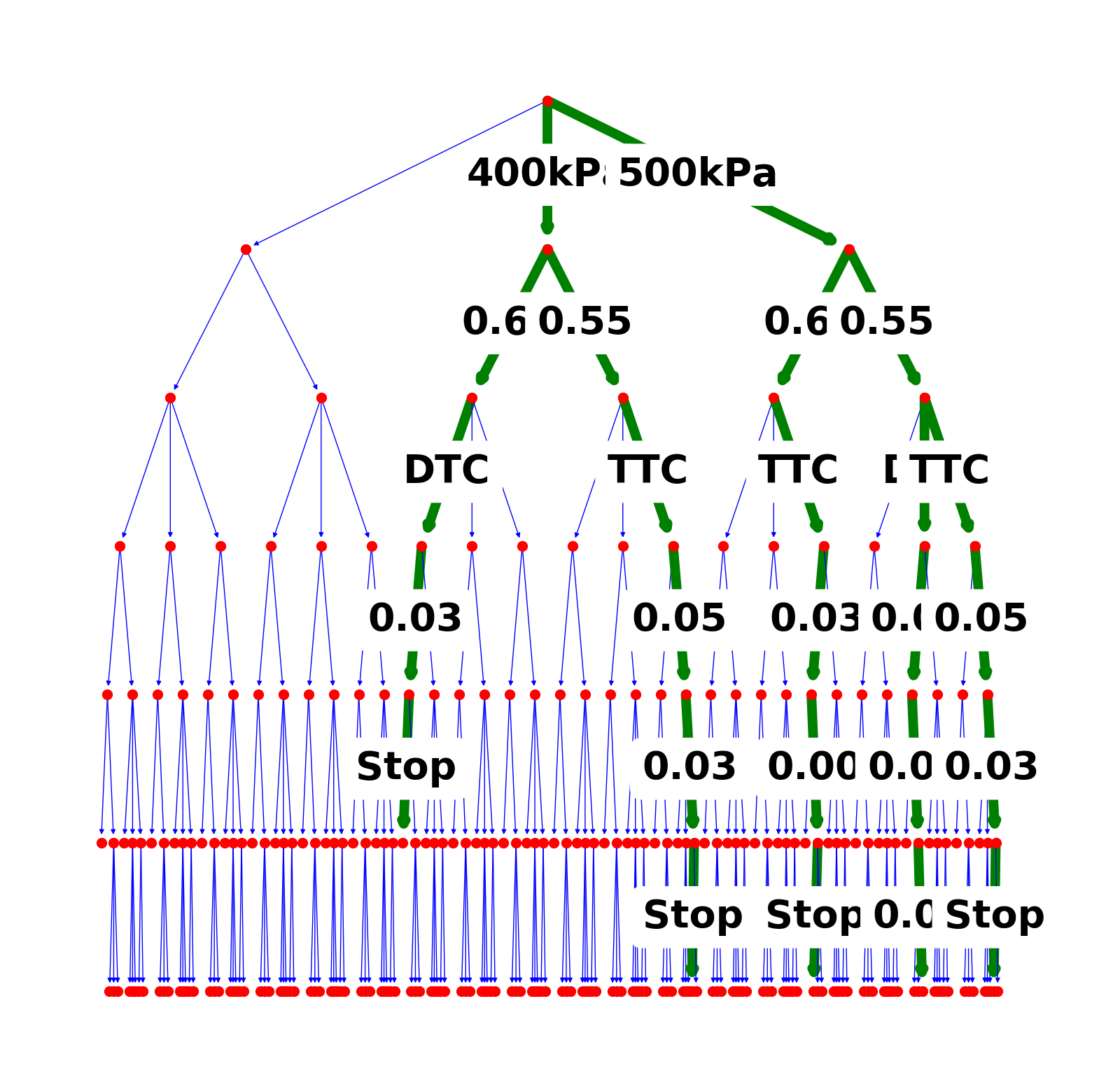}
	}
	\subfigure[Iteration 3, Episode 0, \newline \hspace{\linewidth} Defense Game Score: 0.811]{
		\includegraphics[width=0.235\textwidth]{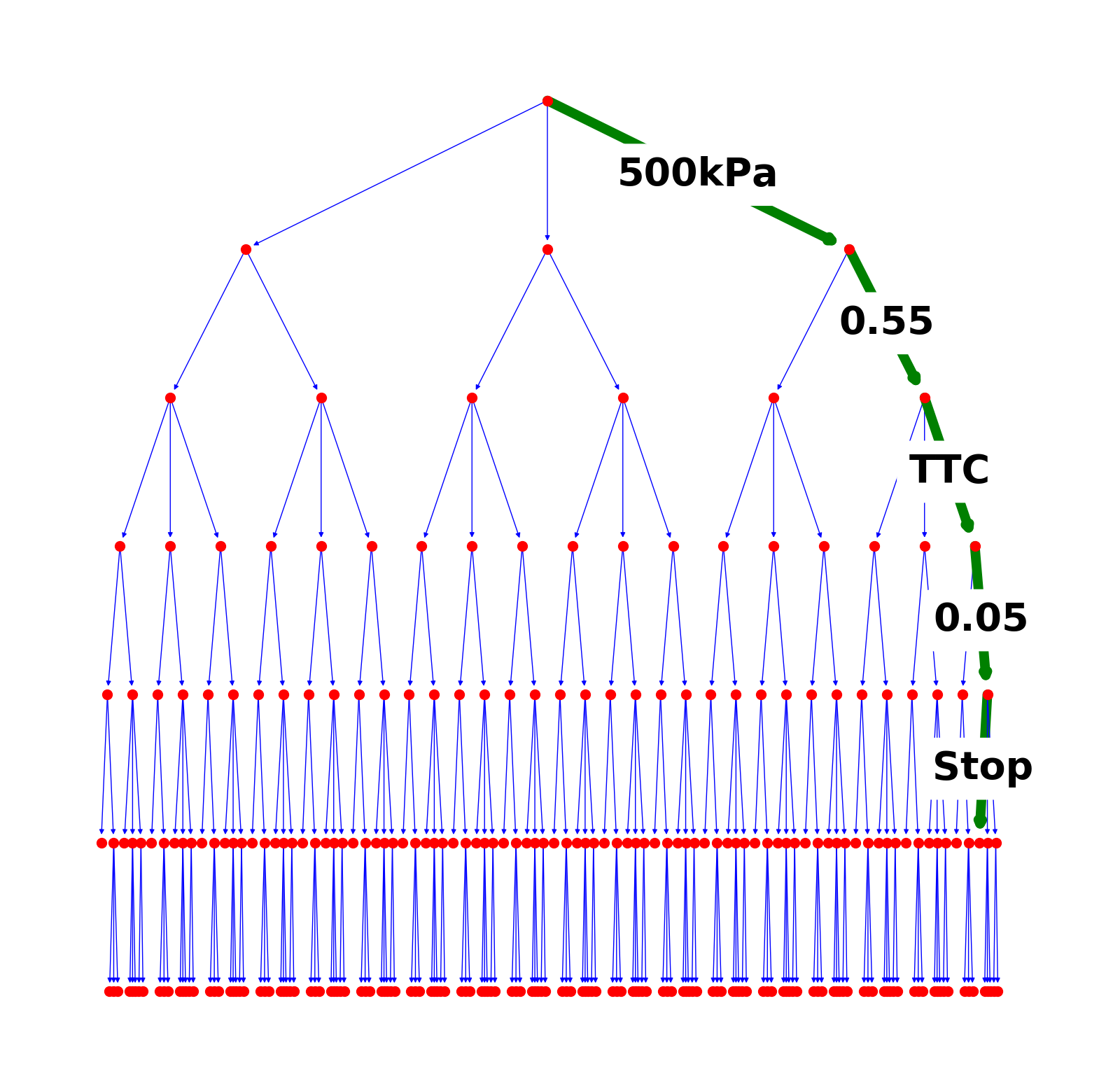}
	}
	\subfigure[Iteration 6, Episode 40, \newline \hspace{\linewidth} Defense Game Score: 0.699]{
		\includegraphics[width=0.235\textwidth]{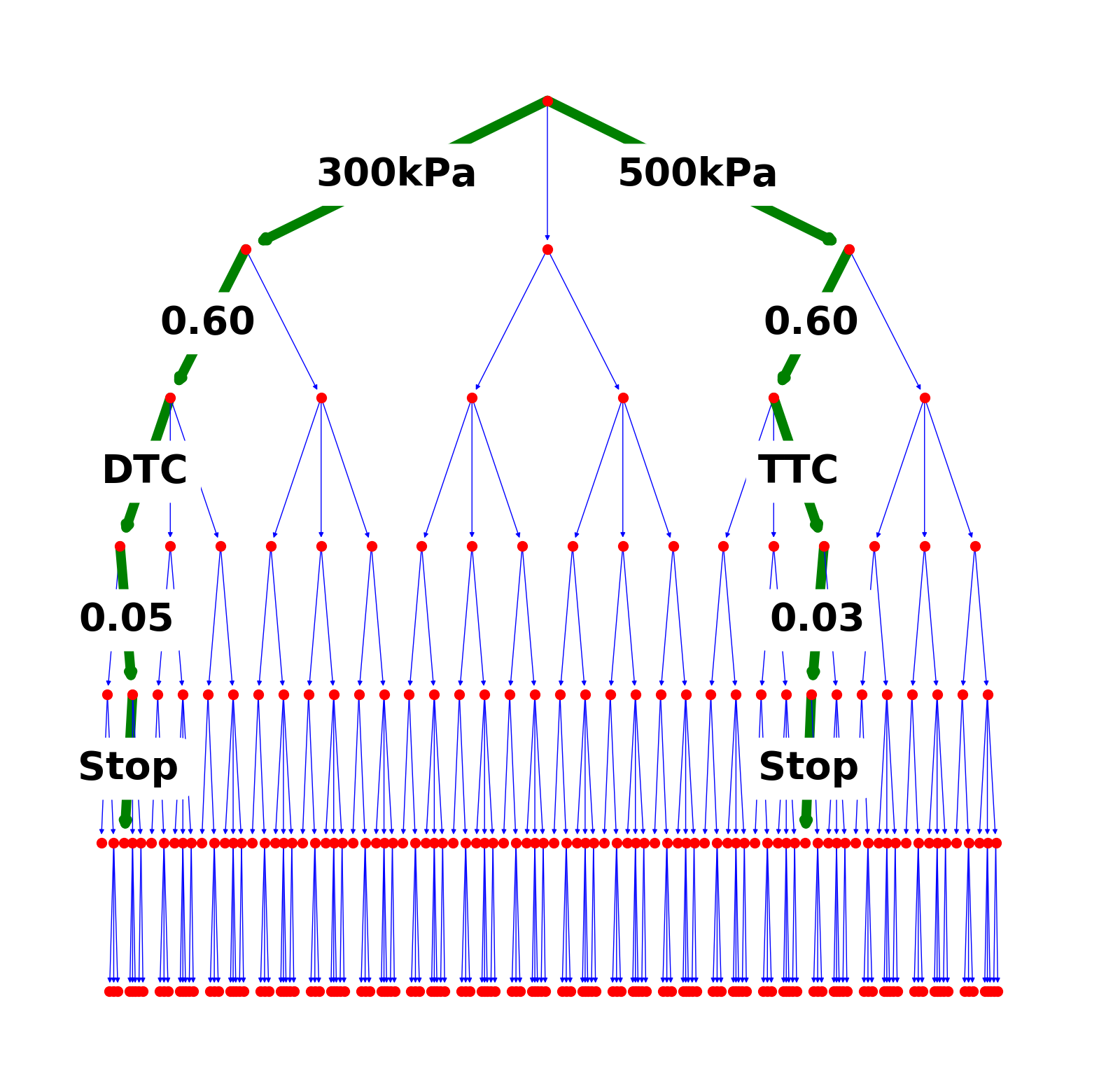}
	}
	\subfigure[Iteration 10, Episode 0, \newline \hspace{\linewidth} Defense Game Score: 0.886]{
		\includegraphics[width=0.235\textwidth]{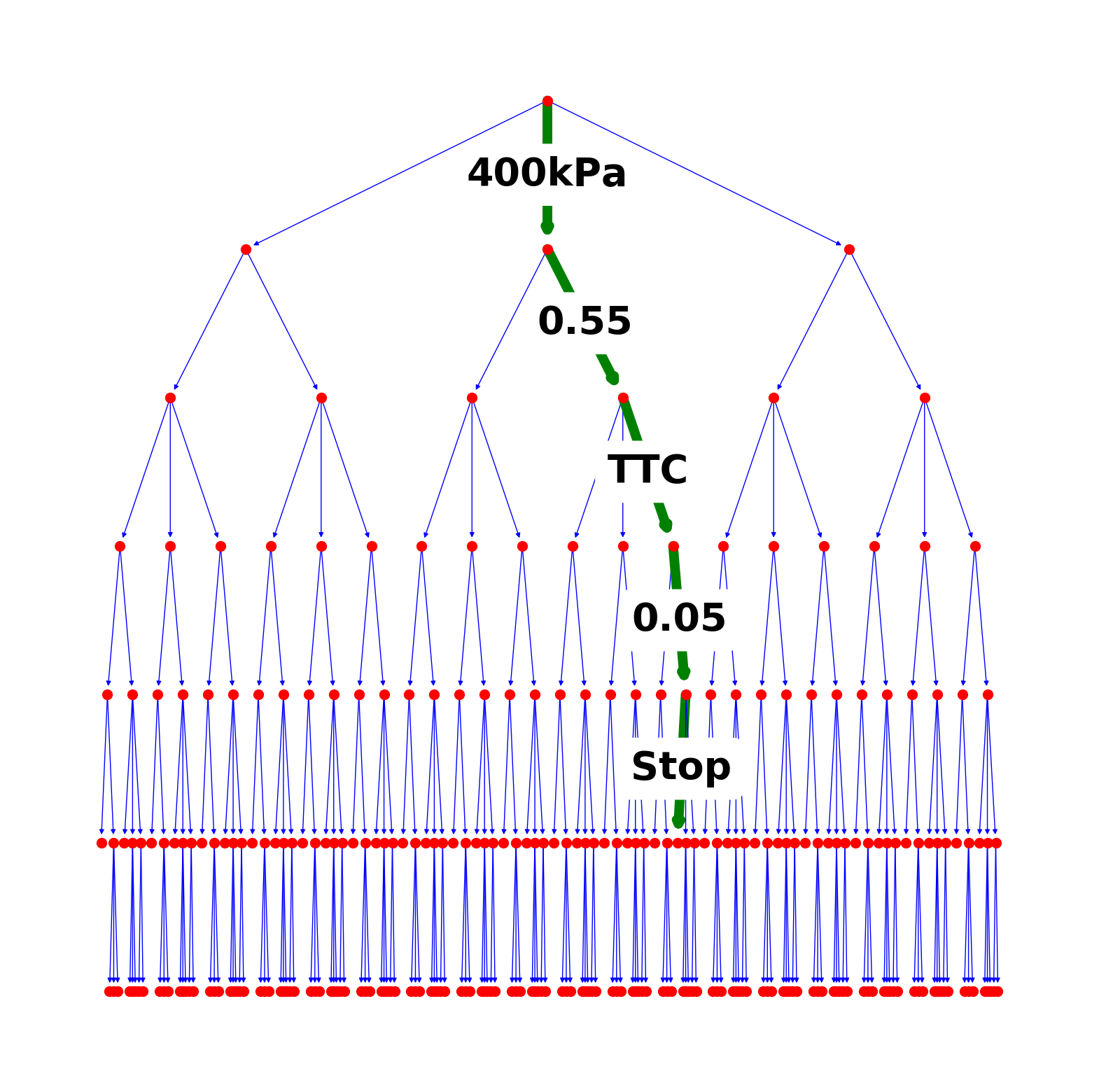}
	}
	\caption{Examples of paths (experiments) in the decision trees selected by the protagonist during the DRL training iterations for Drucker-Prager model.}
	\label{fig:DP_datagame_decisiontree}
\end{figure}

\begin{figure}[h!]\center
	\subfigure[Iteration 0, Episode 10, \newline \hspace{\linewidth} Attack Game Score: -0.151]{
		\includegraphics[width=0.235\textwidth]{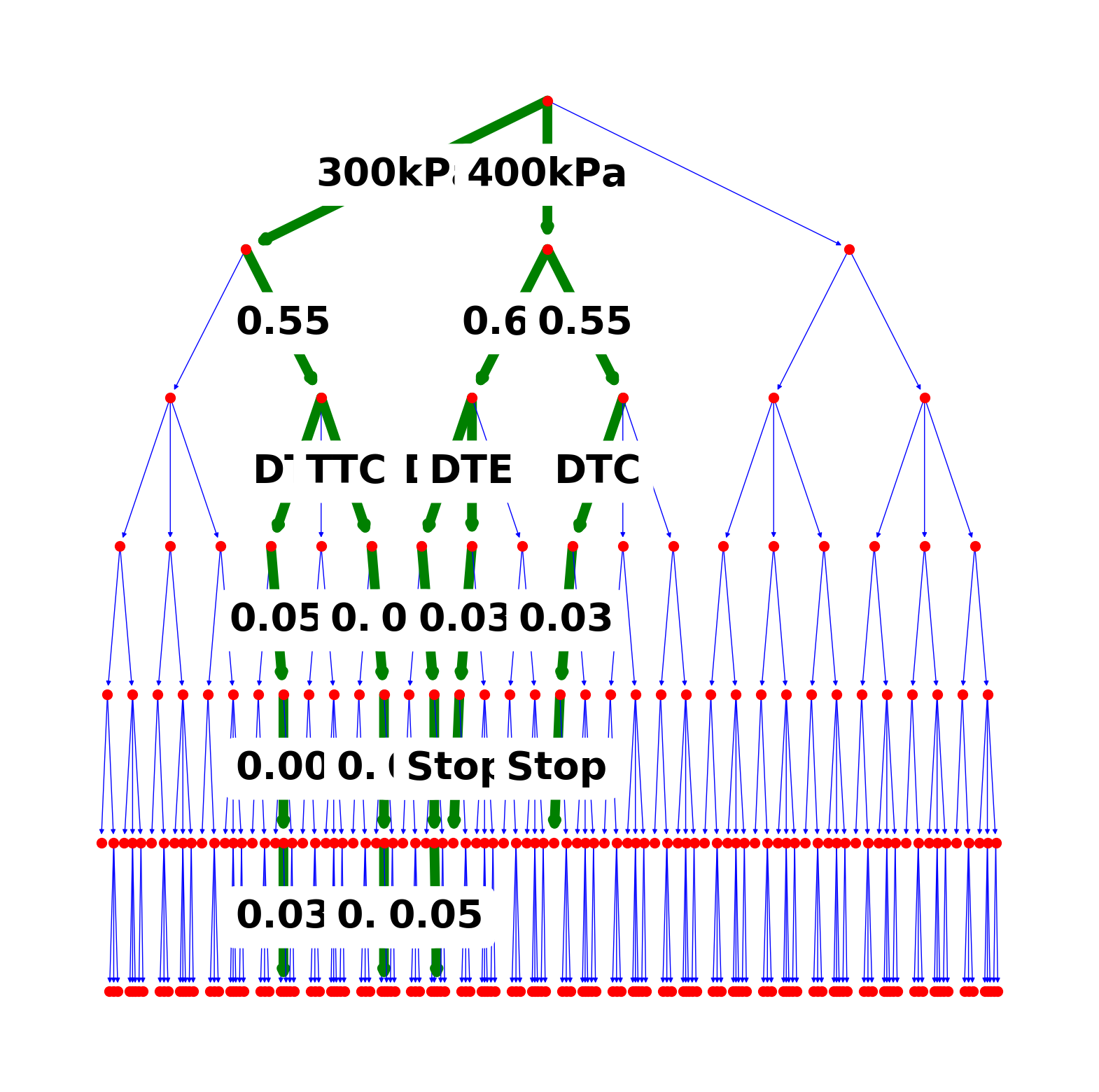}
	}
	\subfigure[Iteration 3, Episode 0, \newline \hspace{\linewidth} Attack Game Score: -0.725]{
		\includegraphics[width=0.235\textwidth]{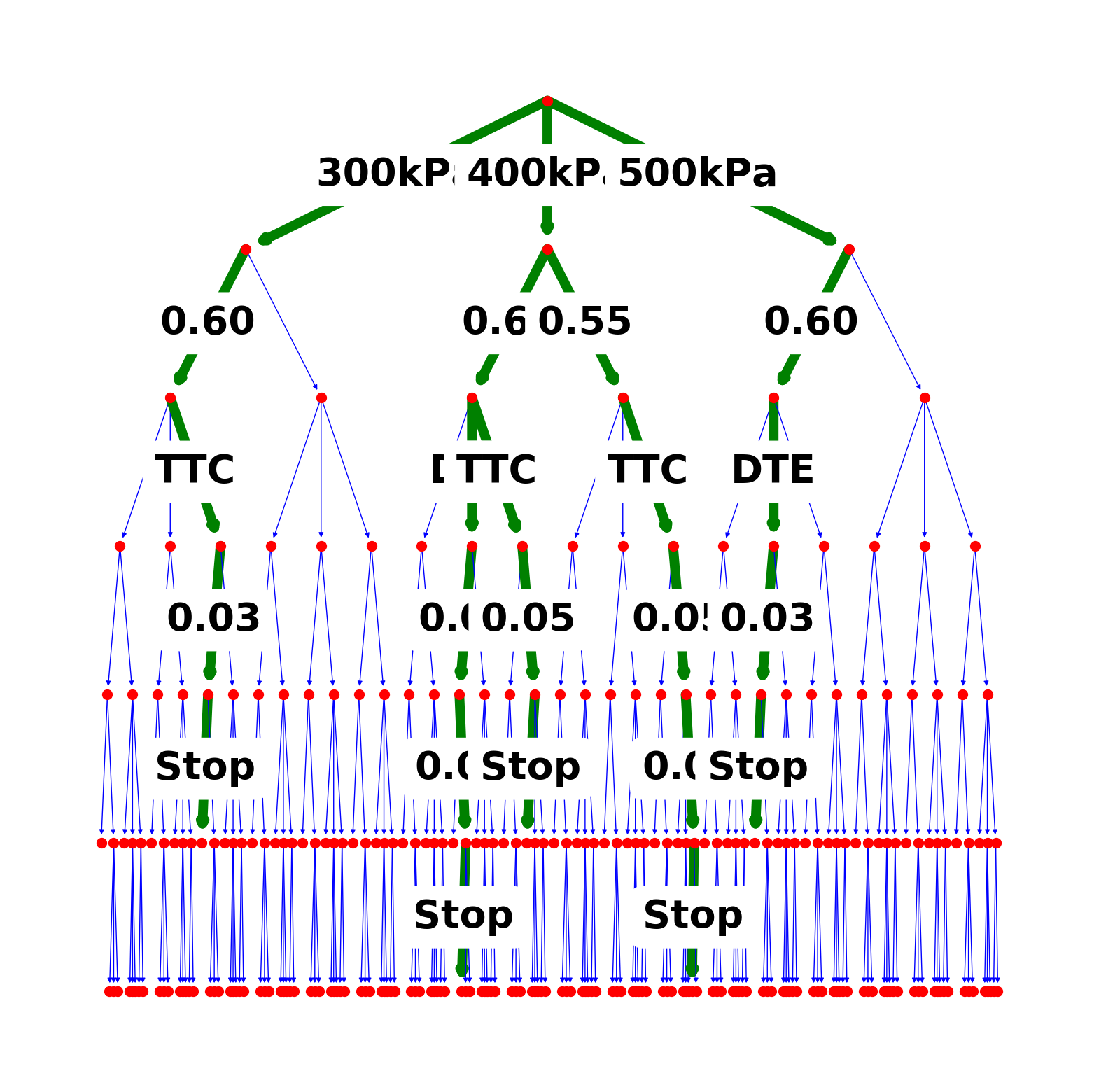}
	}
	\subfigure[Iteration 6, Episode 40, \newline \hspace{\linewidth} Attack Game Score: -0.570]{
		\includegraphics[width=0.235\textwidth]{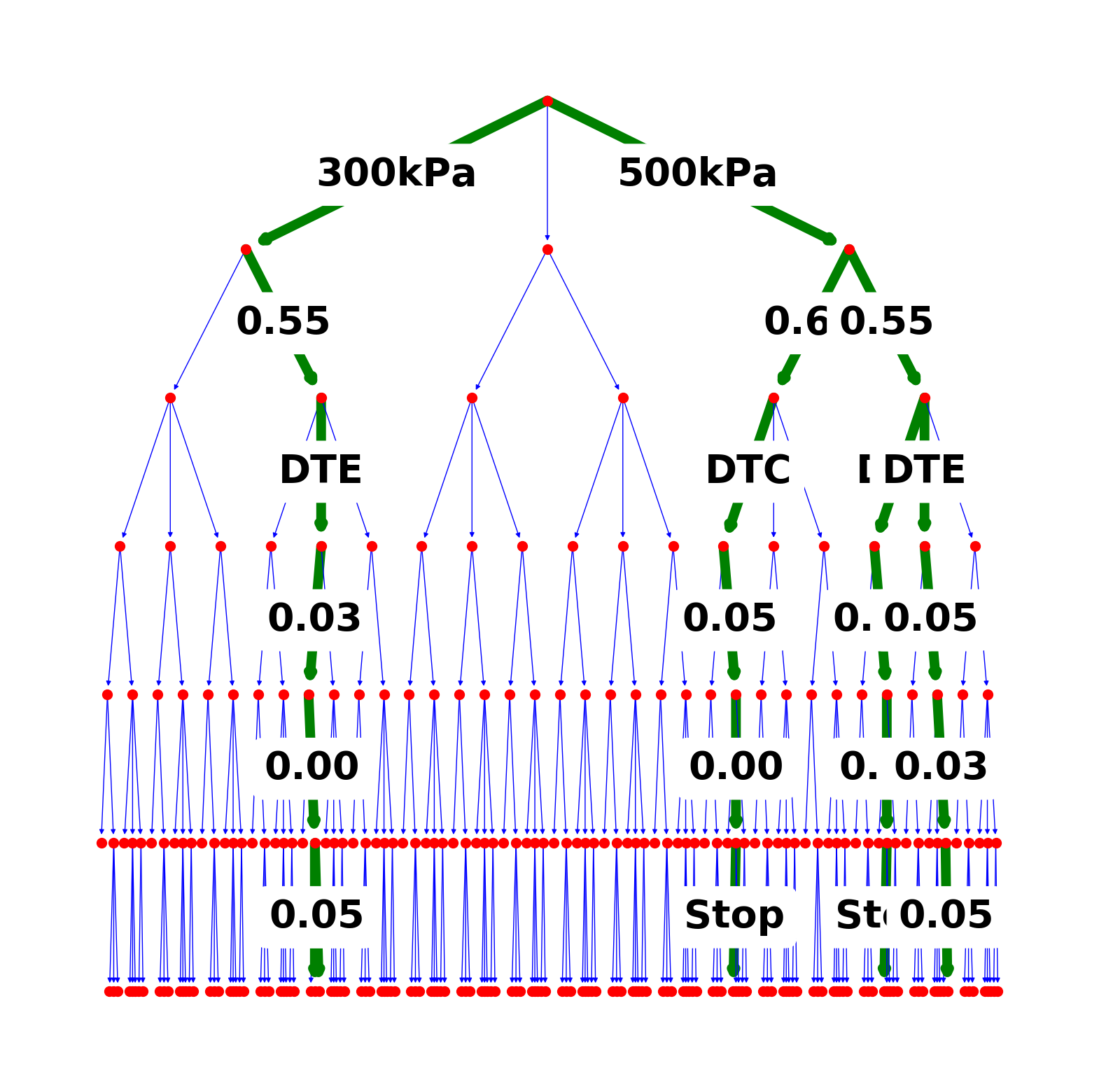}
	}
	\subfigure[Iteration 10, Episode 0, \newline \hspace{\linewidth} Attack Game Score: -0.989]{
		\includegraphics[width=0.235\textwidth]{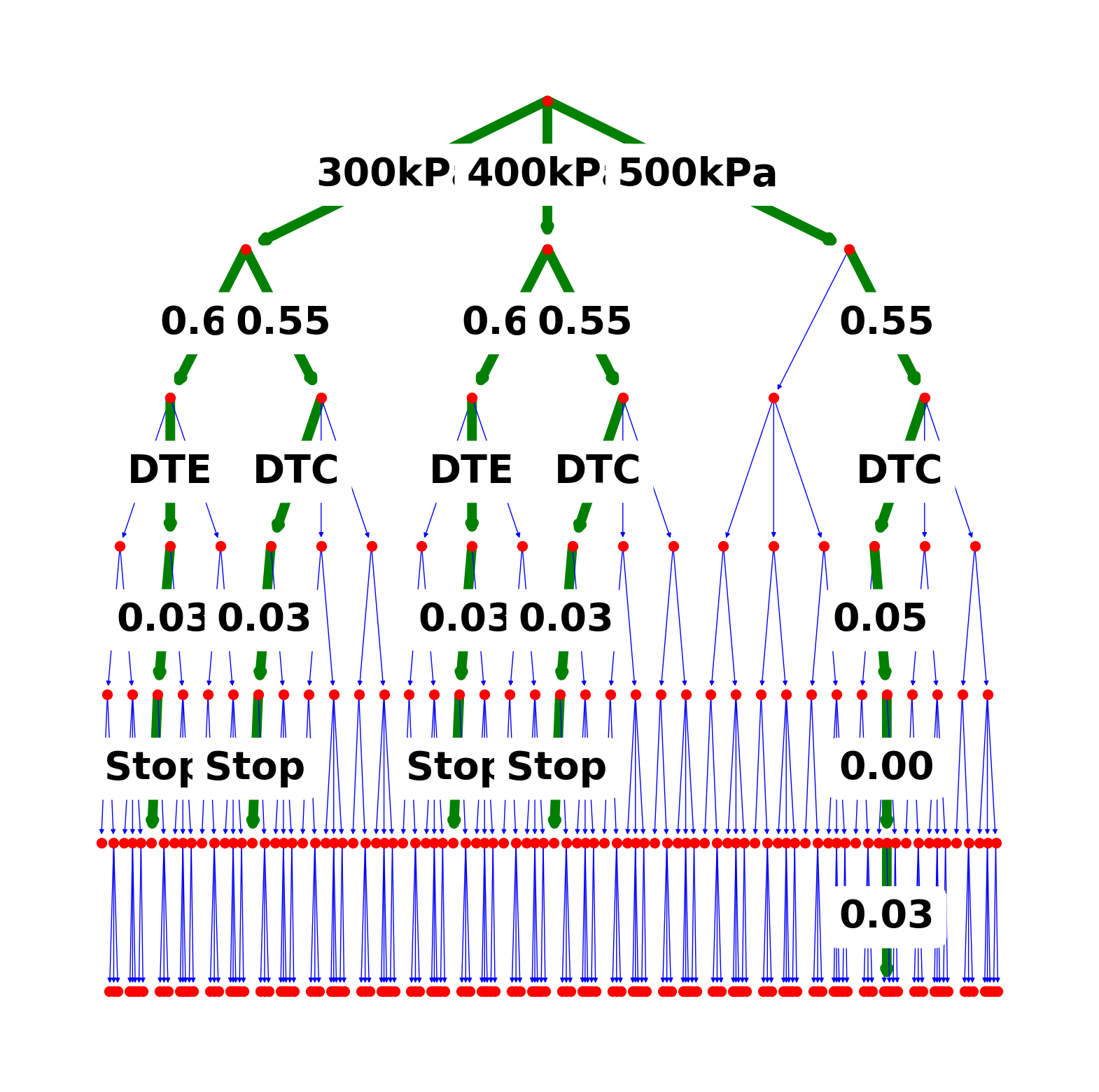}
	}
	\caption{Examples of paths (experiments) in the decision trees selected by the adversary during the DRL training iterations for Drucker-Prager model.}
	\label{fig:DP_attackgame_decisiontree}
\end{figure}

\begin{figure}[h!]\center
	\subfigure[Iteration 0, Episode 10, \newline \hspace{\linewidth} Defense Game Score: 0.262]{
		\includegraphics[width=0.235\textwidth]{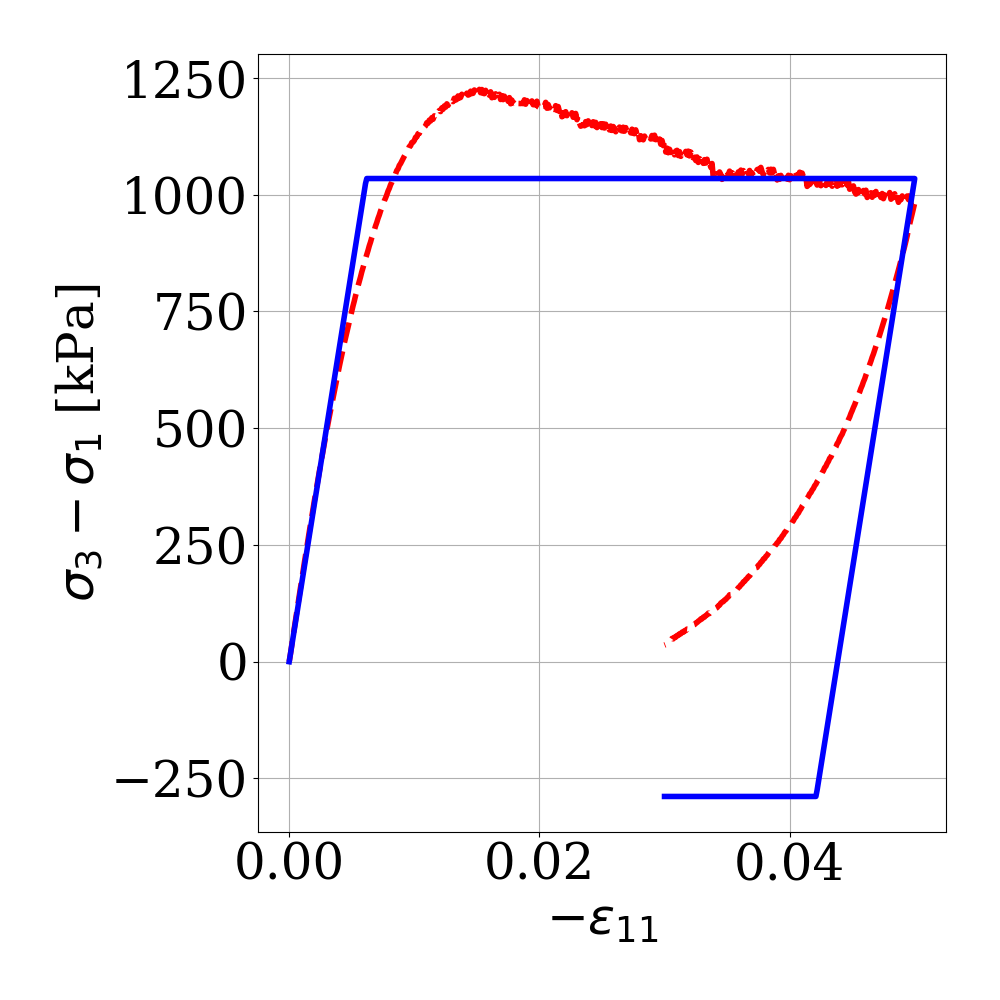}
	}
	\subfigure[Iteration 3, Episode 0, \newline \hspace{\linewidth} Defense Game Score: 0.811]{
		\includegraphics[width=0.235\textwidth]{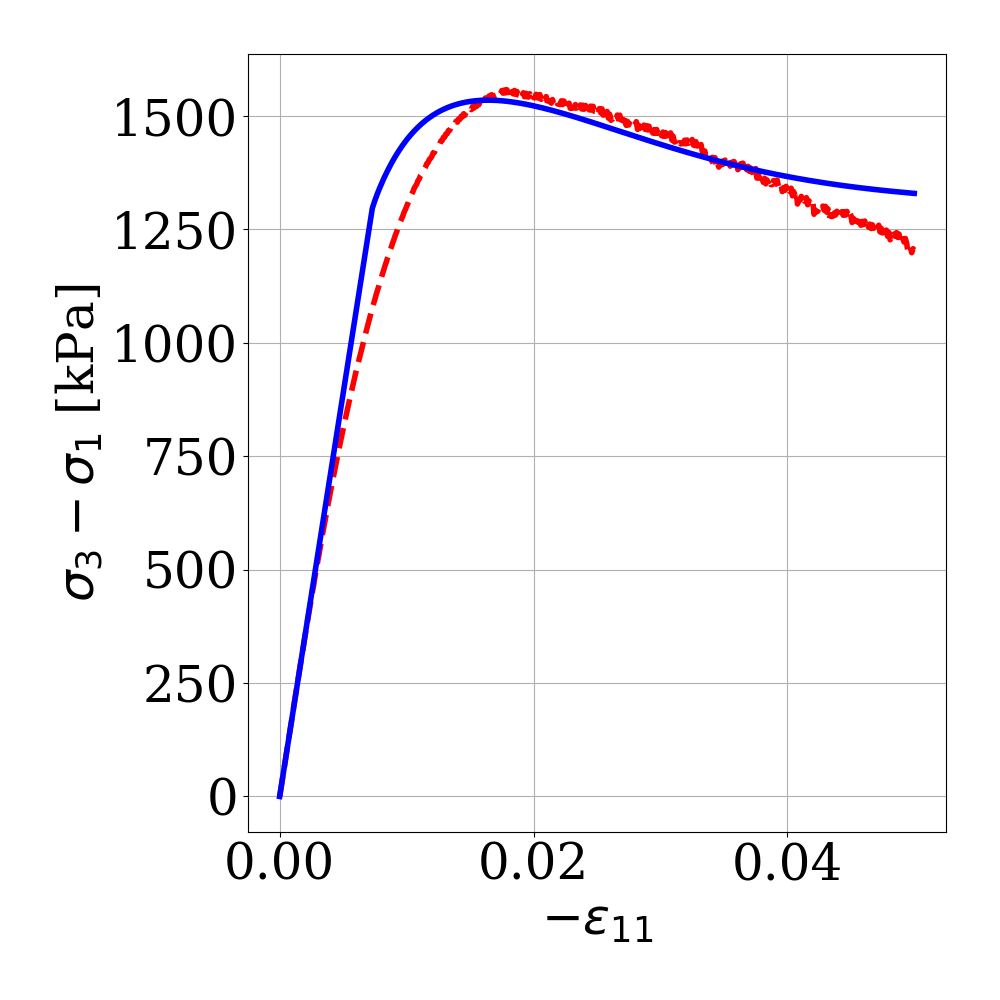}
	}
	\subfigure[Iteration 6, Episode 40, \newline \hspace{\linewidth} Defense Game Score: 0.699]{
		\includegraphics[width=0.235\textwidth]{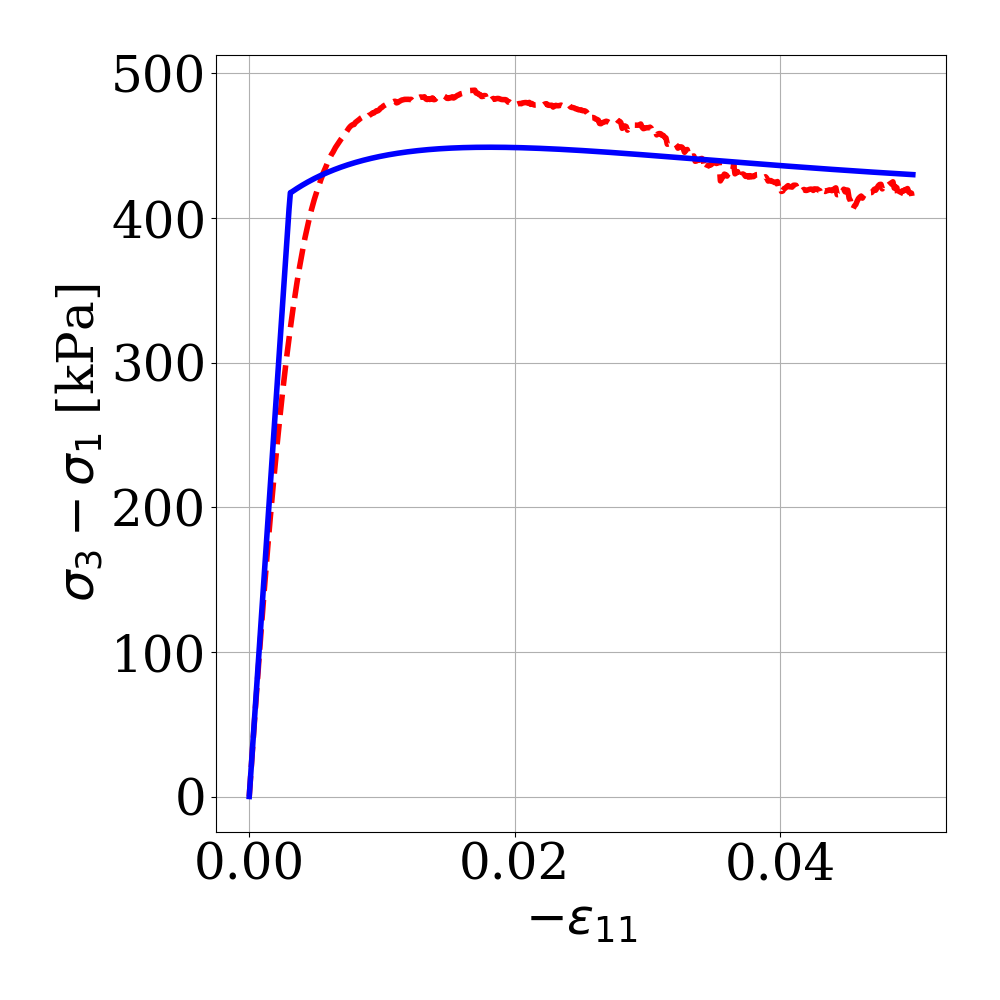}
	}
	\subfigure[Iteration 10, Episode 0, \newline \hspace{\linewidth} Defense Game Score: 0.886]{
		\includegraphics[width=0.235\textwidth]{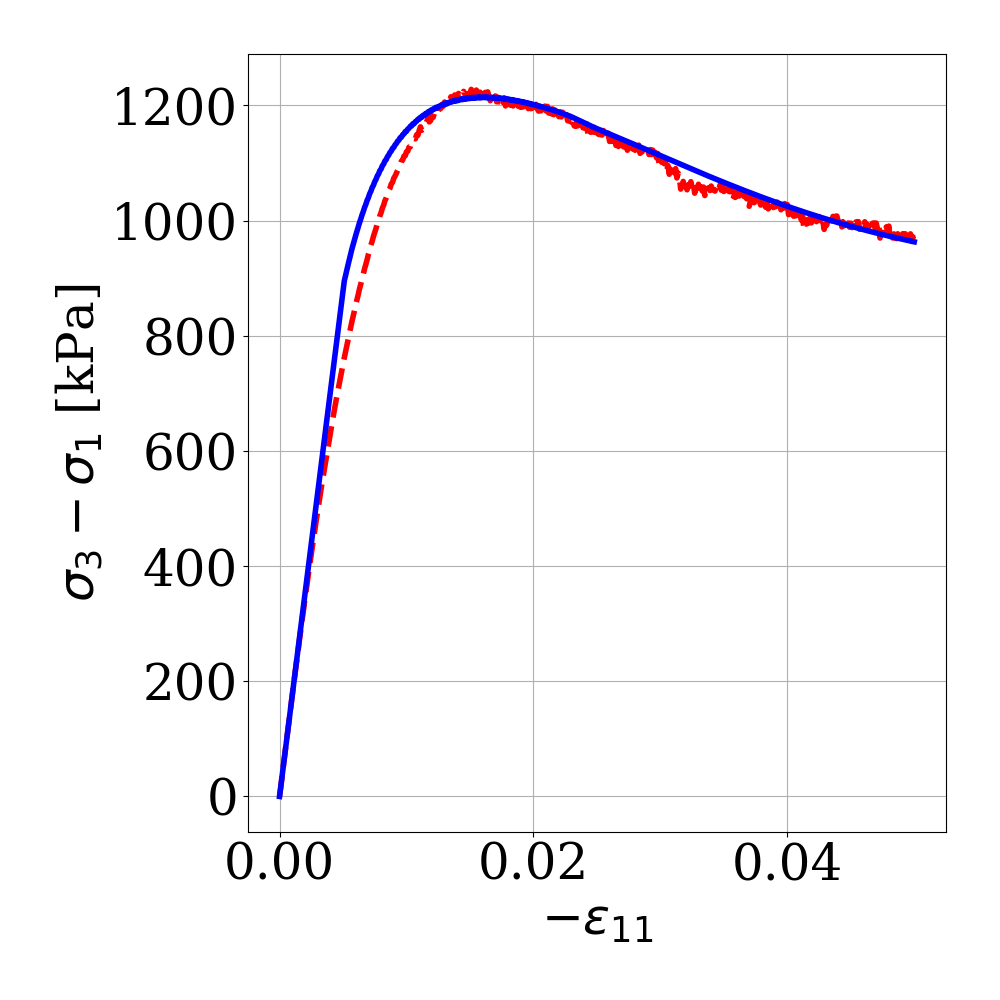}
	}
	\subfigure[Iteration 0, Episode 10, \newline \hspace{\linewidth} Defense Game Score: 0.262]{
		\includegraphics[width=0.235\textwidth]{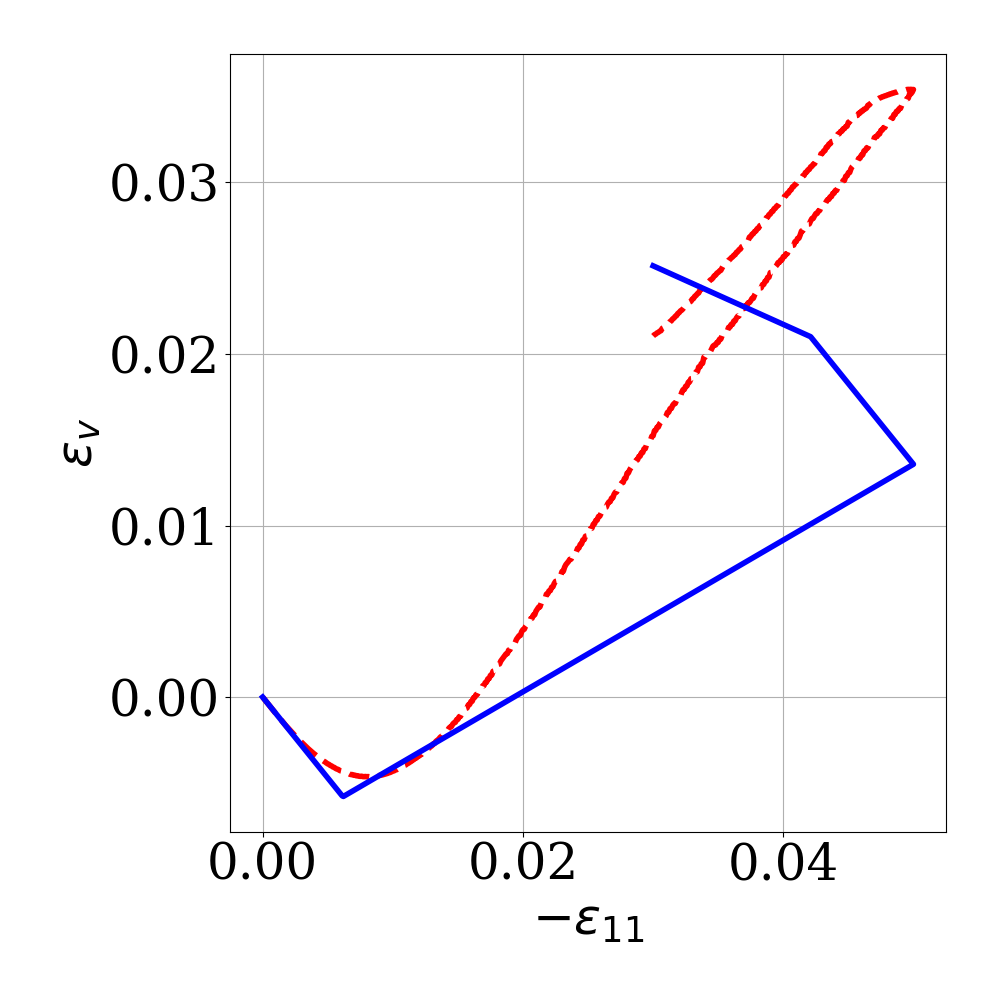}
	}
	\subfigure[Iteration 3, Episode 0, \newline \hspace{\linewidth} Defense Game Score: 0.811]{
		\includegraphics[width=0.235\textwidth]{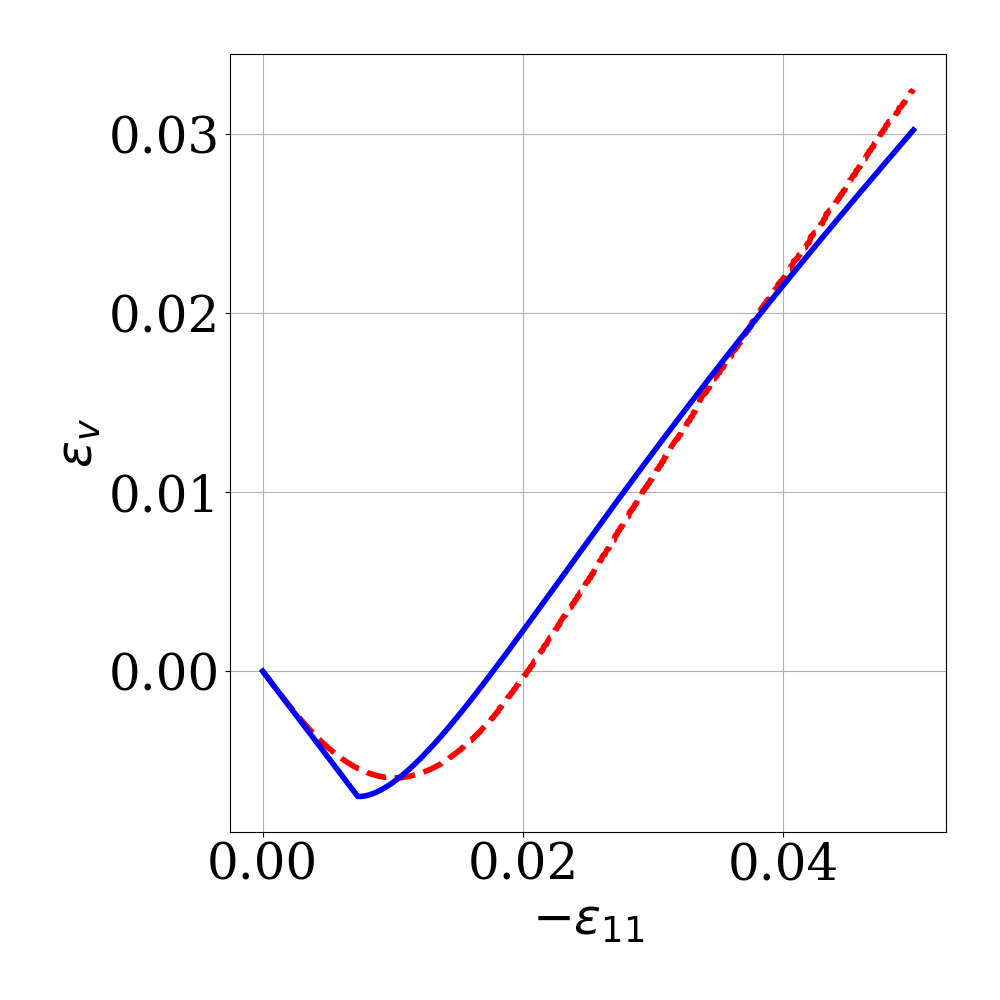}
	}
	\subfigure[Iteration 6, Episode 40, \newline \hspace{\linewidth} Defense Game Score: 0.699]{
		\includegraphics[width=0.235\textwidth]{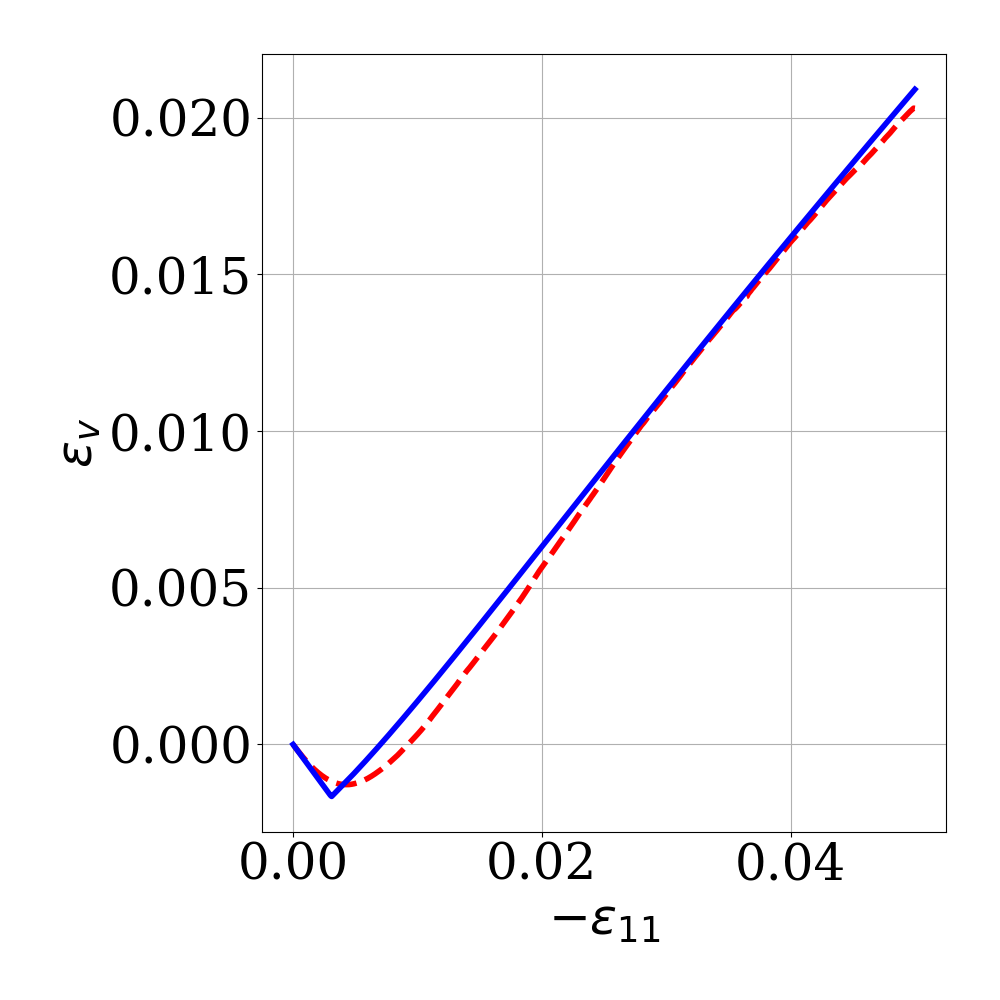}
	}
	\subfigure[Iteration 10, Episode 0, \newline \hspace{\linewidth} Defense Game Score: 0.886]{
		\includegraphics[width=0.235\textwidth]{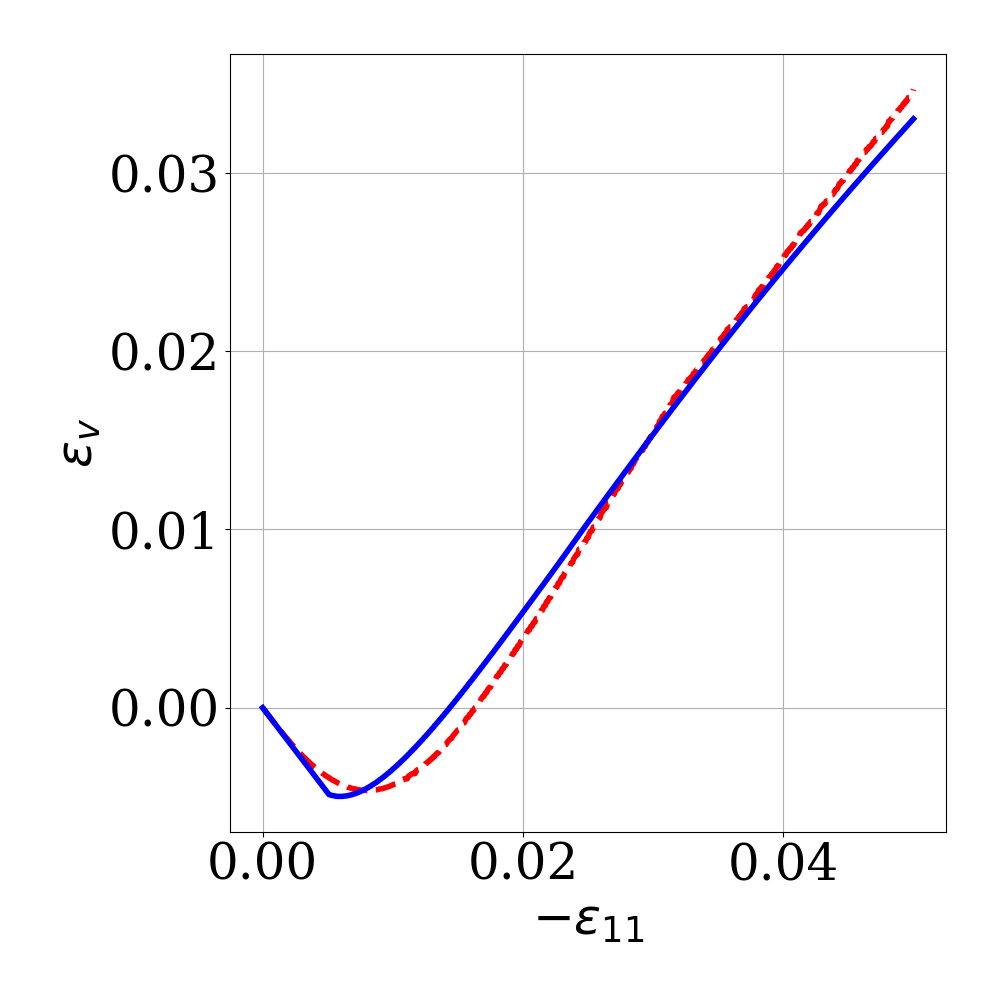}
	}
	\caption{Examples of response curves of the games played by the protagonist during the DRL training iterations for Drucker-Prager model. Experimental data are plotted in red dashed curves, model predictions are plotted in blue solid curves. }
	\label{fig:DP_datagame_curves}
\end{figure}

\begin{figure}[h!]\center
	\subfigure[Iteration 0, Episode 10, \newline \hspace{\linewidth} Attack Game Score: -0.151]{
		\includegraphics[width=0.235\textwidth]{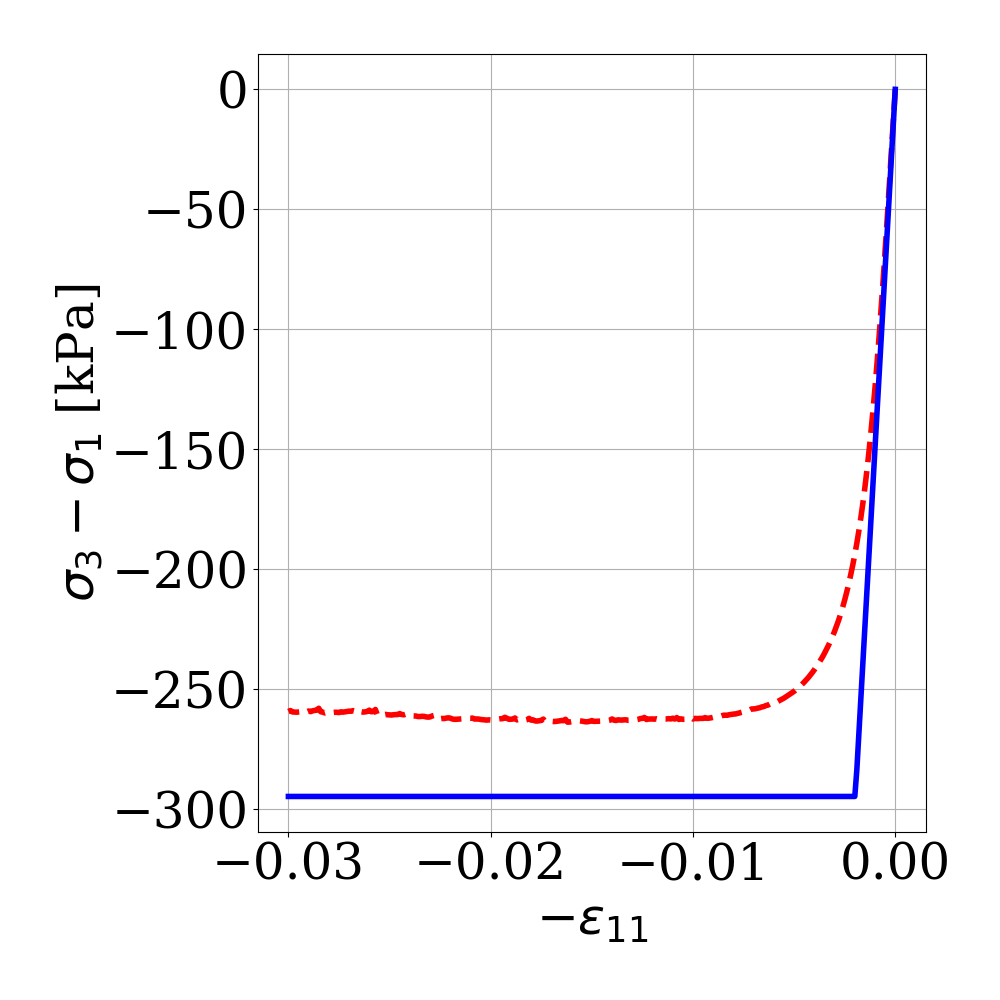}
	}
	\subfigure[Iteration 3, Episode 0, \newline \hspace{\linewidth} Attack Game Score: -0.725]{
		\includegraphics[width=0.235\textwidth]{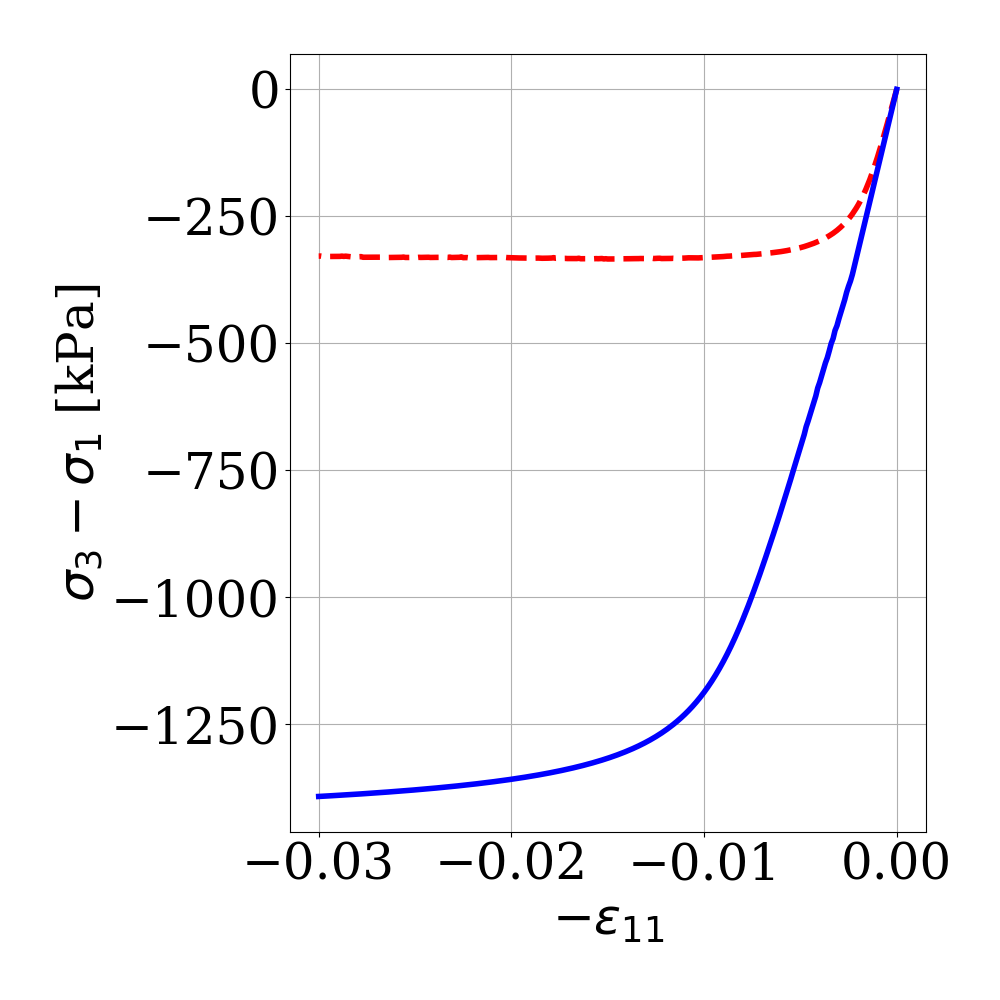}
	}
	\subfigure[Iteration 6, Episode 40, \newline \hspace{\linewidth} Attack Game Score: -0.570]{
		\includegraphics[width=0.235\textwidth]{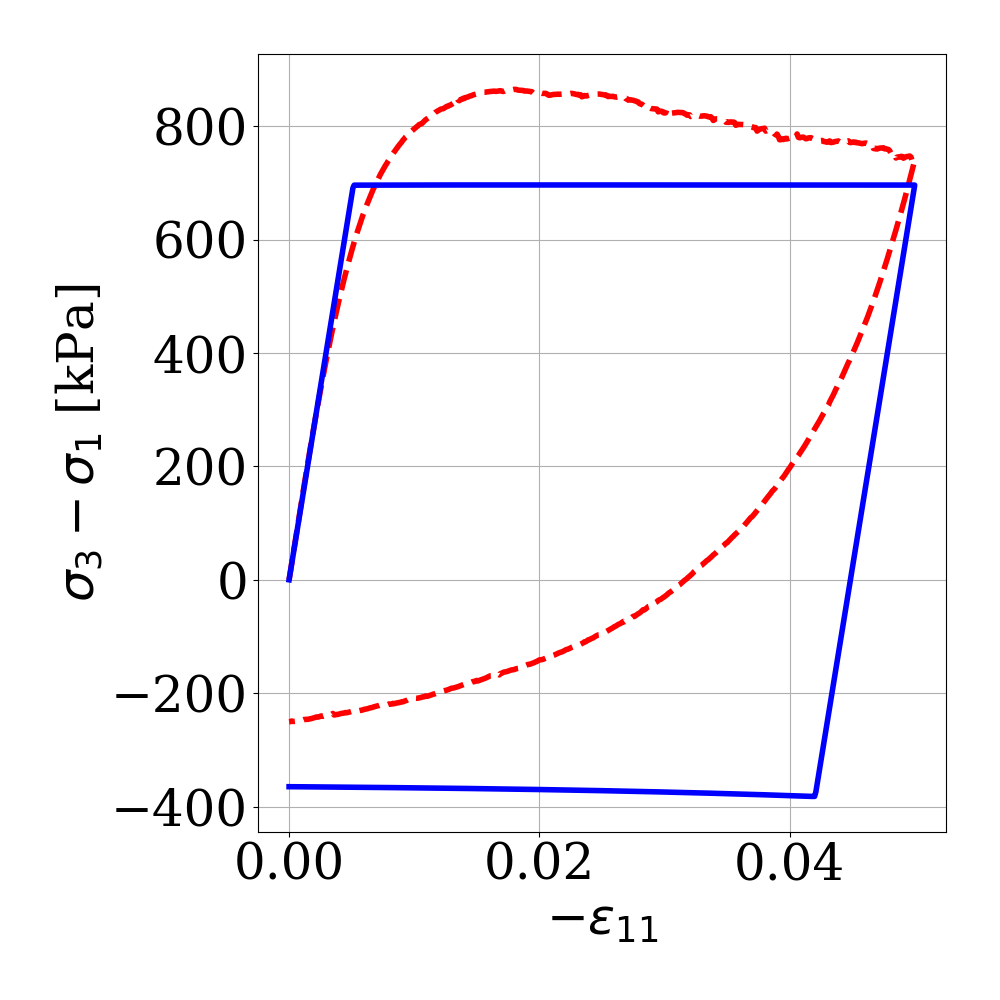}
	}
	\subfigure[Iteration 10, Episode 0, \newline \hspace{\linewidth} Attack Game Score: -0.989]{
		\includegraphics[width=0.235\textwidth]{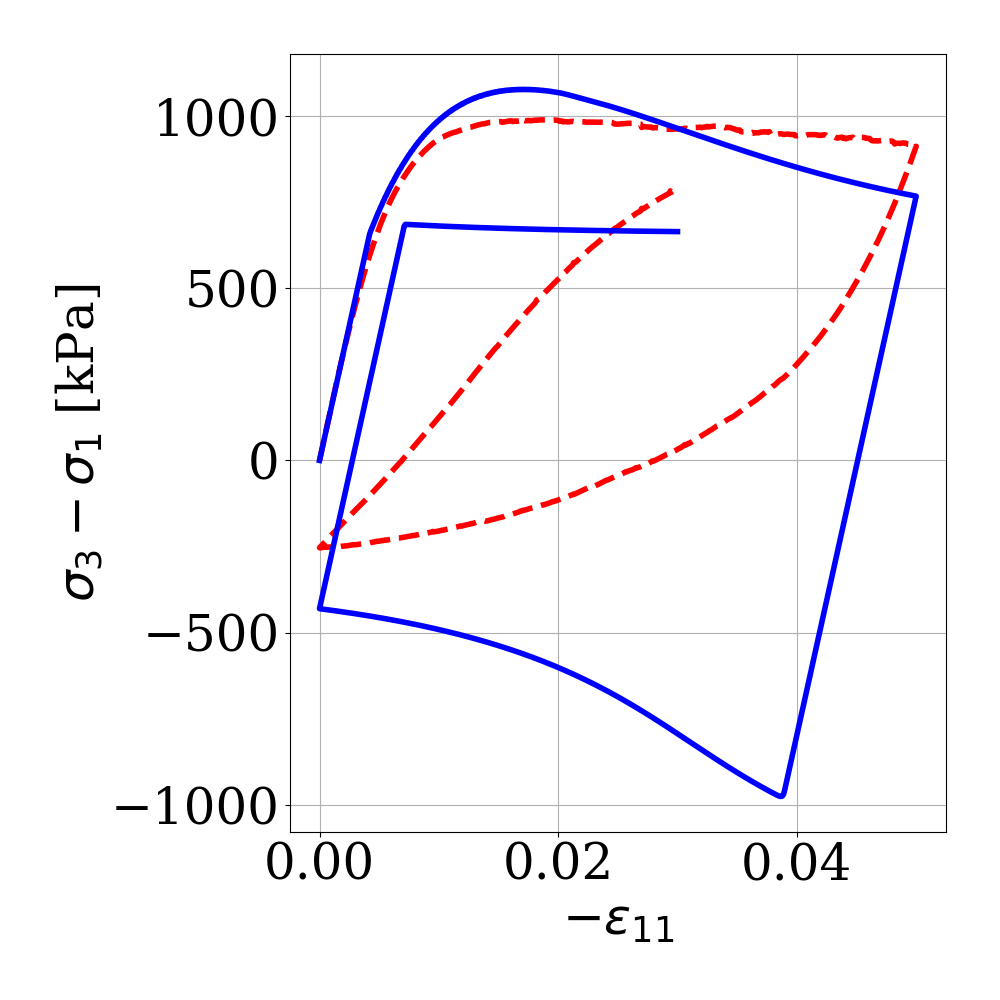}
	}
	\subfigure[Iteration 0, Episode 10, \newline \hspace{\linewidth} Attack Game Score: -0.151]{
		\includegraphics[width=0.235\textwidth]{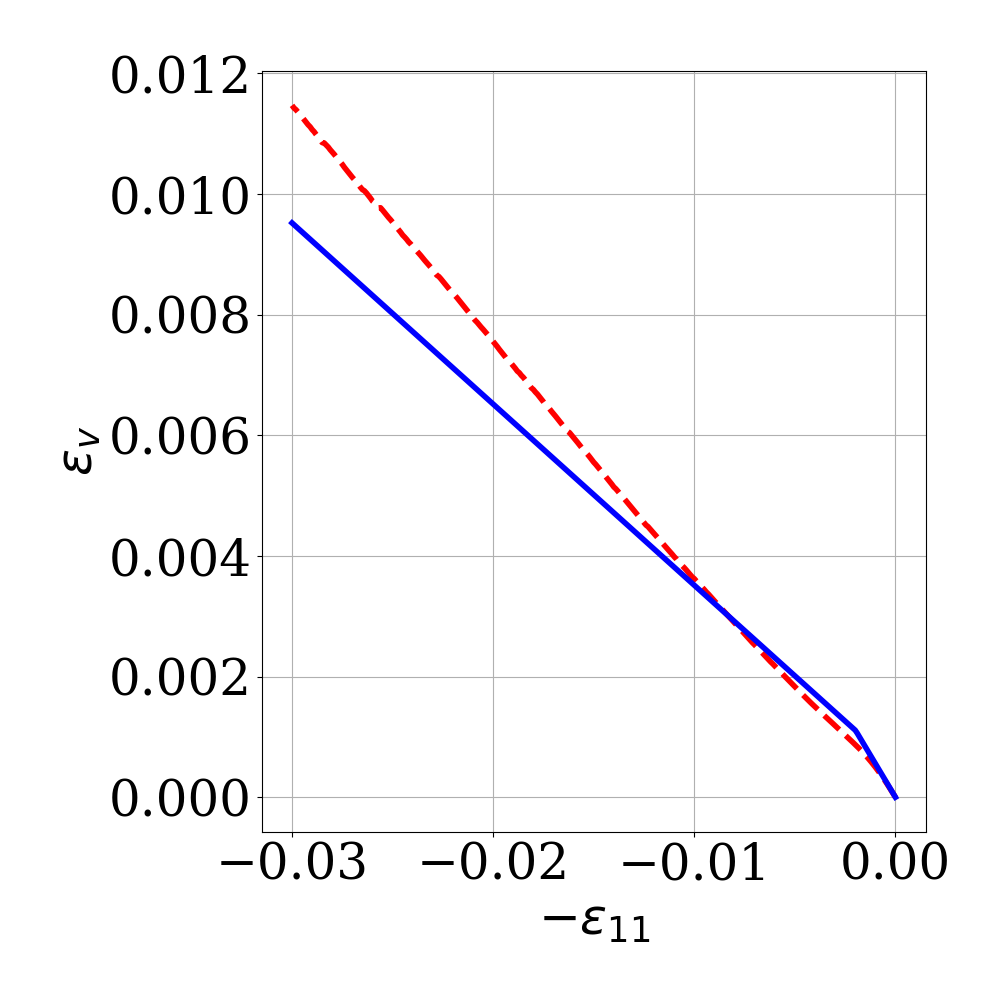}
	}
	\subfigure[Iteration 3, Episode 0, \newline \hspace{\linewidth} Attack Game Score: -0.725]{
		\includegraphics[width=0.235\textwidth]{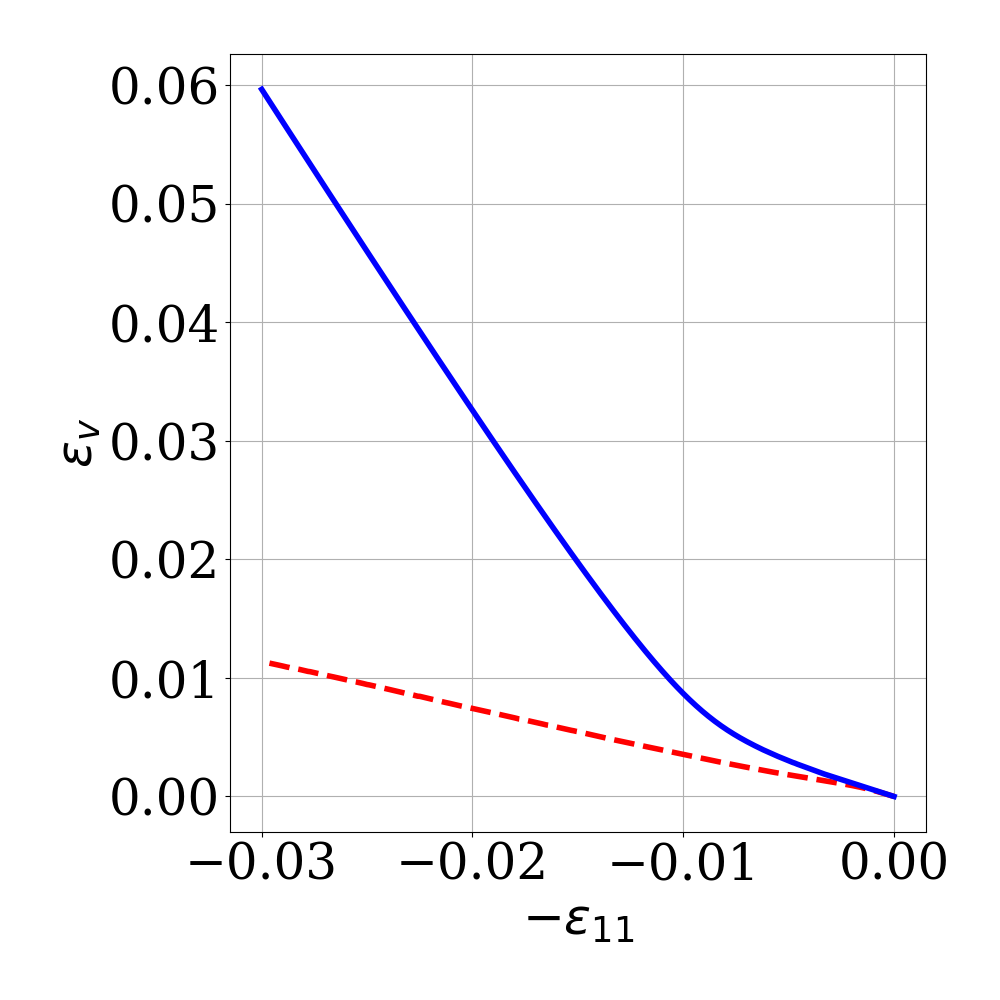}
	}
	\subfigure[Iteration 6, Episode 40, \newline \hspace{\linewidth} Attack Game Score: -0.570]{
		\includegraphics[width=0.235\textwidth]{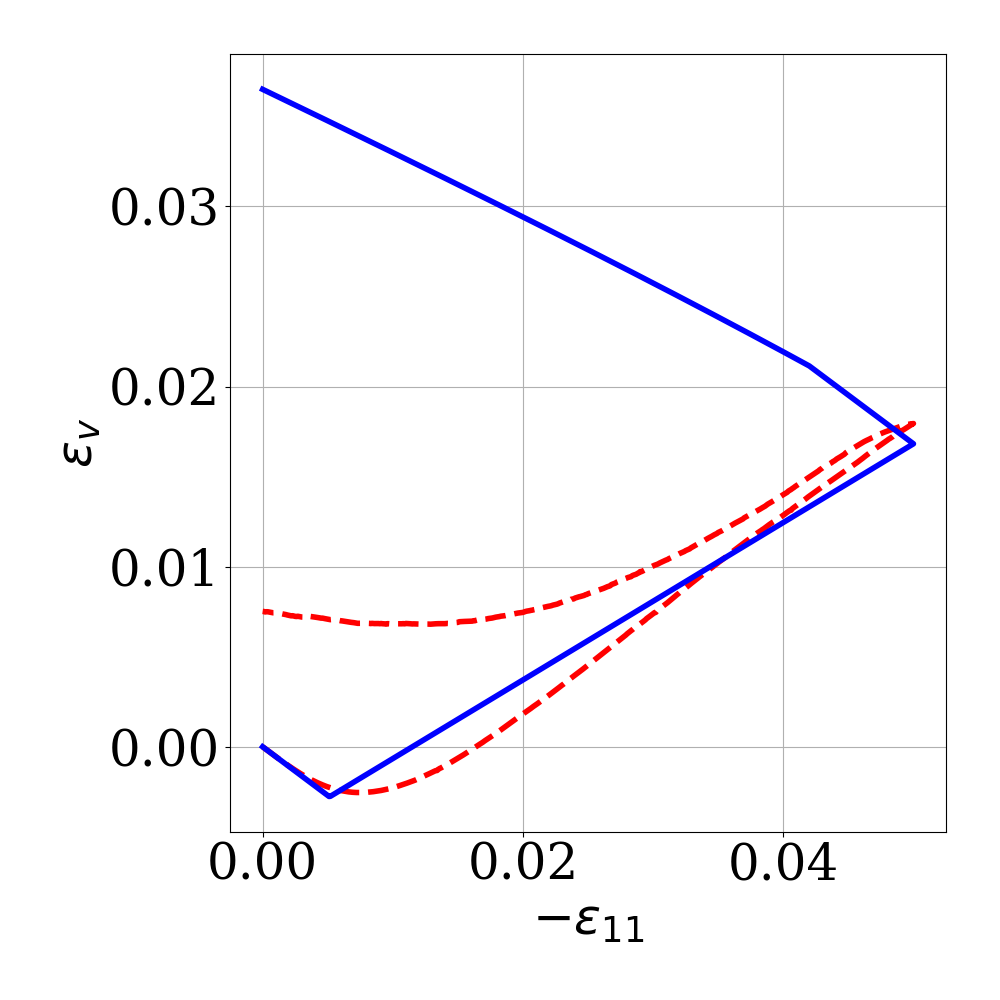}
	}
	\subfigure[Iteration 10, Episode 0, \newline \hspace{\linewidth} Attack Game Score: -0.989]{
		\includegraphics[width=0.235\textwidth]{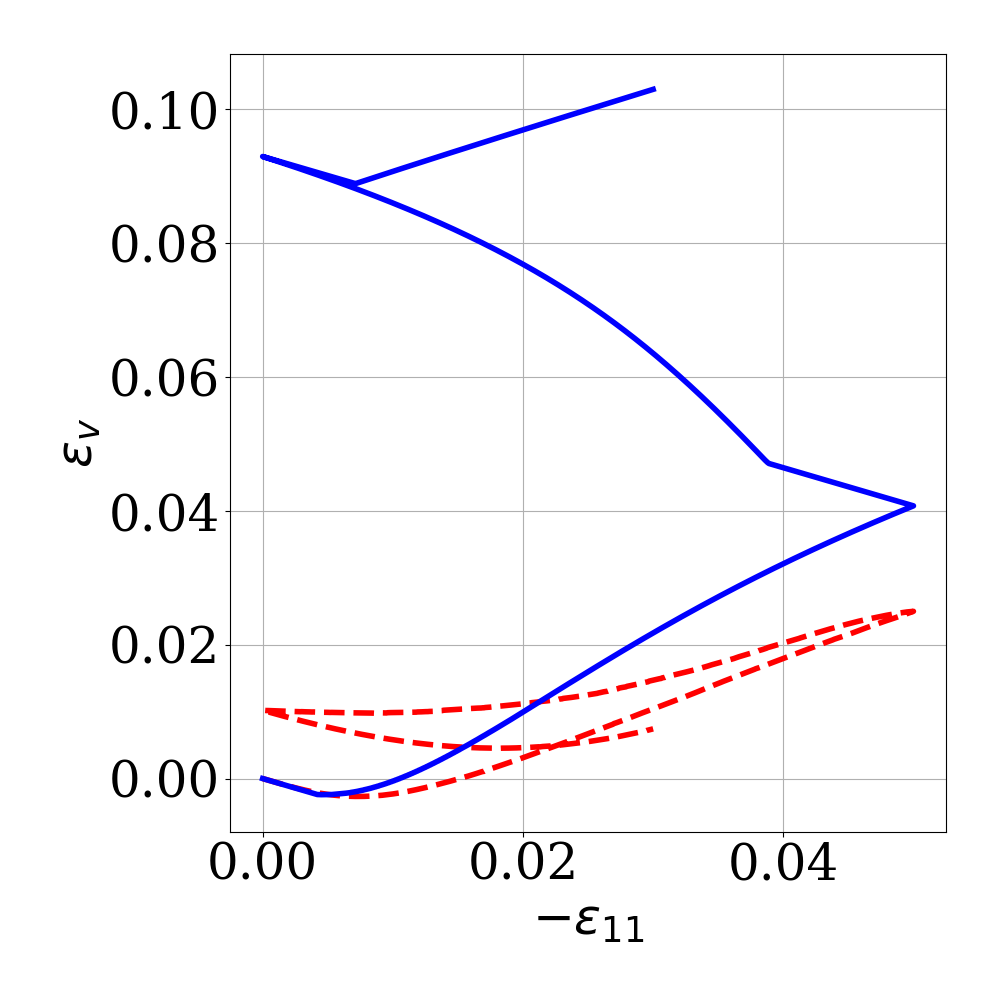}
	}
	\caption{Examples of response curves of the games played by the adversary during the DRL training iterations for Drucker-Prager model. Experimental data are plotted in red dashed curves, model predictions are plotted in blue solid curves. }
	\label{fig:DP_attackgame_curves}
\end{figure}

Examples of paths (experiments) selected by the protagonist during the DRL iterations are shown in Fig. \ref{fig:DP_datagame_decisiontree}. 
Based on all the evidences such as  ones shown in these examples, the agent realizes that the Drucker-Prager model is not designed to simultaneously replicate data from samples with different initial confinement, initial void ratio, test types, and unloading-reloading paths.  In the end, the agent concludes that the model is only accurate in modeling the mechanical behaviour of a single sample in TTC test with monotonic loading. 
Meanwhile, the adversary tries to attack the models calibrated by the protagonist using the experiments as shown in Fig. \ref{fig:DP_attackgame_decisiontree}.  The agent progressively comes to the conclusion that, when calibrated with monotonic TTC data, the model fails to predict DTC or DTE experiments on other samples with unloading-reloading. 
Fig. \ref{fig:DP_datagame_curves} and Fig. \ref{fig:DP_attackgame_curves} give example response curves associated to the example decision tree paths shown in Fig. \ref{fig:DP_datagame_decisiontree} and Fig.\ref{fig:DP_attackgame_decisiontree}, respectively. They illustrate the strength of the model in replicating the hardening-softening and contraction-dilation behavior of a densely compressed granular material. 
They also expose the model's weakness in predicting the unloading-reloading behaviour, regardless of the calibration data. 
These conclusions on the Drucker-Prager model by the AI agents are consistent with the judgements from human experts, but they  are drawn from the  reinforcement learning on the two-player game without human knowledge.

\subsection{Experiment 2: SANISAND model}
\label{subsec:sanisandmodel_example}
The two-player non-cooperative game is played by DRL-based AI experimentalists for SANISAND model. 
The formulations of the model are detailed by Eq. (\ref{eq:sanisand_1}), (\ref{eq:sanisand_2}), (\ref{eq:sanisand_3}), (\ref{eq:sanisand_4}), (\ref{eq:sanisand_5}). 
The initial guesses, upper and lower bounds of the material parameters for Dakota calibration are presented in Table \ref{tab:sanisand_parameters_guess}. 
The game settings are $N^{max}_{path}=5$ for both the protagonist and the adversary, $E^{max}_{NS}=1.0$, $E^{min}_{NS}=-1.0$. 
The hyperparameters for the DRL algorithm are $numIters=10$, $numEpisodes=40$, $numMCTSSims=50$, $\alpha_{\text{SCORE}}=1.0$, $\alpha_{range} = 0.2$, $i_{lookback}=4$, $\tau_{train} = 1.0$, $\tau_{test} = 0.1$. 
The policy/value networks for both AI agents are identical to the ones used in the previous example. 
In order to help explore $\min (\{E^1_{NS}\})$ in Eq. (\ref{eq:reward_protagonist}), we also manually pre-select 5 experiments that have unloading-reloading paths which need to be predicted by all calibrated SANISAND models, along with the test data selected by the adversary.

\begin{figure}[h!]\center
	\subfigure[Protagonist]{
		\includegraphics[width=0.45\textwidth]{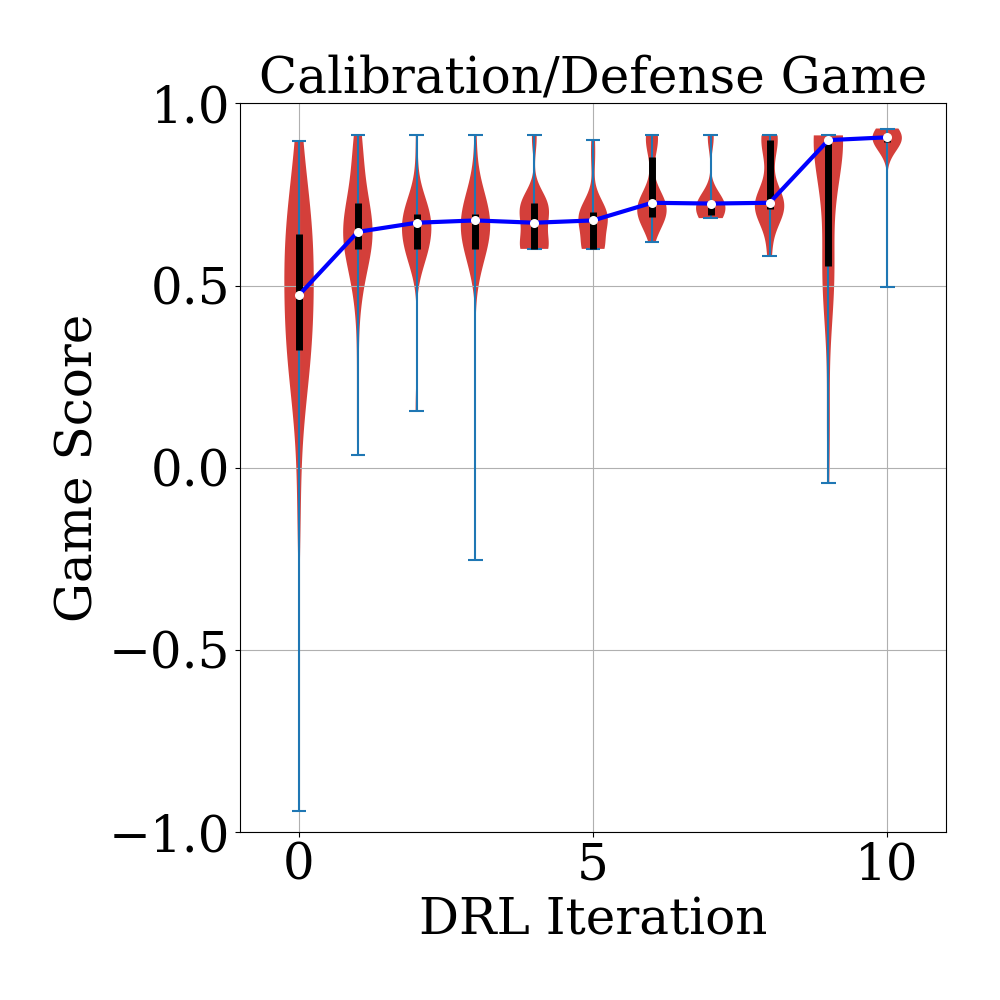}
	}
	\subfigure[Adversary]{
		\includegraphics[width=0.45\textwidth]{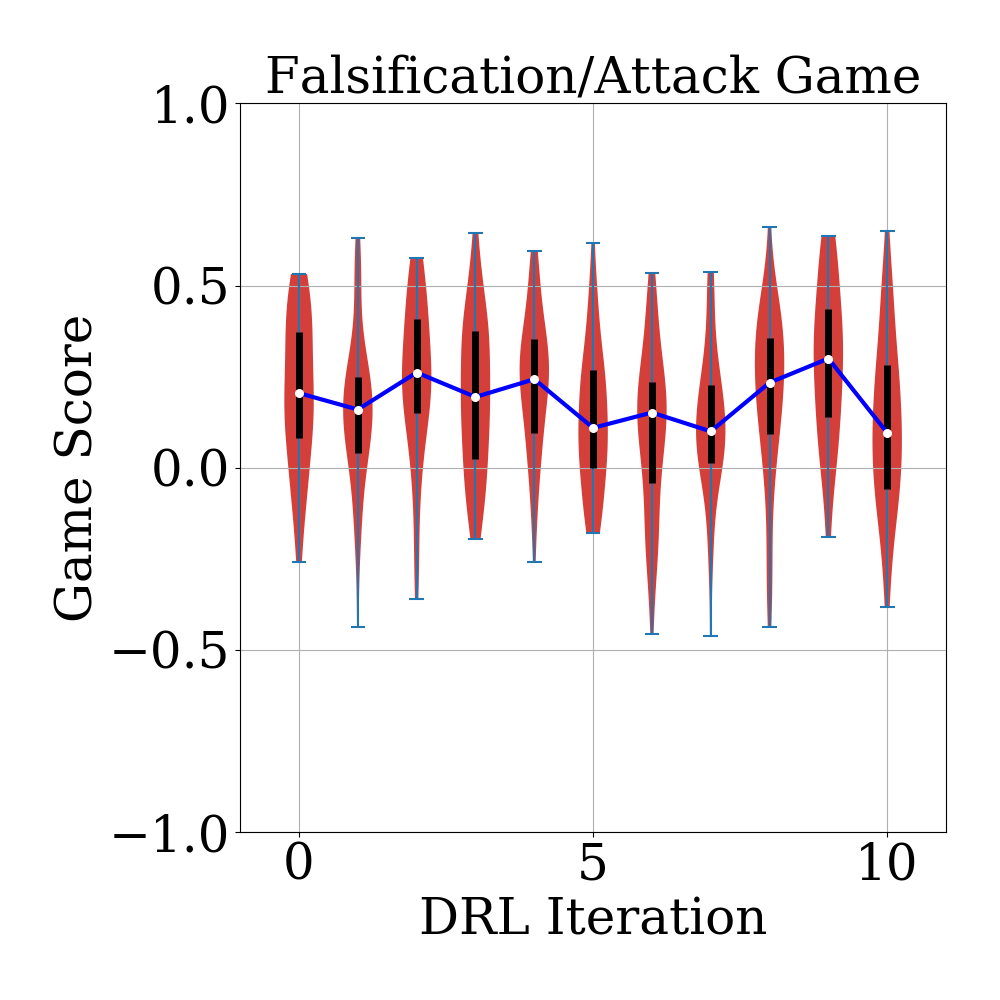}
	}
	\caption{Violin plots of the density distributions of game scores in each DRL iteration in SANISAND model. The shaded area represents the density distribution of scores. The white point represents the median. The thick black bar represents the inter-quantile range between 25\% quantile and 75\% quantile. The maximum and minimum scores played in each iteration are marked.}
	\label{fig:SANISAND_learn_violinplot}
\end{figure}

\begin{figure}[h!]\center
	\subfigure[Iteration 0, Episode 24, \newline \hspace{\linewidth} Defense Game Score: 0.162]{
		\includegraphics[width=0.235\textwidth]{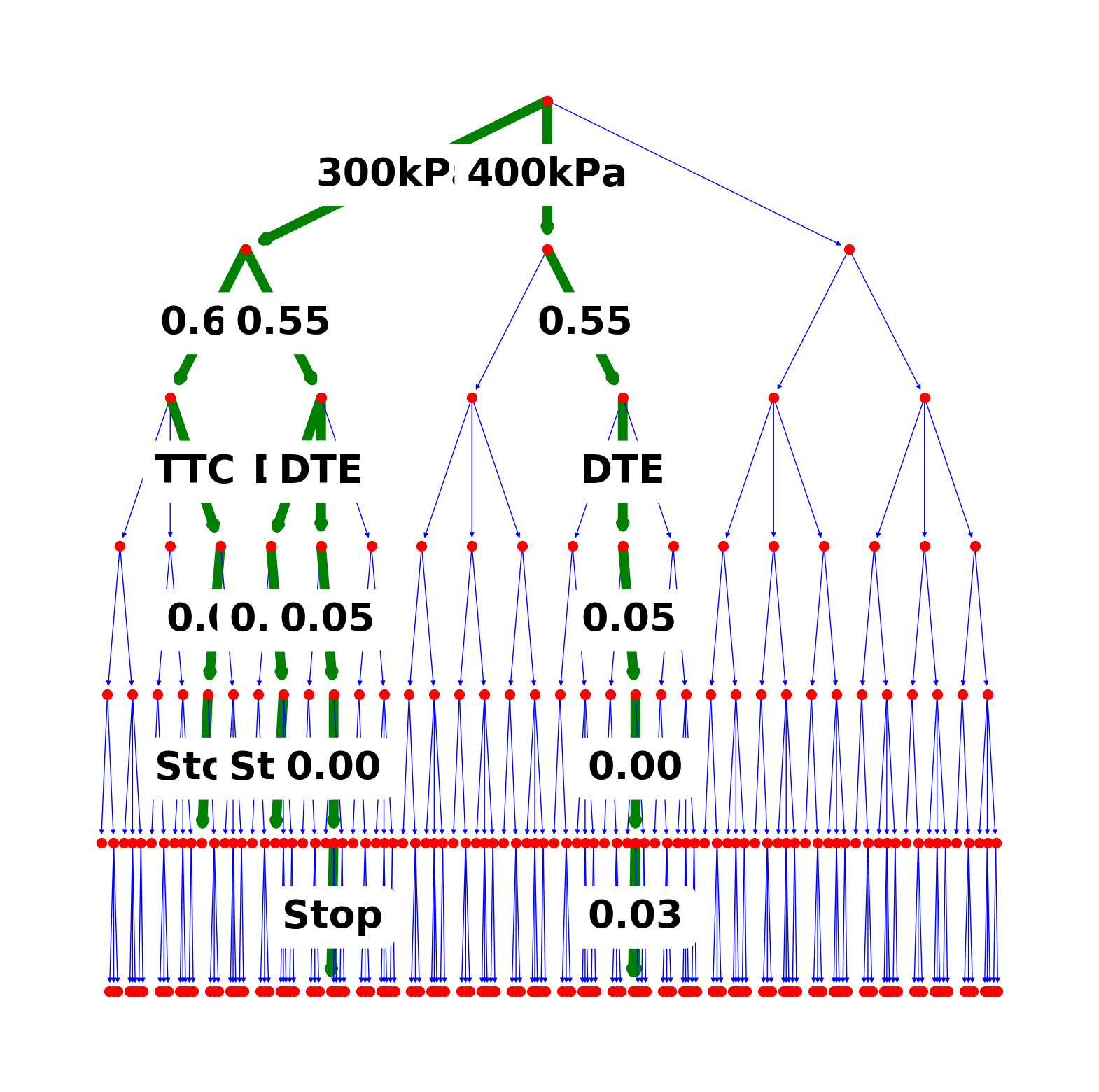}
	}
	\subfigure[Iteration 3, Episode 33, \newline \hspace{\linewidth} Defense Game Score: 0.626]{
		\includegraphics[width=0.235\textwidth]{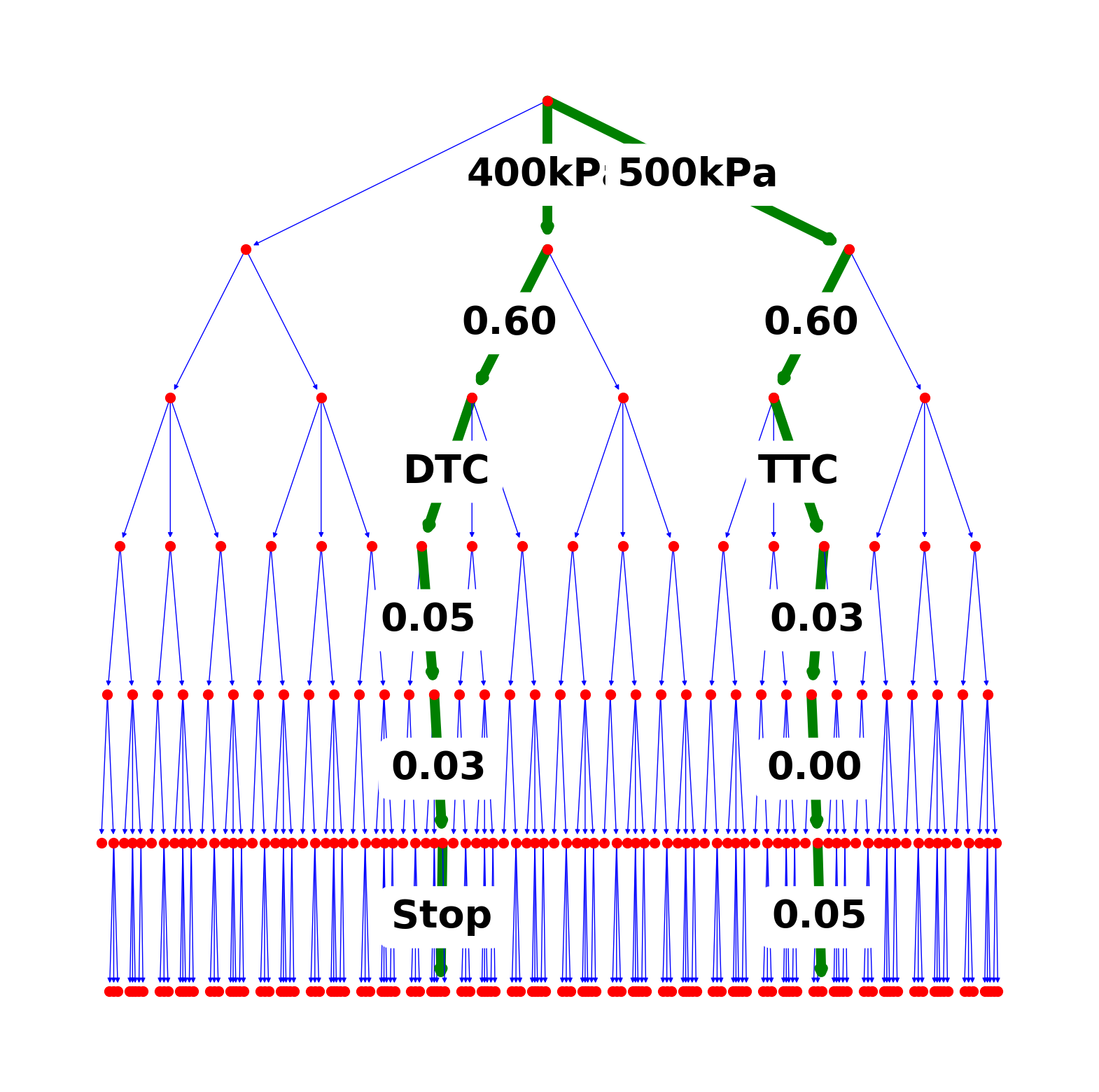}
	}
	\subfigure[Iteration 6, Episode 10, \newline \hspace{\linewidth} Defense Game Score: 0.687]{
		\includegraphics[width=0.235\textwidth]{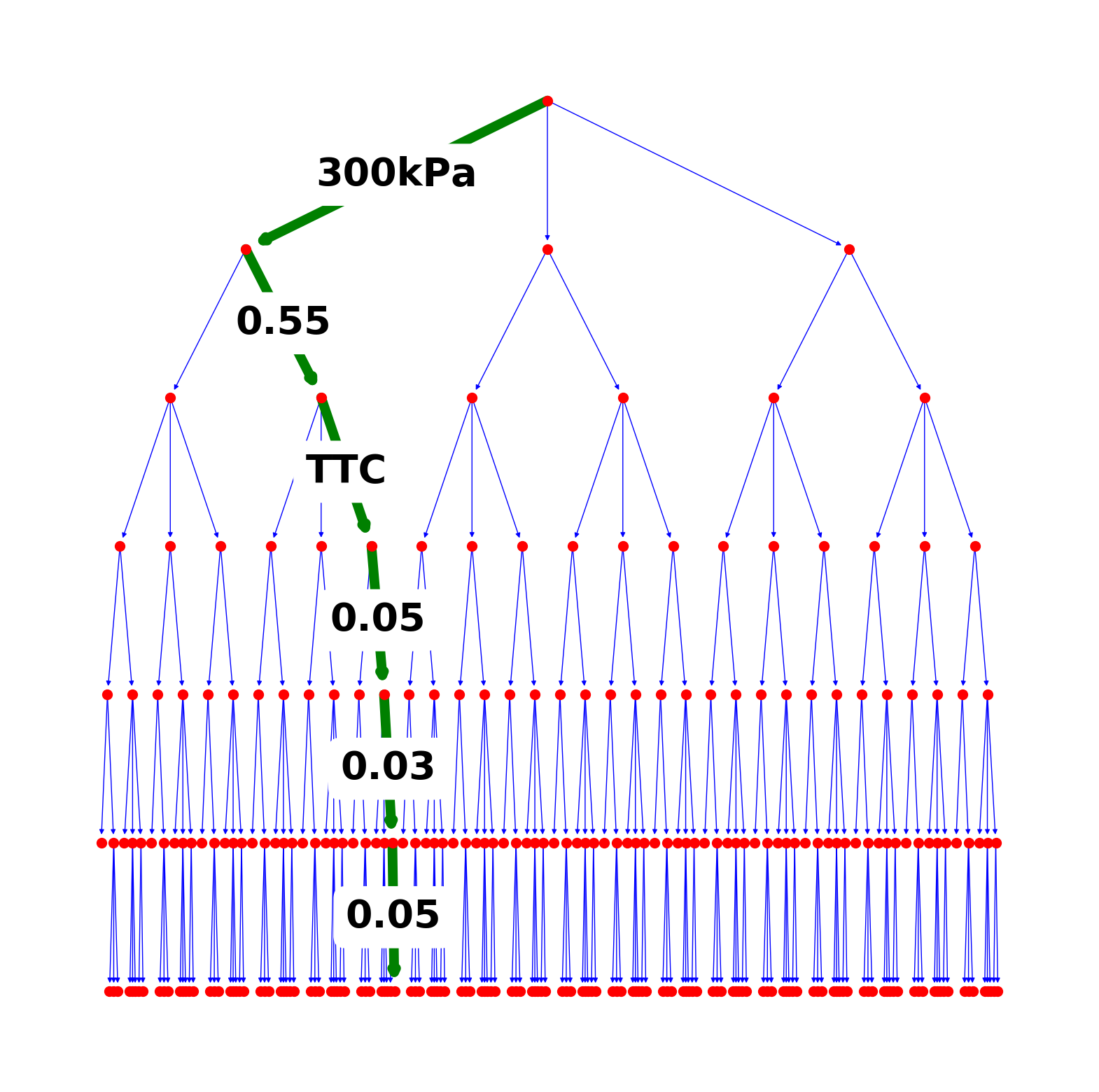}
	}
	\subfigure[Iteration 10, Episode 30, \newline \hspace{\linewidth} Defense Game Score: 0.912]{
		\includegraphics[width=0.235\textwidth]{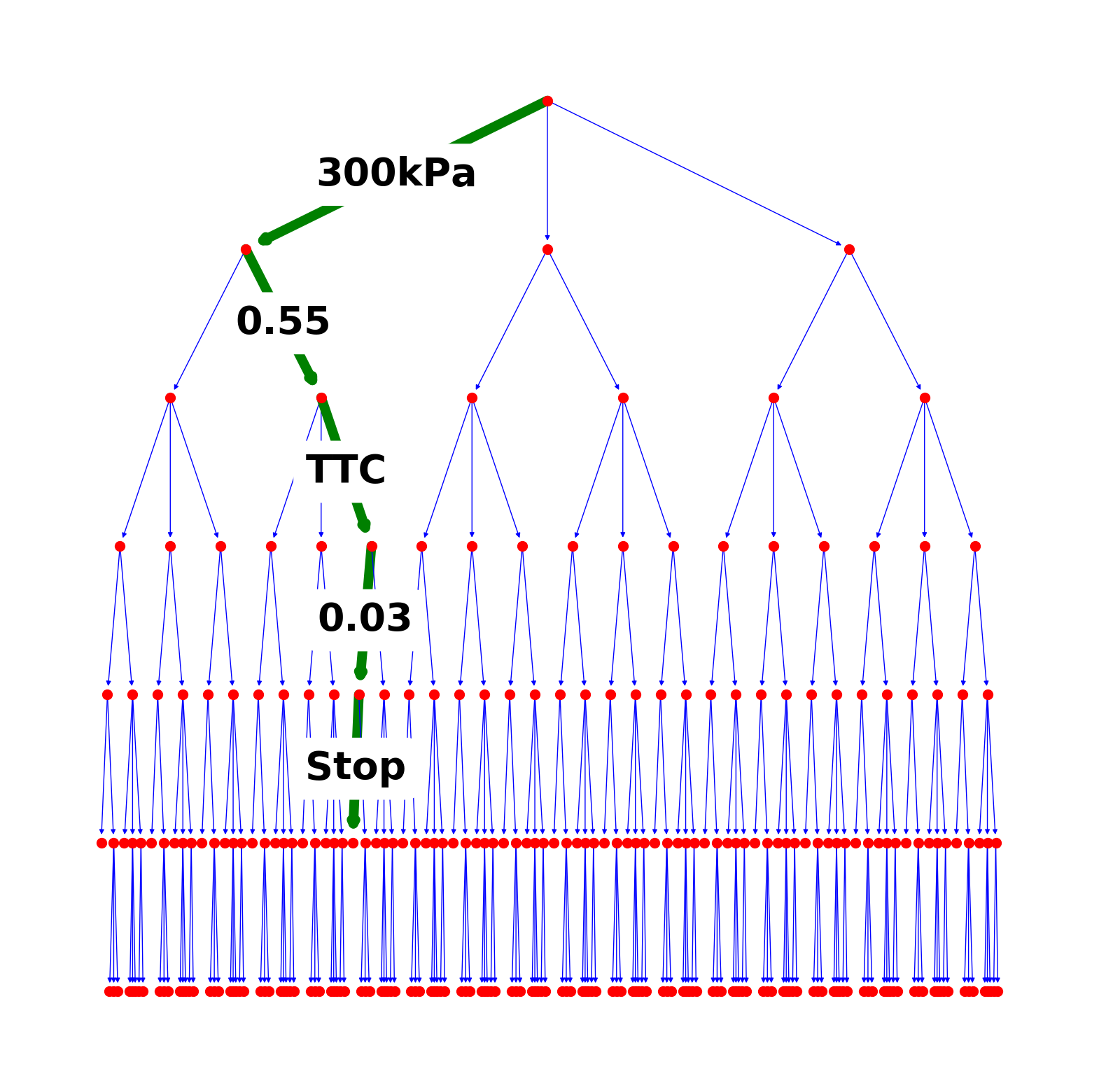}
	}
	\caption{Examples of paths (experiments) in the decision trees selected by the protagonist during the DRL training iterations for SANISAND model.}
	\label{fig:SANISAND_datagame_decisiontree}
\end{figure}

\begin{figure}[h!]\center
	\subfigure[Iteration 0, Episode 24, \newline \hspace{\linewidth} Attack Game Score: 0.206]{
		\includegraphics[width=0.235\textwidth]{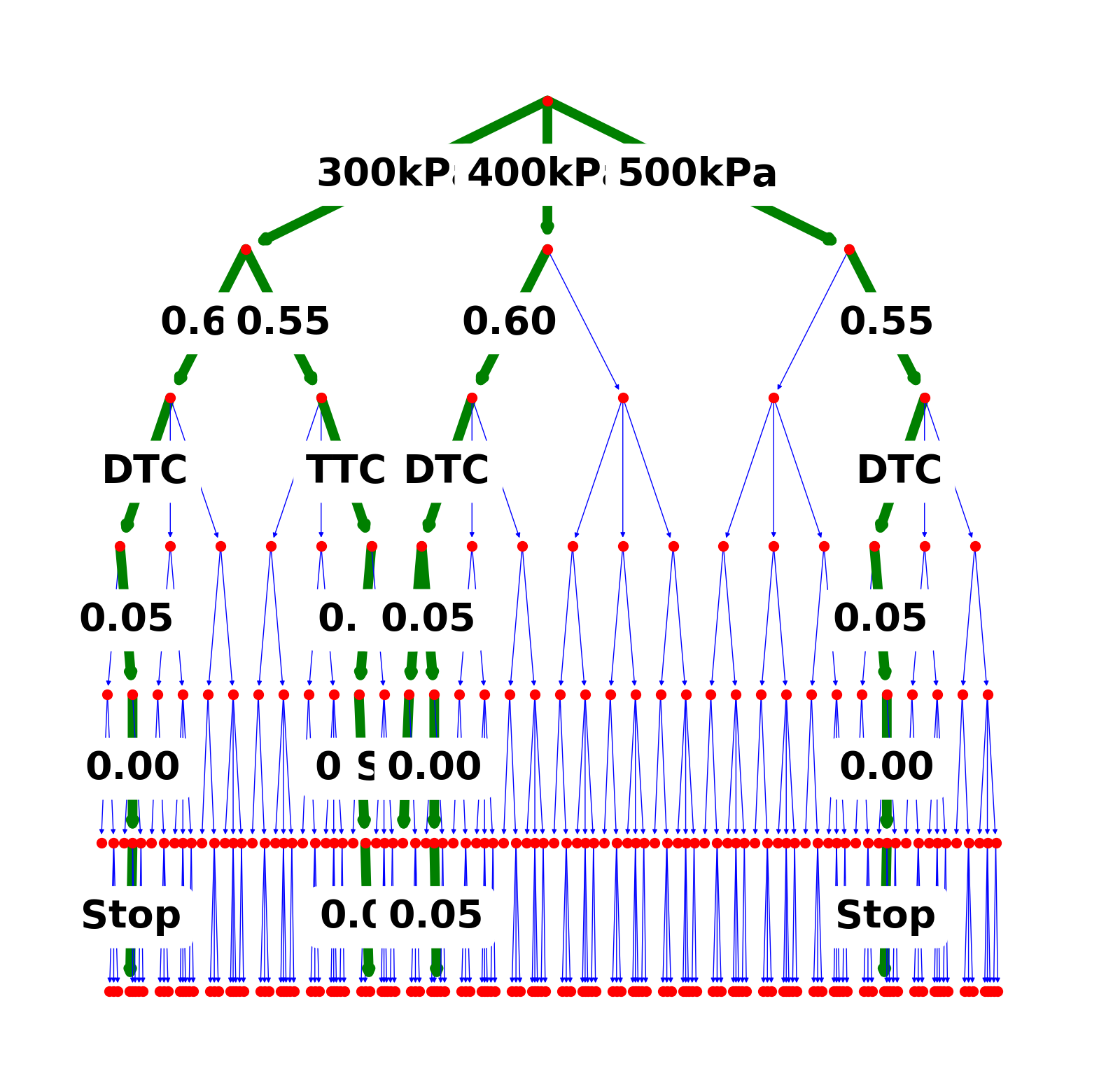}
	}
	\subfigure[Iteration 3, Episode 33, \newline \hspace{\linewidth} Attack Game Score: 0.323]{
		\includegraphics[width=0.235\textwidth]{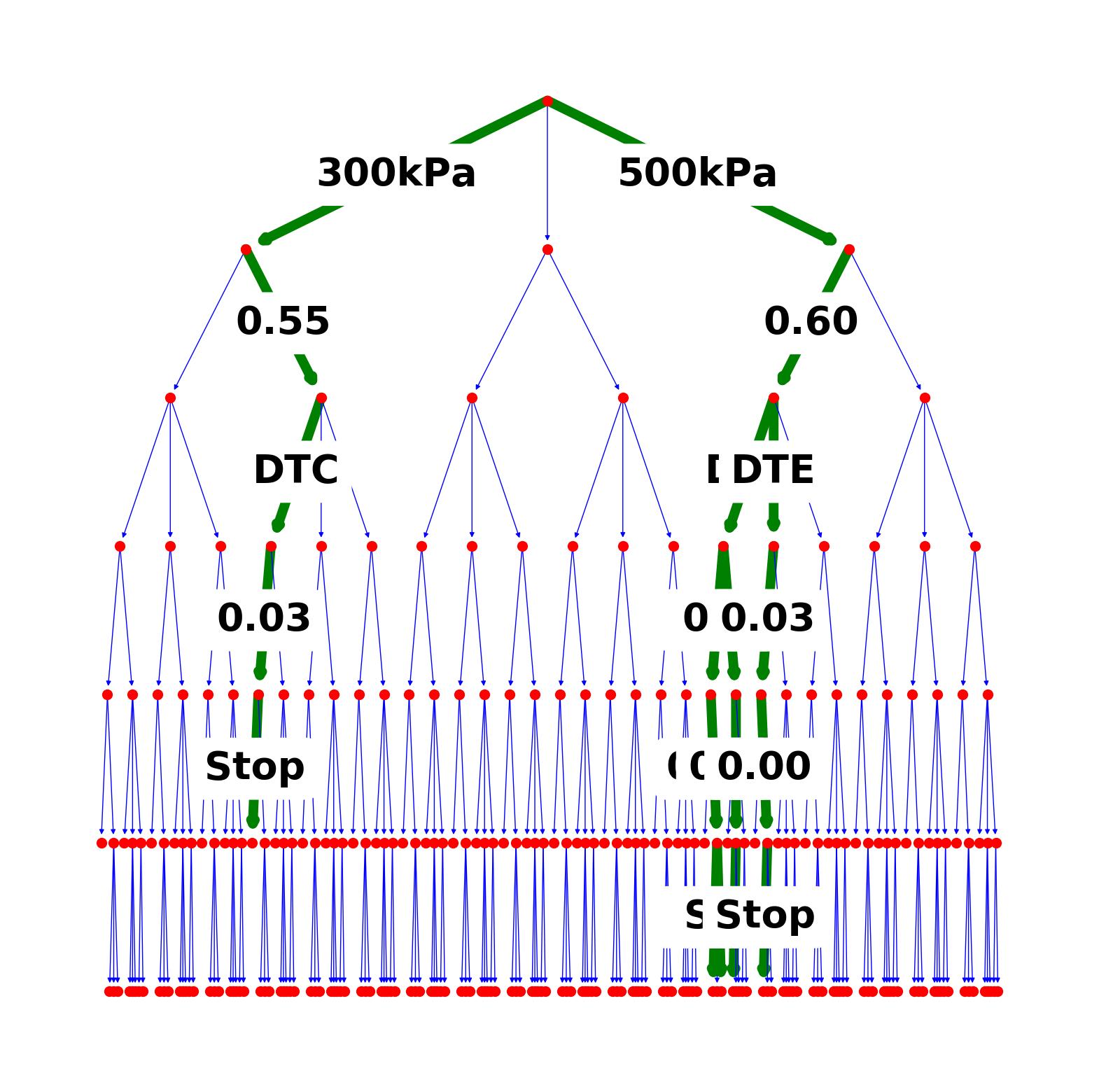}
	}
	\subfigure[Iteration 6, Episode 10, \newline \hspace{\linewidth} Attack Game Score: 0.295]{
		\includegraphics[width=0.235\textwidth]{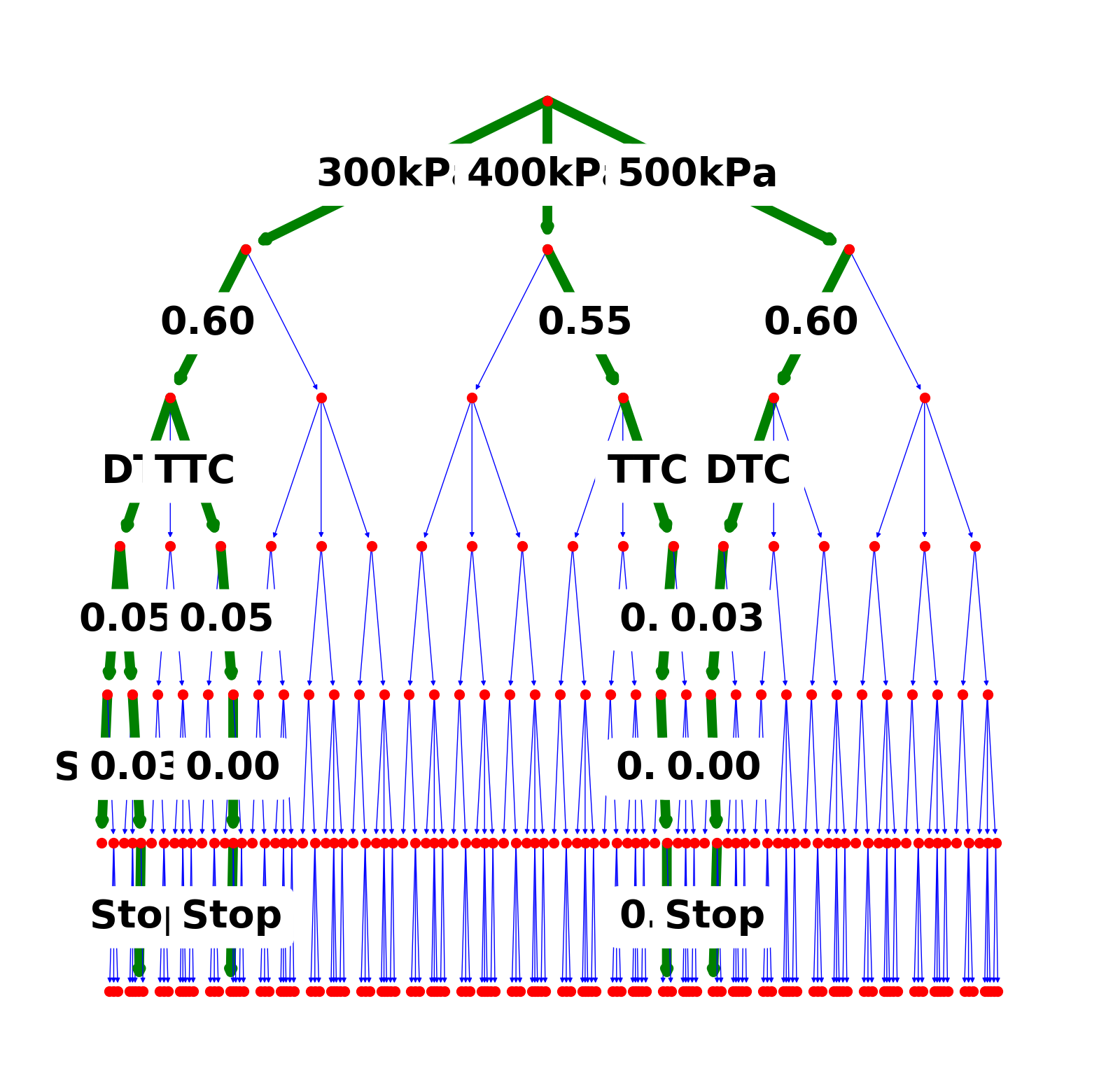}
	}
	\subfigure[Iteration 10, Episode 30, \newline \hspace{\linewidth} Attack Game Score: -0.290]{
		\includegraphics[width=0.235\textwidth]{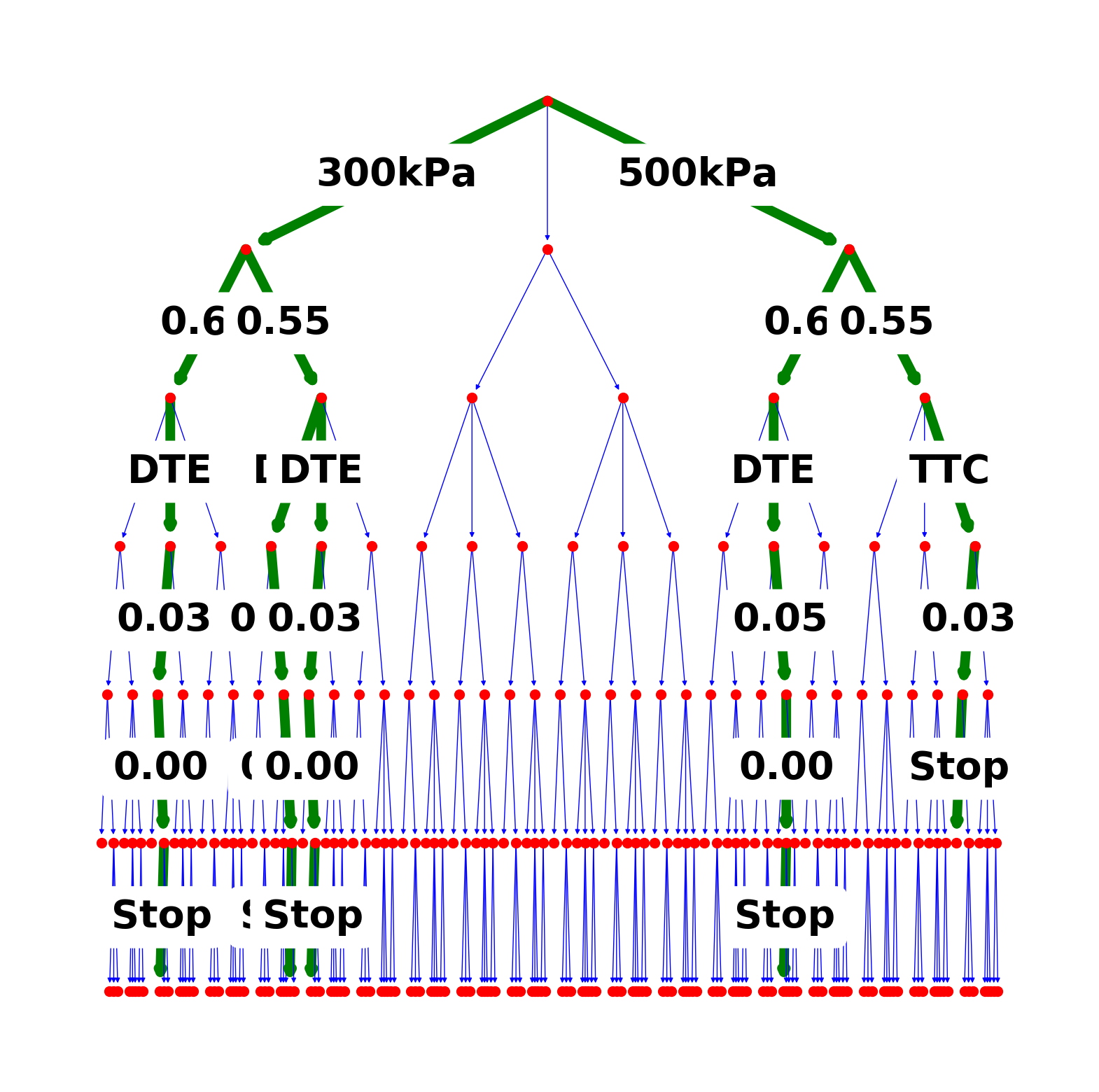}
	}
	\caption{Examples of paths (experiments) in the decision trees selected by the adversary during the DRL training iterations for SANISAND model.}
	\label{fig:SANISAND_attackgame_decisiontree}
\end{figure}

\begin{figure}[h!]\center
	\subfigure[Iteration 0, Episode 24, \newline \hspace{\linewidth} Defense Game Score: 0.162]{
		\includegraphics[width=0.235\textwidth]{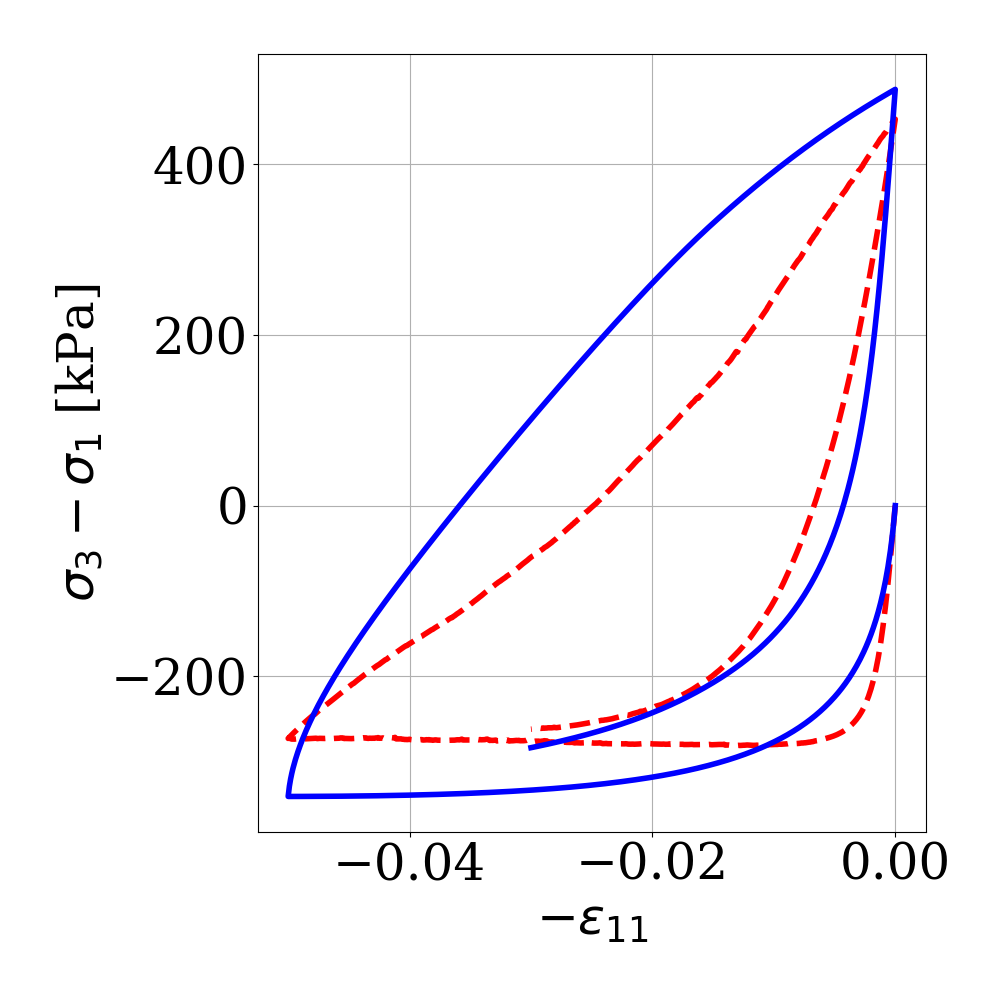}
	}
	\subfigure[Iteration 3, Episode 33, \newline \hspace{\linewidth} Defense Game Score: 0.626]{
		\includegraphics[width=0.235\textwidth]{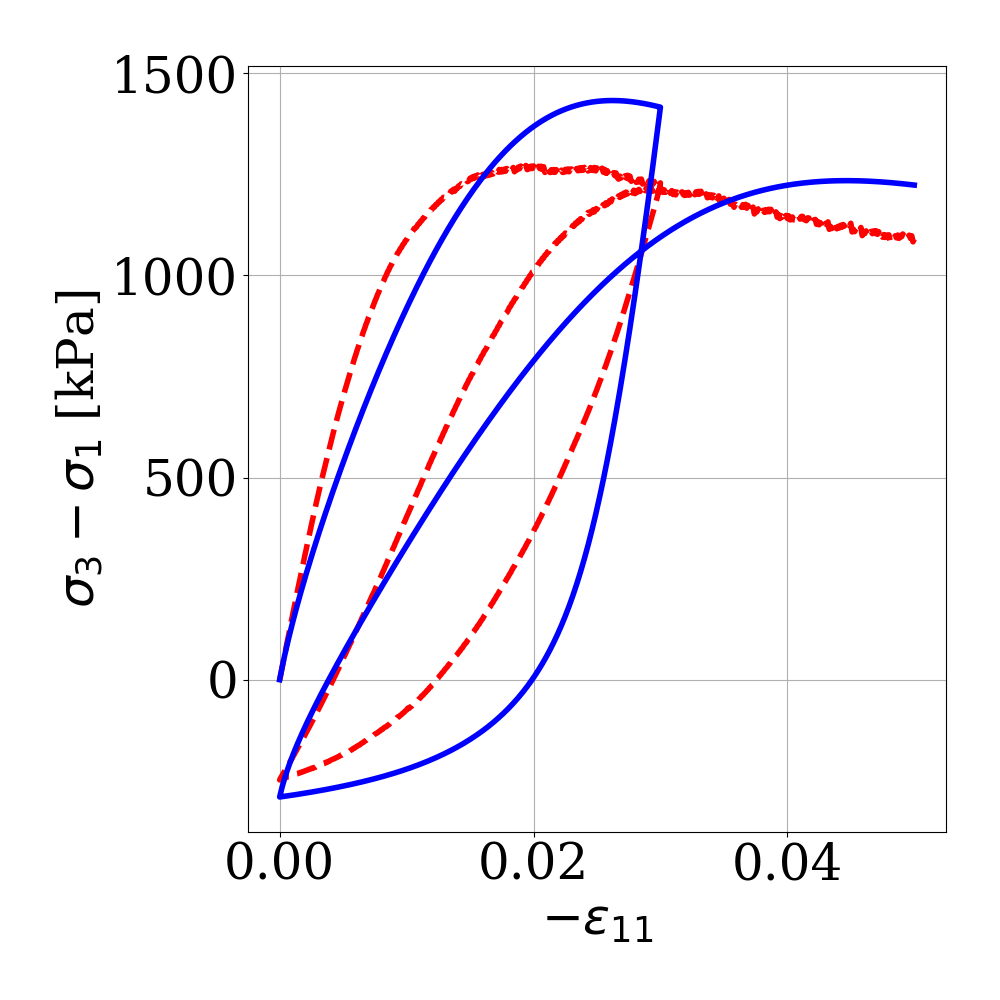}
	}
	\subfigure[Iteration 6, Episode 10, \newline \hspace{\linewidth} Defense Game Score: 0.687]{
		\includegraphics[width=0.235\textwidth]{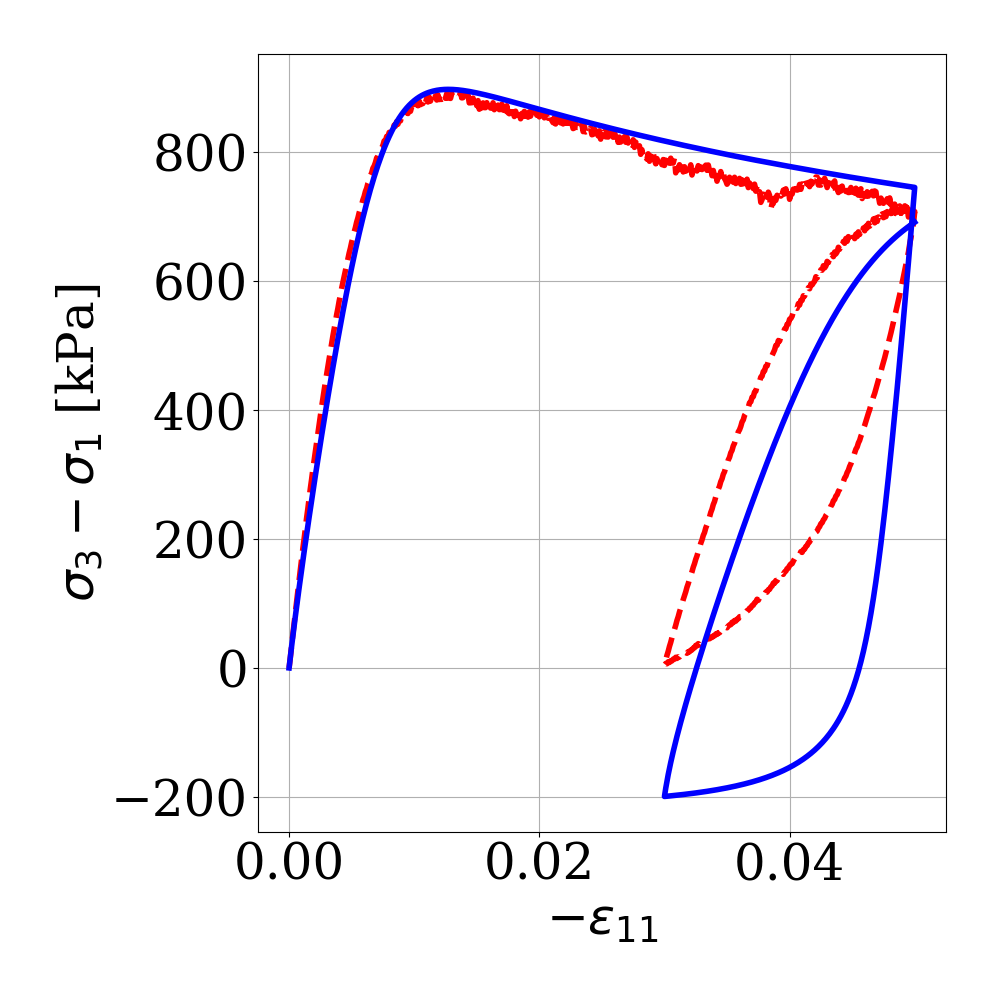}
	}
	\subfigure[Iteration 10, Episode 30, \newline \hspace{\linewidth} Defense Game Score: 0.912]{
		\includegraphics[width=0.235\textwidth]{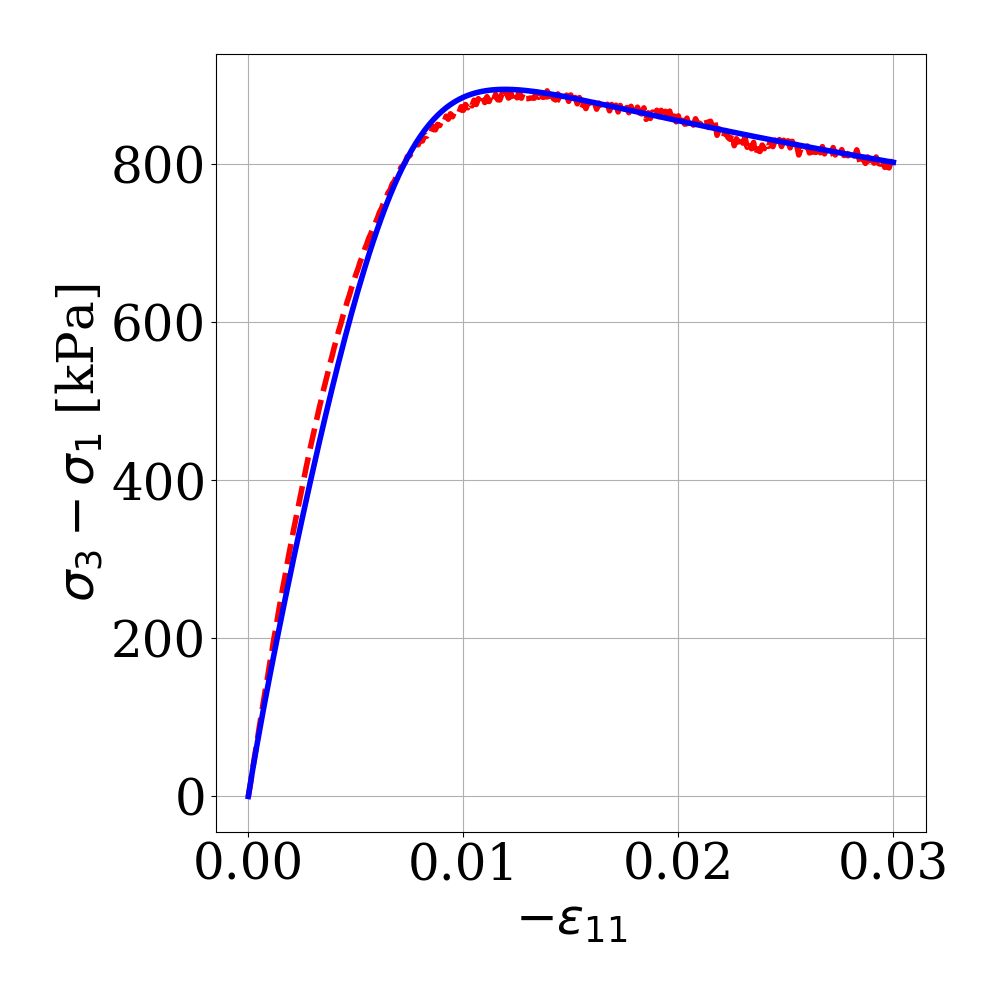}
	}
	\subfigure[Iteration 0, Episode 24, \newline \hspace{\linewidth} Defense Game Score: 0.162]{
		\includegraphics[width=0.235\textwidth]{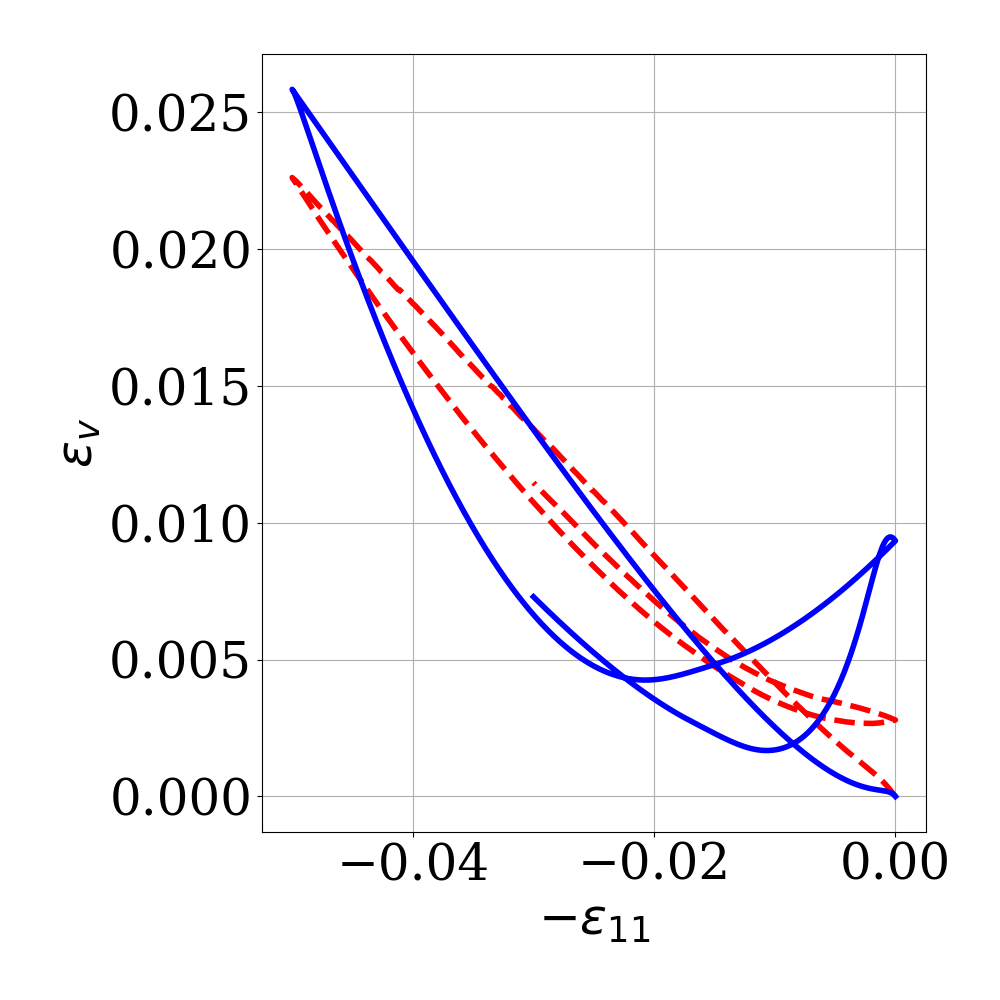}
	}
	\subfigure[Iteration 3, Episode 33, \newline \hspace{\linewidth} Defense Game Score: 0.626]{
		\includegraphics[width=0.235\textwidth]{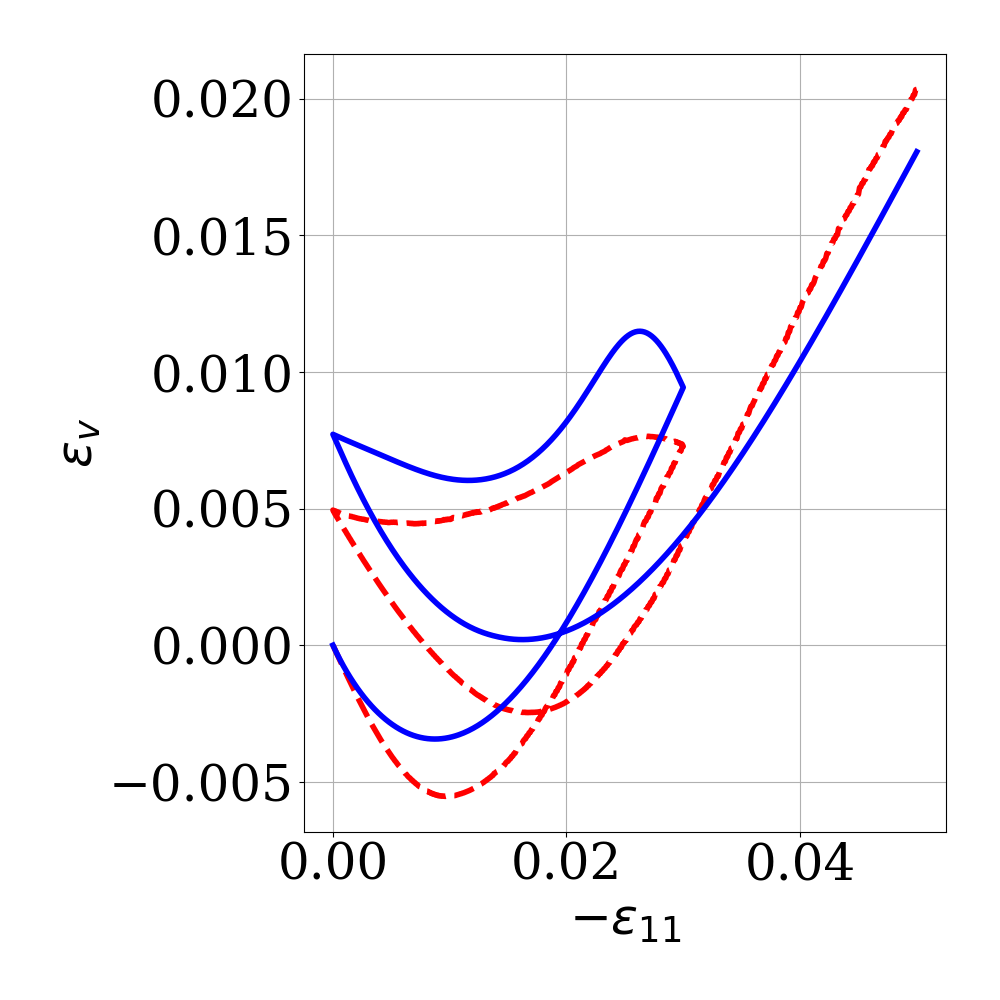}
	}
	\subfigure[Iteration 6, Episode 10, \newline \hspace{\linewidth} Defense Game Score: 0.687]{
		\includegraphics[width=0.235\textwidth]{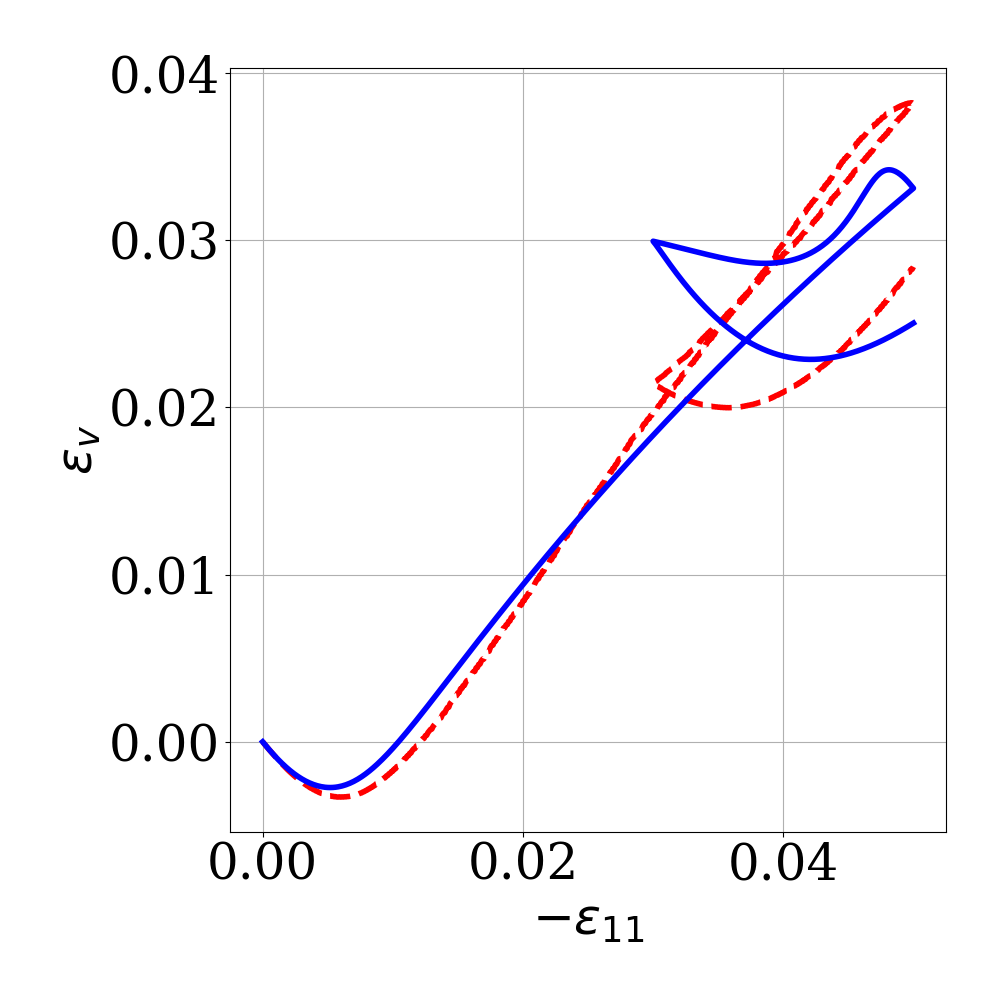}
	}
	\subfigure[Iteration 10, Episode 30, \newline \hspace{\linewidth} Defense Game Score: 0.912]{
		\includegraphics[width=0.235\textwidth]{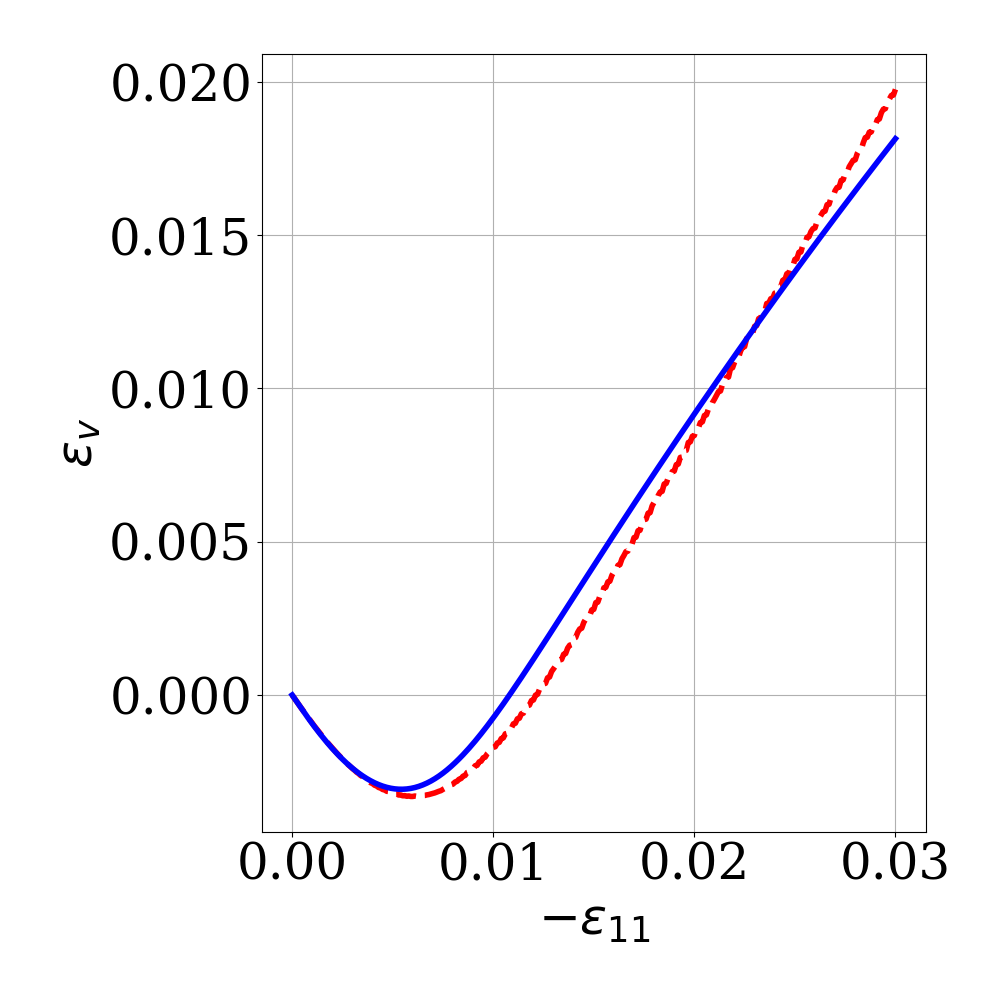}
	}
	\caption{Examples of response curves of the games played by the protagonist during the DRL training iterations for SANISAND model. Experimental data are plotted in red dashed curves, model predictions are plotted in blue solid curves. }
	\label{fig:SANISAND_datagame_curves}
\end{figure}

\begin{figure}[h!]\center
	\subfigure[Iteration 0, Episode 24, \newline \hspace{\linewidth} Attack Game Score: 0.206]{
		\includegraphics[width=0.235\textwidth]{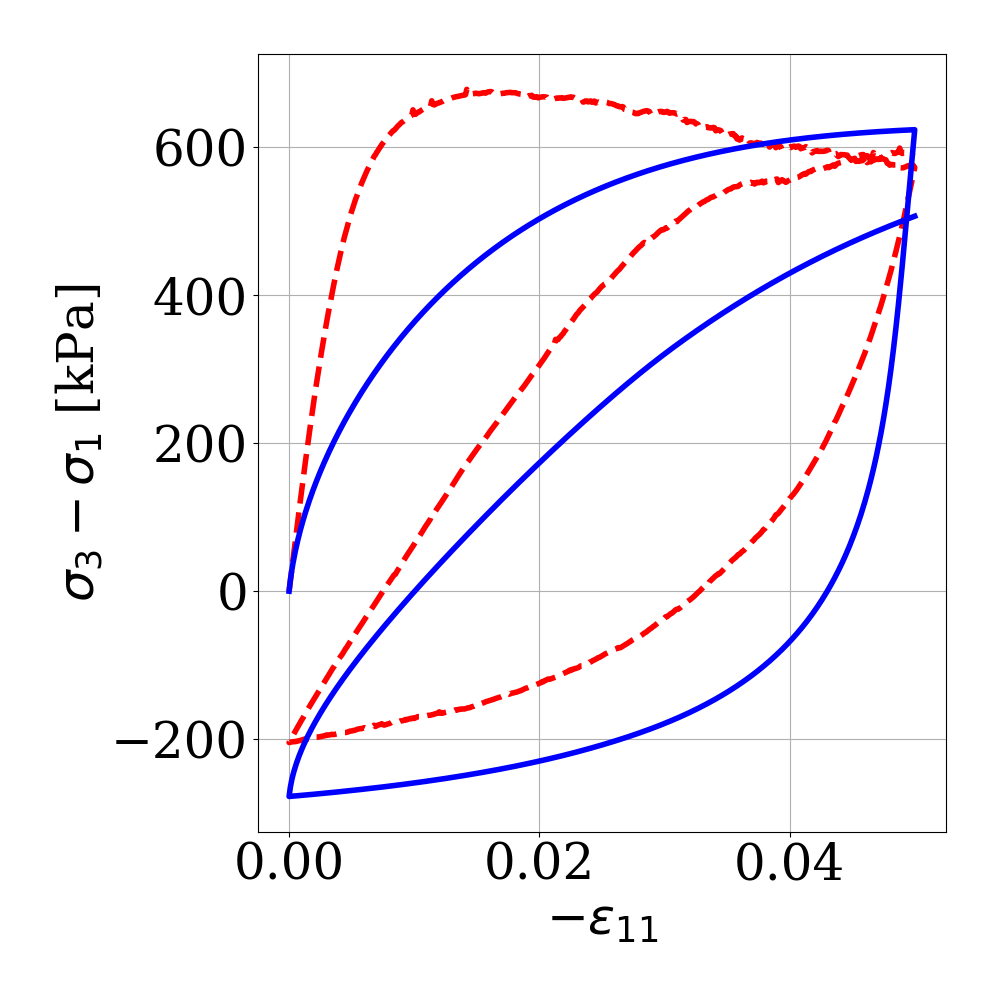}
	}
	\subfigure[Iteration 3, Episode 33, \newline \hspace{\linewidth} Attack Game Score: 0.323]{
		\includegraphics[width=0.235\textwidth]{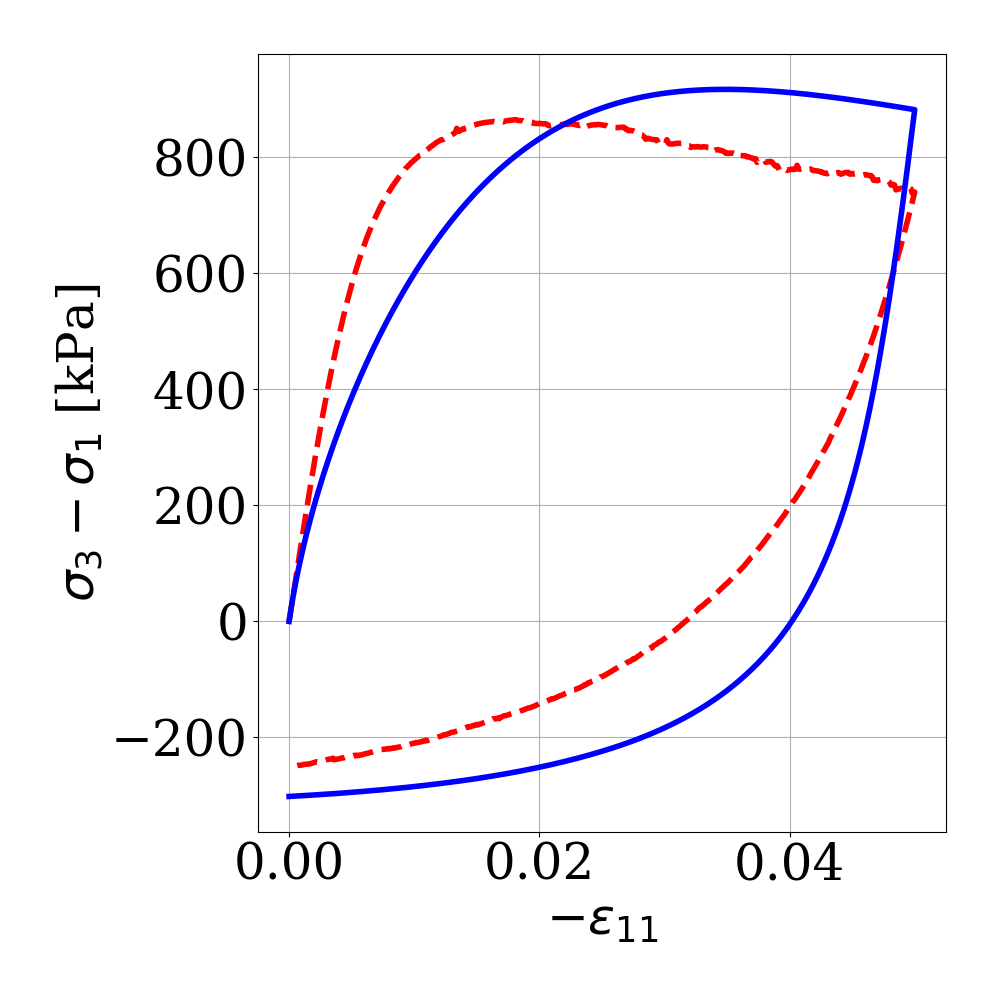}
	}
	\subfigure[Iteration 6, Episode 10, \newline \hspace{\linewidth} Attack Game Score: 0.295]{
		\includegraphics[width=0.235\textwidth]{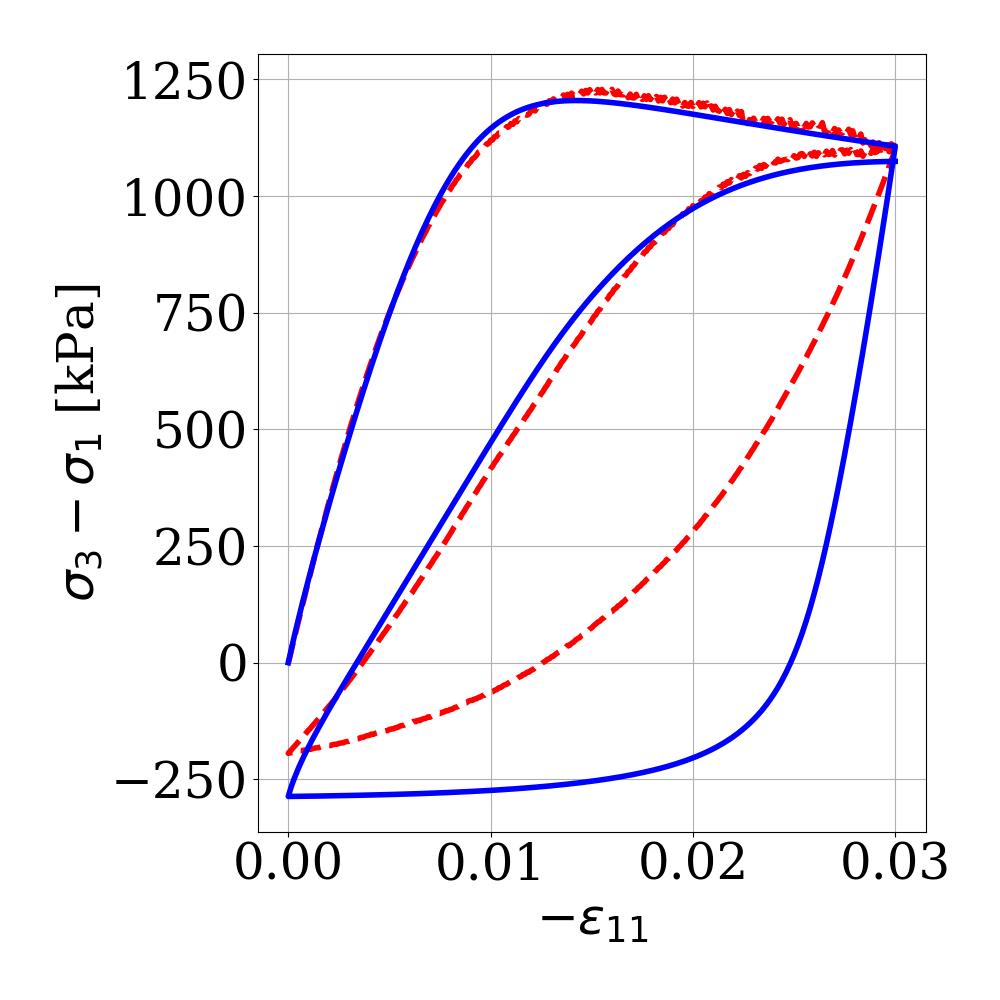}
	}
	\subfigure[Iteration 10, Episode 30, \newline \hspace{\linewidth} Attack Game Score: -0.290]{
		\includegraphics[width=0.235\textwidth]{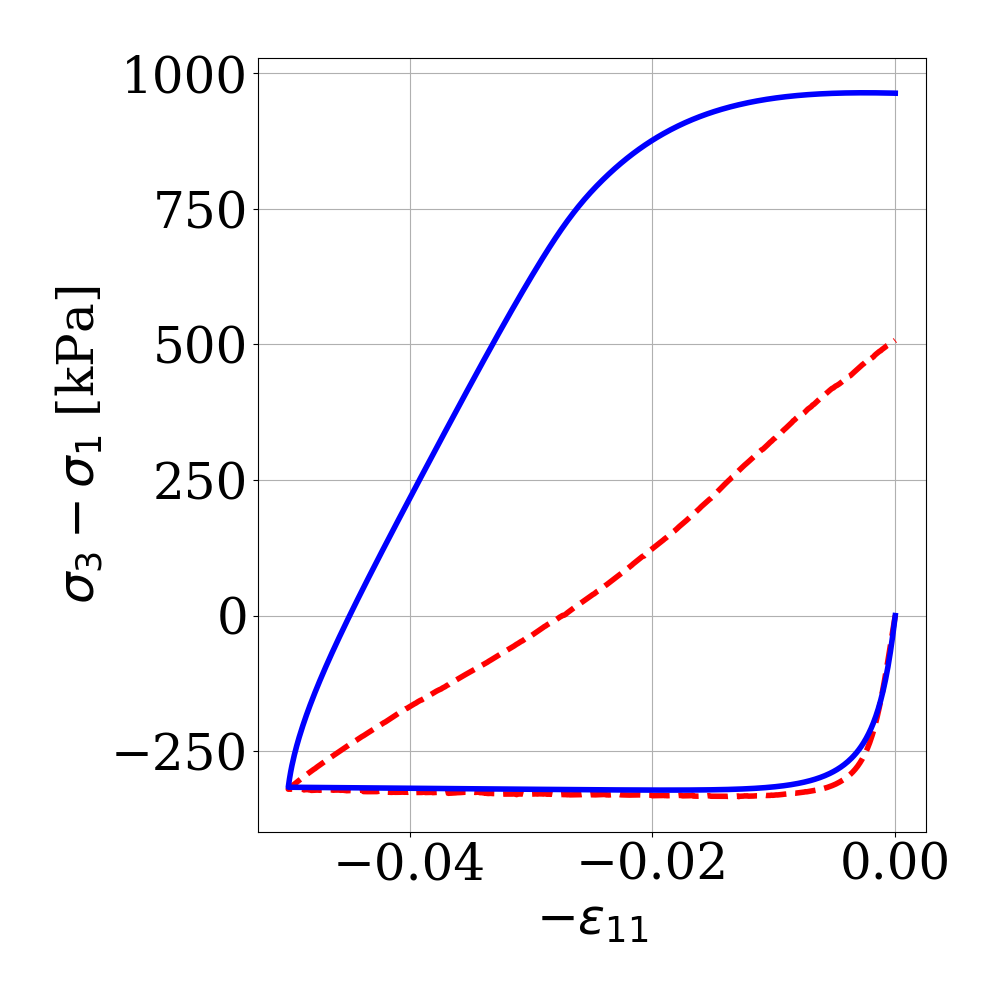}
	}
	\subfigure[Iteration 0, Episode 24, \newline \hspace{\linewidth} Attack Game Score: 0.206]{
		\includegraphics[width=0.235\textwidth]{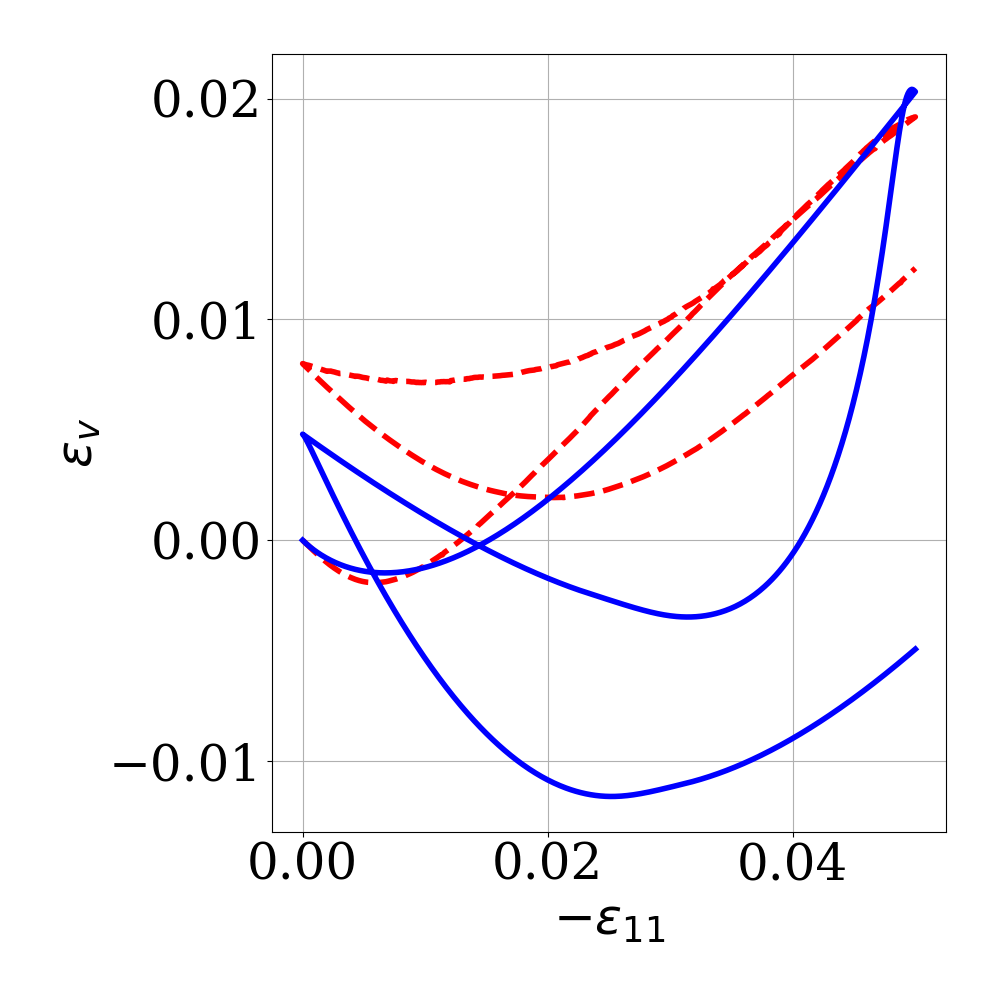}
	}
	\subfigure[Iteration 3, Episode 33, \newline \hspace{\linewidth} Attack Game Score: 0.323]{
		\includegraphics[width=0.235\textwidth]{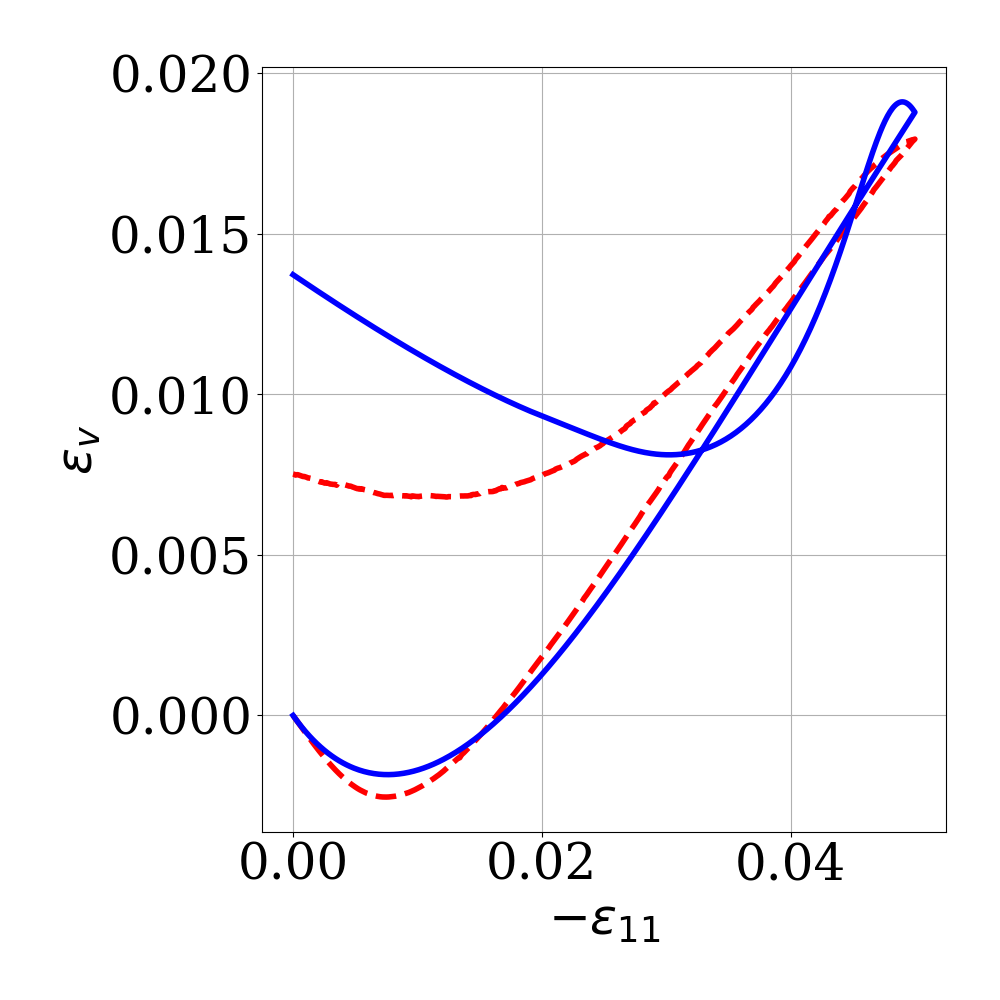}
	}
	\subfigure[Iteration 6, Episode 10, \newline \hspace{\linewidth} Attack Game Score: 0.295]{
		\includegraphics[width=0.235\textwidth]{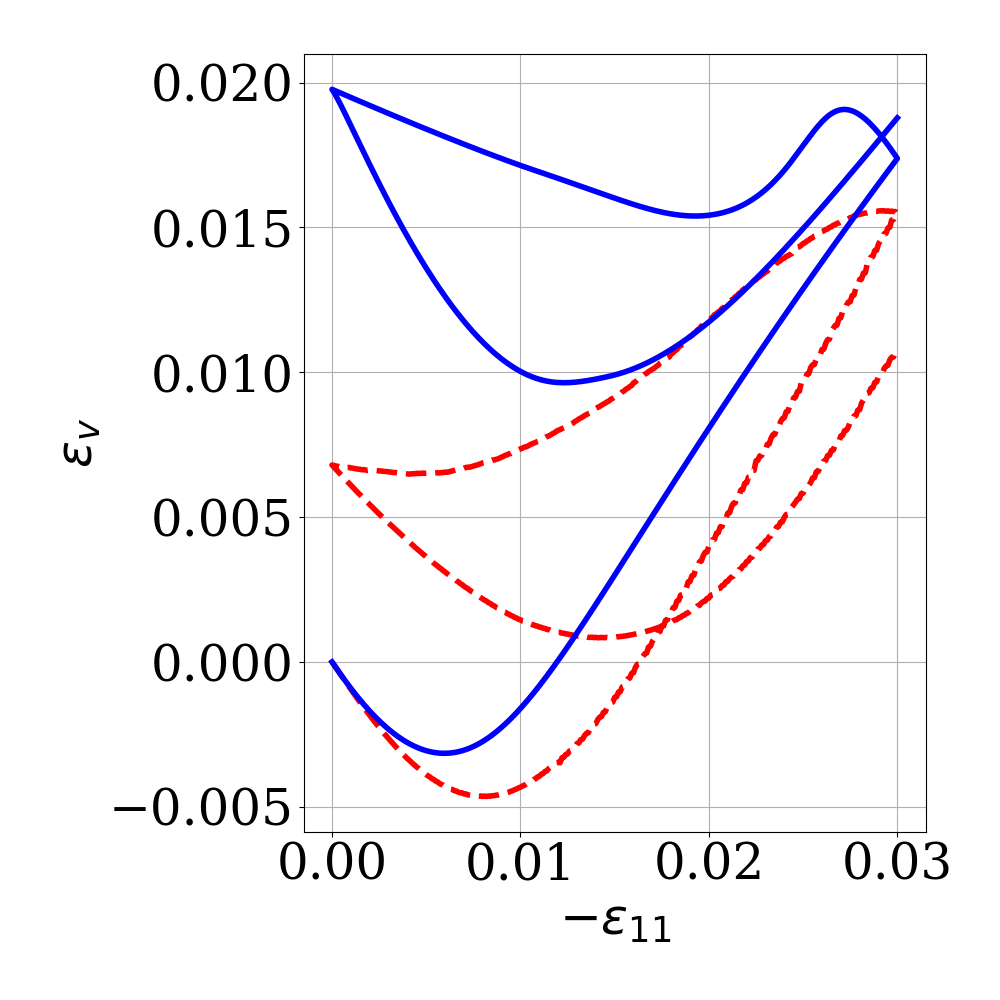}
	}
	\subfigure[Iteration 10, Episode 30, \newline \hspace{\linewidth} Attack Game Score: -0.290]{
		\includegraphics[width=0.235\textwidth]{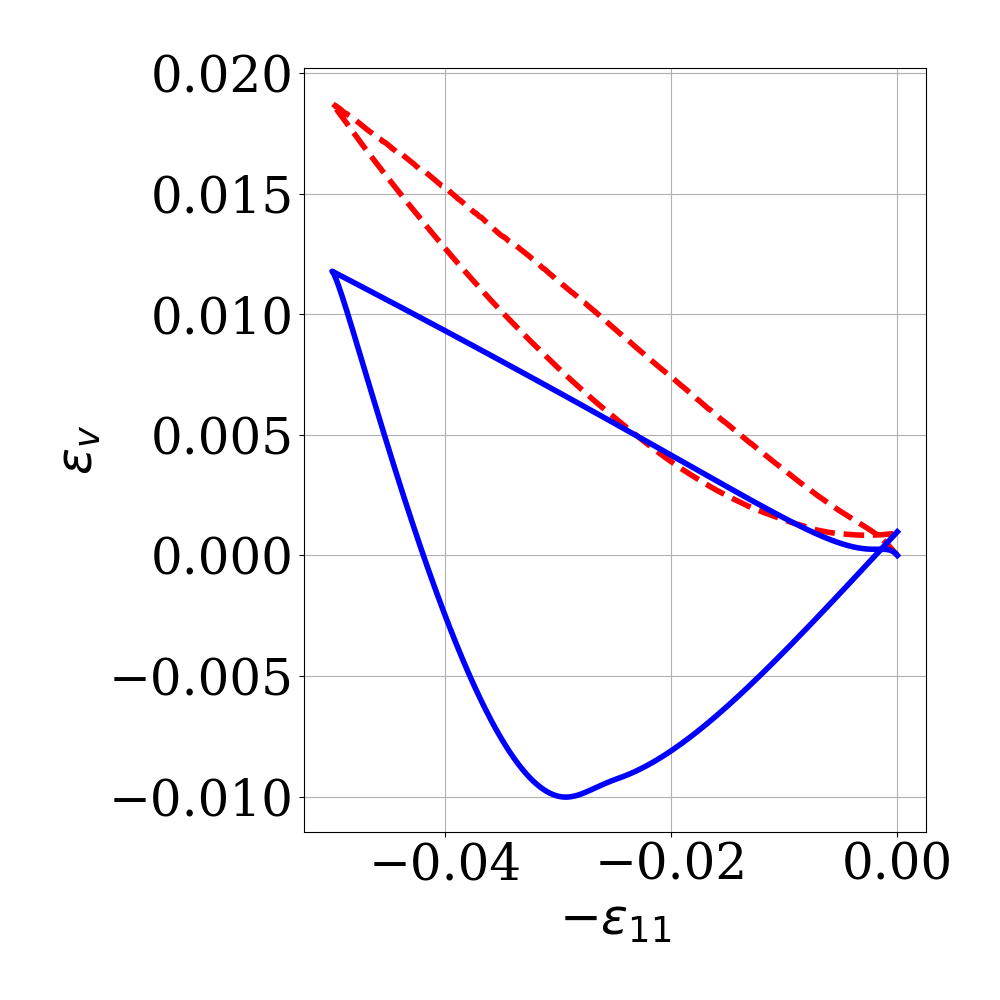}
	}
	\caption{Examples of response curves of the games played by the adversary during the DRL training iterations for SANISAND model. Experimental data are plotted in red dashed curves, model predictions are plotted in blue solid curves. }
	\label{fig:SANISAND_attackgame_curves}
\end{figure}

The statistics of the game scores played for the "Calibration/Defense" by the protagonist and the "Falsification/Attack" by the adversary during the DRL iterations are shown in Fig. \ref{fig:SANISAND_learn_violinplot}. 
The AI agents only know the experimental decision tree and the rules of the two-player game without any prior knowledge on the strengths and weaknesses of the SANISAND model. 
The improvement of the protagonist's policy is shown by the increase of the median of game scores, and also the narrowing of inter-quantile ranges. 
Some lower scores encountered during the later iterations of the DRL are due to random explorations of game strategies by the agent. 
Fig. \ref{fig:SANISAND_datagame_decisiontree} provides some example experiments selected by the protagonist for calibration data and Fig. \ref{fig:SANISAND_datagame_curves} provides some example response curves associated to these experiments. 
Meanwhile, the adversary tries to attack the models calibrated by the protagonist using some experiments as shown in Fig. \ref{fig:SANISAND_attackgame_decisiontree}. 
Fig. \ref{fig:SANISAND_attackgame_curves} gives example response curves associated to these adversarial decision tree paths. 
These attacks inform and drive the protagonist to find more adequate calibration data via the score systems in this game. 

In the end, the protagonist concludes that the model is accurate in modeling the mechanical behaviour of a single sample in TTC test with monotonic loading. 
In this case, the final value explored for $\min (\{E^1_{NS}\})$ is $-0.933$, which is slightly above the lower bound $E^{min}_{NS}=-1.0$ in the game score setting. 
Hence the decay coefficient in Eq. (\ref{eq:reward_protagonist}) is not activated and the protagonist score is equal to the calibration score. 
The adversary concludes that, when calibrated with this monotonic TTC data, the model is not accurate in predicting DTC, DTE, TTC experiments on other samples with unloading-reloading. 
Nevertheless, based on all the game episodes played during the DRL, the agents learn that the SANISAND model is capable of replicating the hardening-softening and contraction-dilation behavior of a densely compressed granular material. 
They also learn that SANISAND is more powerful than Drucker-Prager in replicating data from samples with different initial confinement, initial void ratio, test types, and unloading-reloading paths. 
The not perfect calibrations and forward predictions as shown in Fig. \ref{fig:SANISAND_datagame_curves} and Fig. \ref{fig:SANISAND_attackgame_curves} may be due to the choice of calibration procedures, or because the formulations of SANISAND are designed based on the real sand behaviour, but our data comes from DEM numerical samples. 

\subsection{Experiment 3: Deep learning graph-based traction-separation model}
\label{subsec:rnntsmodel_example}
The two-player non-cooperative game is played by DRL-based AI experimentalists for the data-driven traction-separation model. 
The neural network architectures and the calibration of the model are detailed in Section \ref{subsec:rnntsmodel}. 
The game settings are $N^{max}_{path}=15$ for the protagonist and $N^{max}_{path}=10$ for the adversary, $E^{max}_{NS}=1.0$, $E^{min}_{NS}=0.8$. Hence the combination number of the selected experimental decision tree paths in this example is about $1e23$. 
The hyperparameters for the DRL algorithm are $numIters=10$, $numEpisodes=40$, $numMCTSSims=50$, $\alpha_{\text{SCORE}}=2.5$, $\alpha_{range} = 0.1$, $i_{lookback}=4$, $\tau_{train} = 1.0$, $\tau_{test} = 0.1$. 
The policy/value networks are identical to the ones used in the previous examples. 
In this example, since the number of the possible game configurations is enormous, we constrain the policy of the protagonist to play only the winning games (those who have got their rewards $\text{Reward}_{\text{protagonist}} \geq R^{max}_{p} - 0.1*R^{range}_{p}*\alpha_{range}$) in 1/5 of the 40 game episodes in each training iteration and in the last iteration of "competitive gameplays". 
Also, we manually pre-select 10 experiments that have two loading cycles and regard them as the "shared" test data. 
They need to be predicted by all calibrated traction-separation models, along with the test data selected by the adversary, in order to help explore $\min (\{E^1_{NS}\})$ in Eq. (\ref{eq:reward_protagonist}). 
These methods are applied in addition to the general reinforcement learning framework in Section \ref{sec:gamedrl} to enhance the convergence of the agents' gameplay strategies. 

\begin{figure}[h!]\center
	\subfigure[Protagonist]{
		\includegraphics[width=0.45\textwidth]{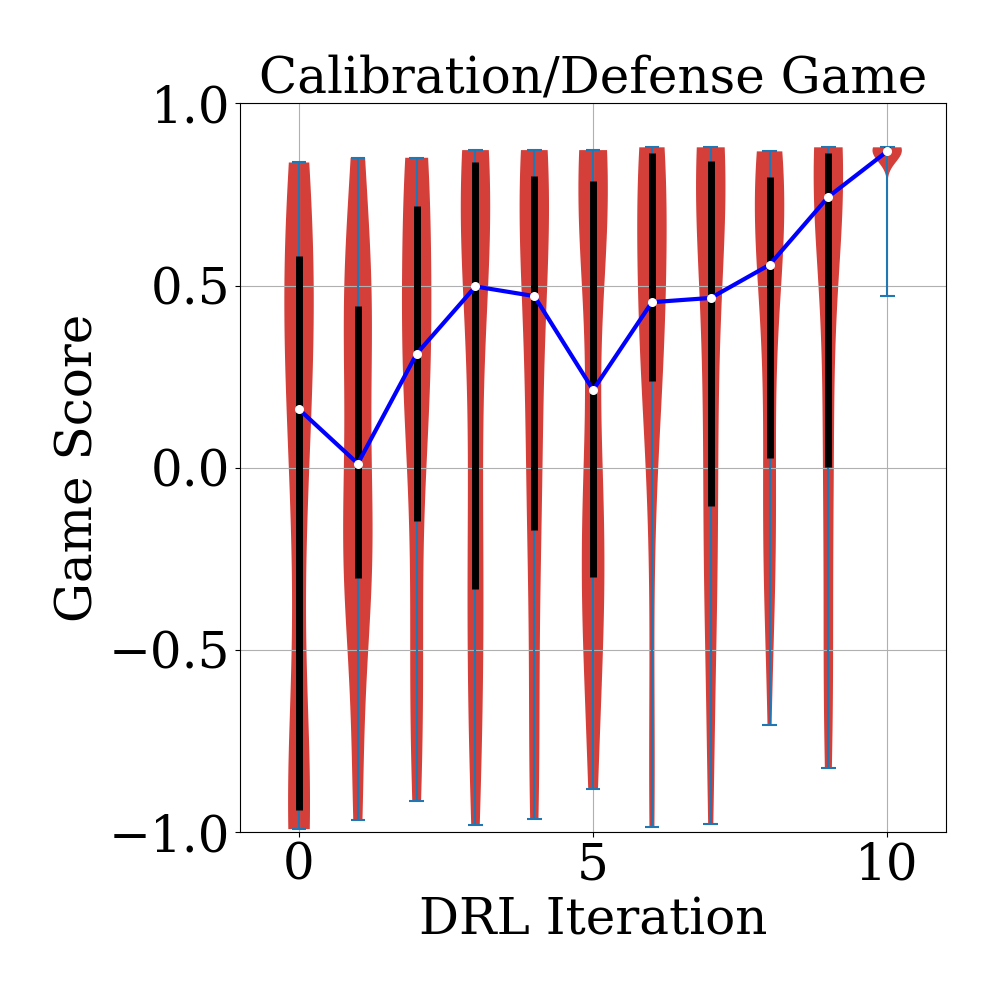}
	}
	\subfigure[Adversary]{
		\includegraphics[width=0.45\textwidth]{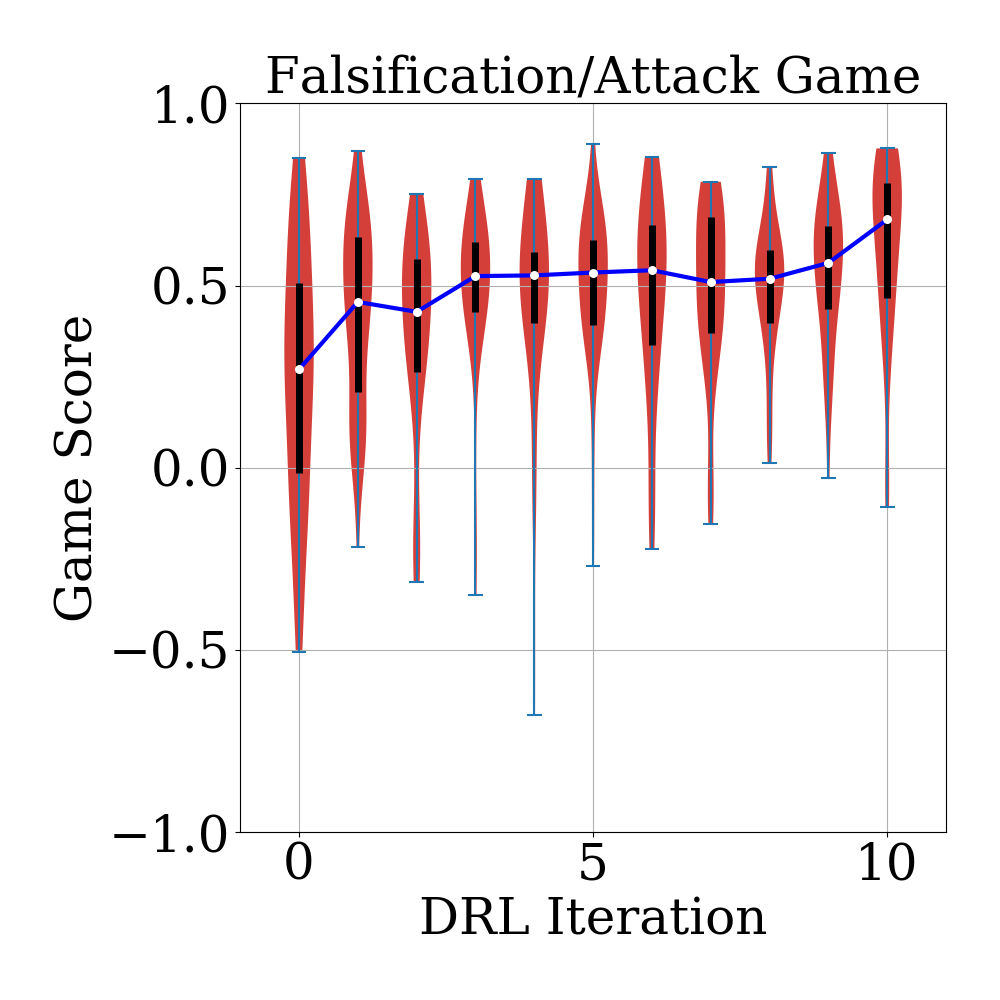}
	}
	\caption{Violin plots of the density distributions of game scores in each DRL iteration in data-driven traction-separation model. The shaded area represents the density distribution of scores. The white point represents the median. The thick black bar represents the inter-quantile range between 25\% quantile and 75\% quantile. The maximum and minimum scores played in each iteration are marked.}
	\label{fig:TS_learn_violinplot}
\end{figure}

\begin{figure}[h!]\center
	\subfigure[Iteration 0, Episode 11, \newline \hspace{\linewidth} Defense Game Score: -0.992]{
		\includegraphics[width=0.235\textwidth]{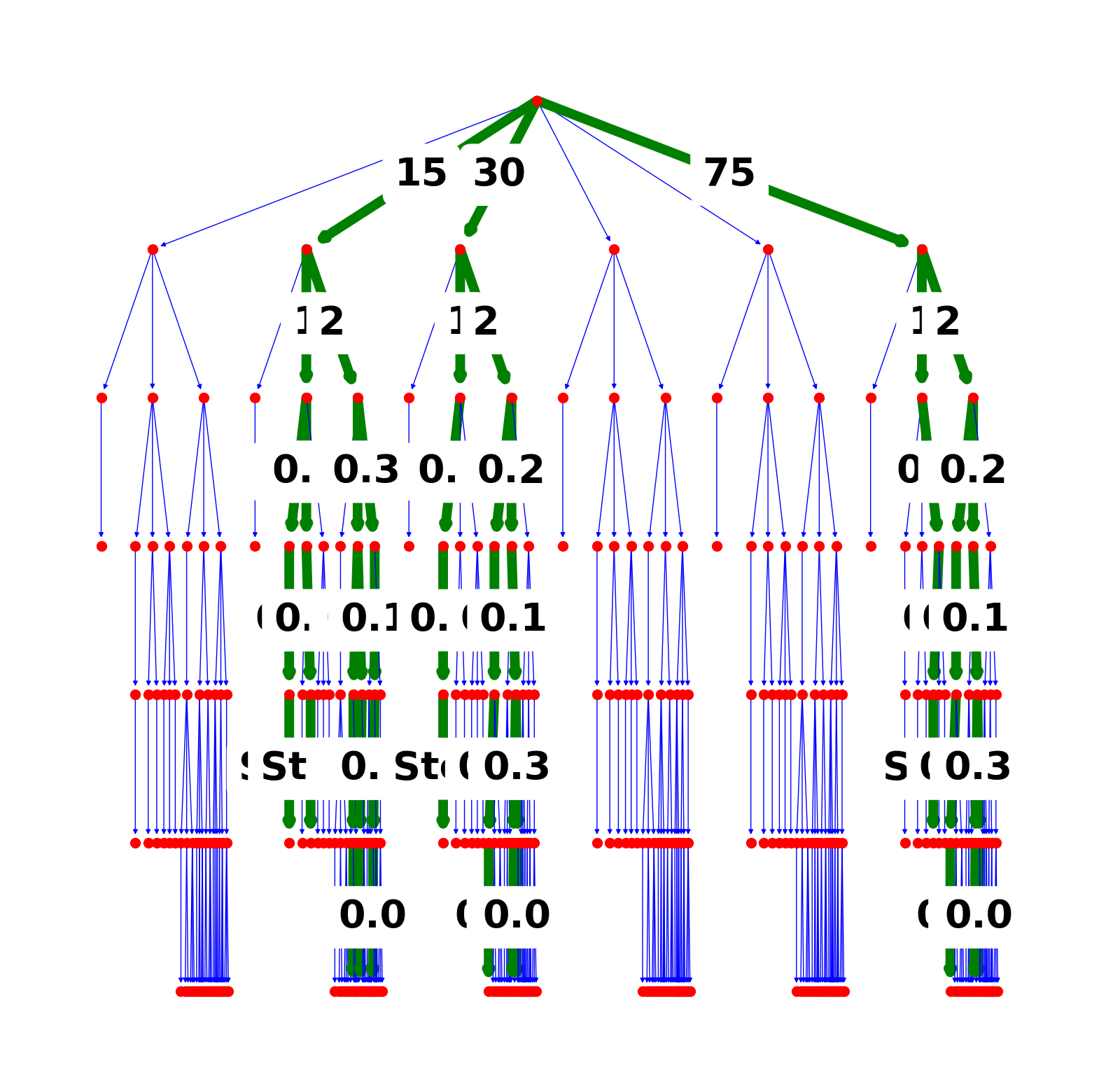}
	}
	\subfigure[Iteration 3, Episode 7, \newline \hspace{\linewidth} Defense Game Score: -0.331]{
		\includegraphics[width=0.235\textwidth]{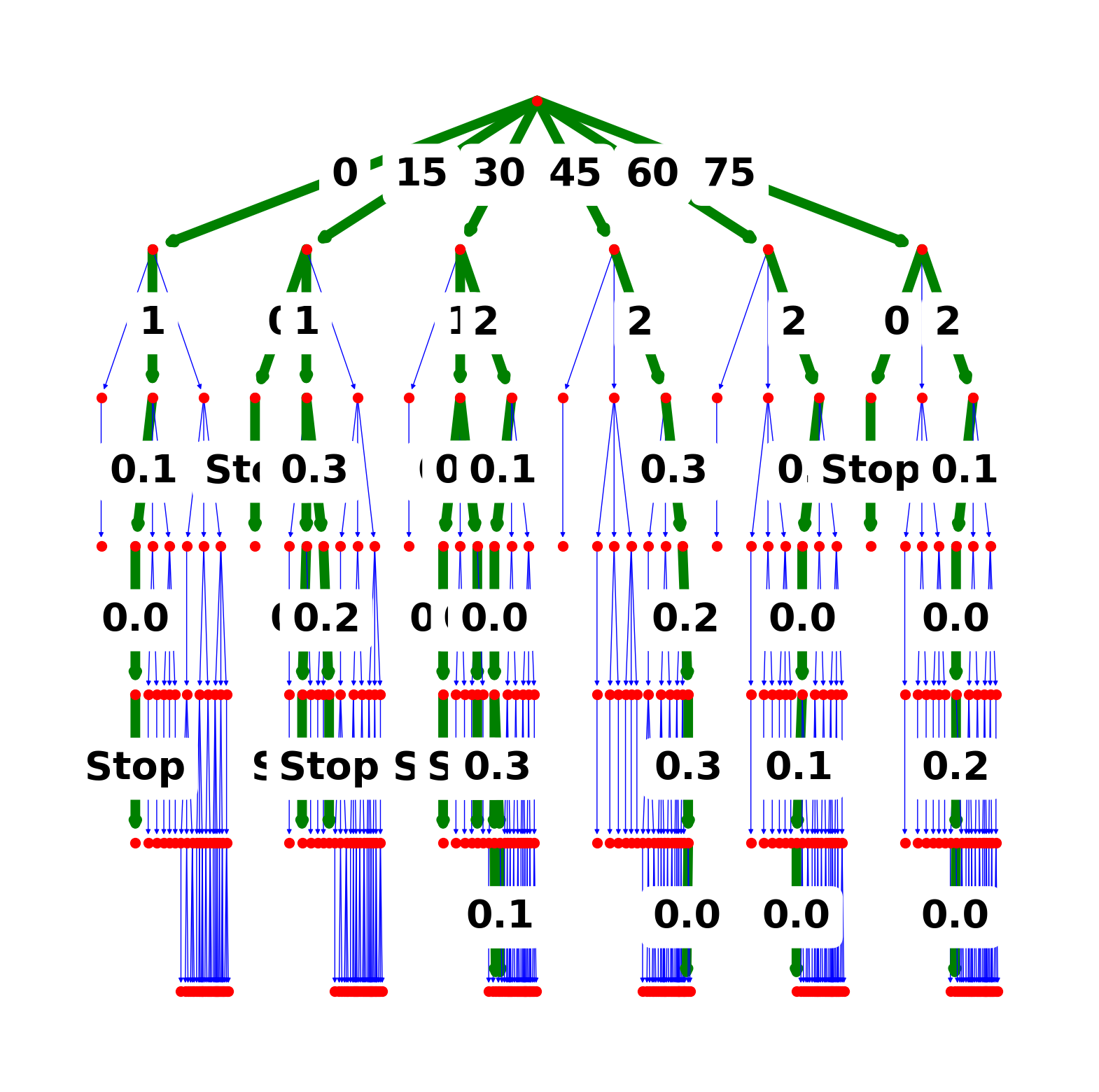}
	}
	\subfigure[Iteration 6, Episode 31, \newline \hspace{\linewidth} Defense Game Score: 0.269]{
		\includegraphics[width=0.235\textwidth]{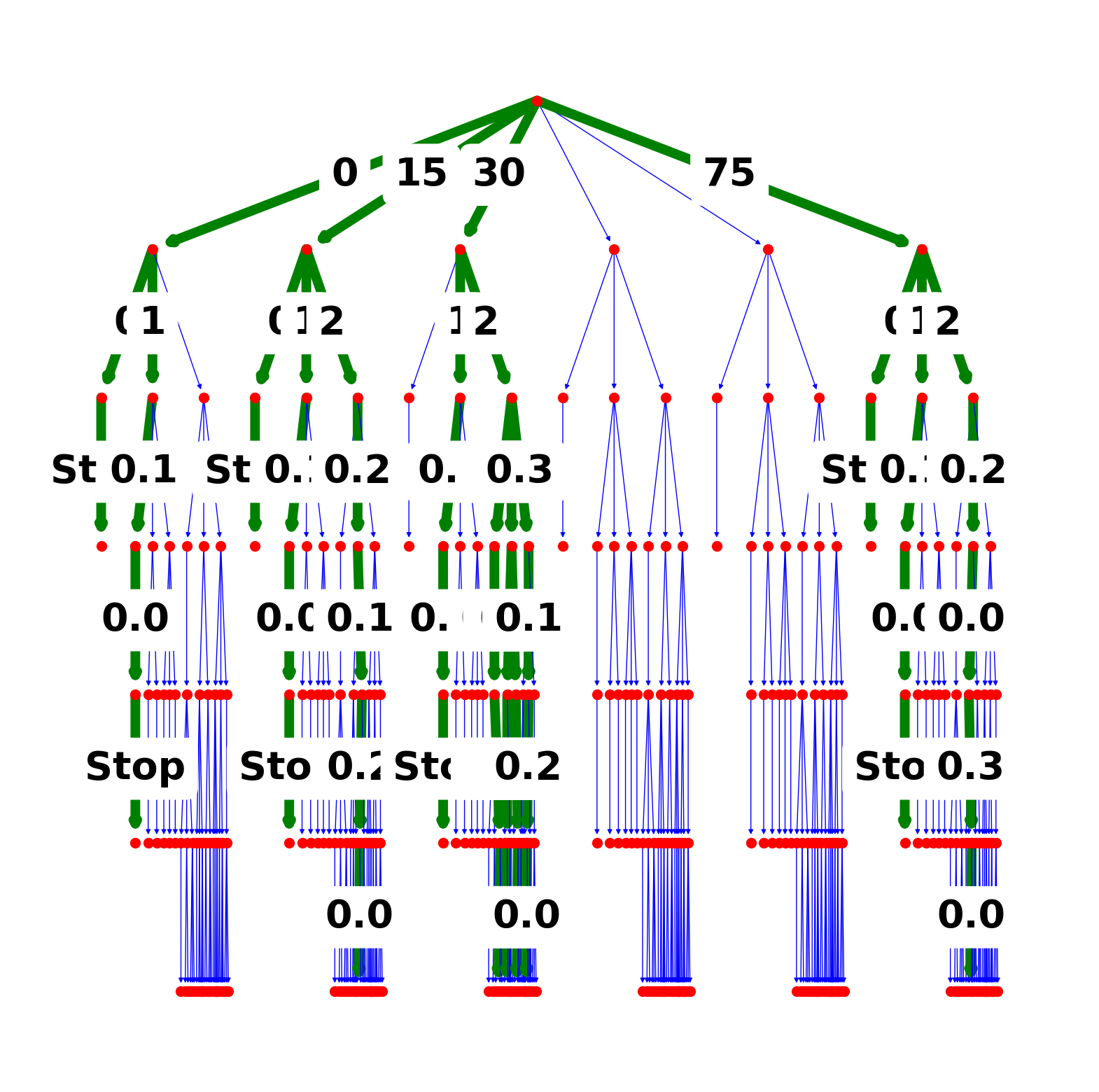}
	}
	\subfigure[Iteration 10, Episode 20, \newline \hspace{\linewidth} Defense Game Score: 0.879]{
		\includegraphics[width=0.235\textwidth]{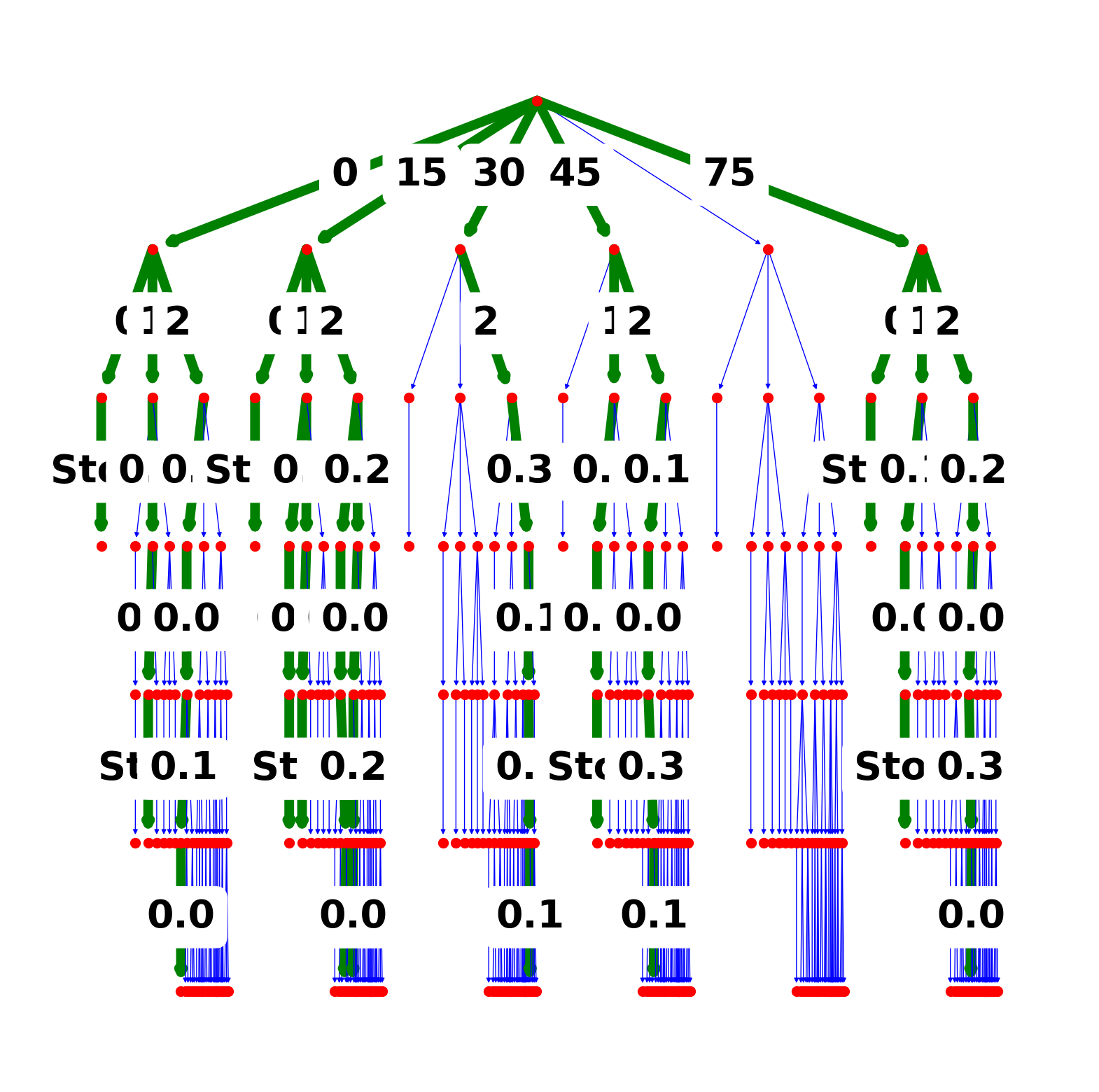}
	}
	\caption{Examples of paths (experiments) in the decision trees selected by the protagonist during the DRL training iterations for traction-separation model.}
	\label{fig:TS_datagame_decisiontree}
\end{figure}

\begin{figure}[h!]\center
	\subfigure[Iteration 0]{
		\includegraphics[width=0.235\textwidth]{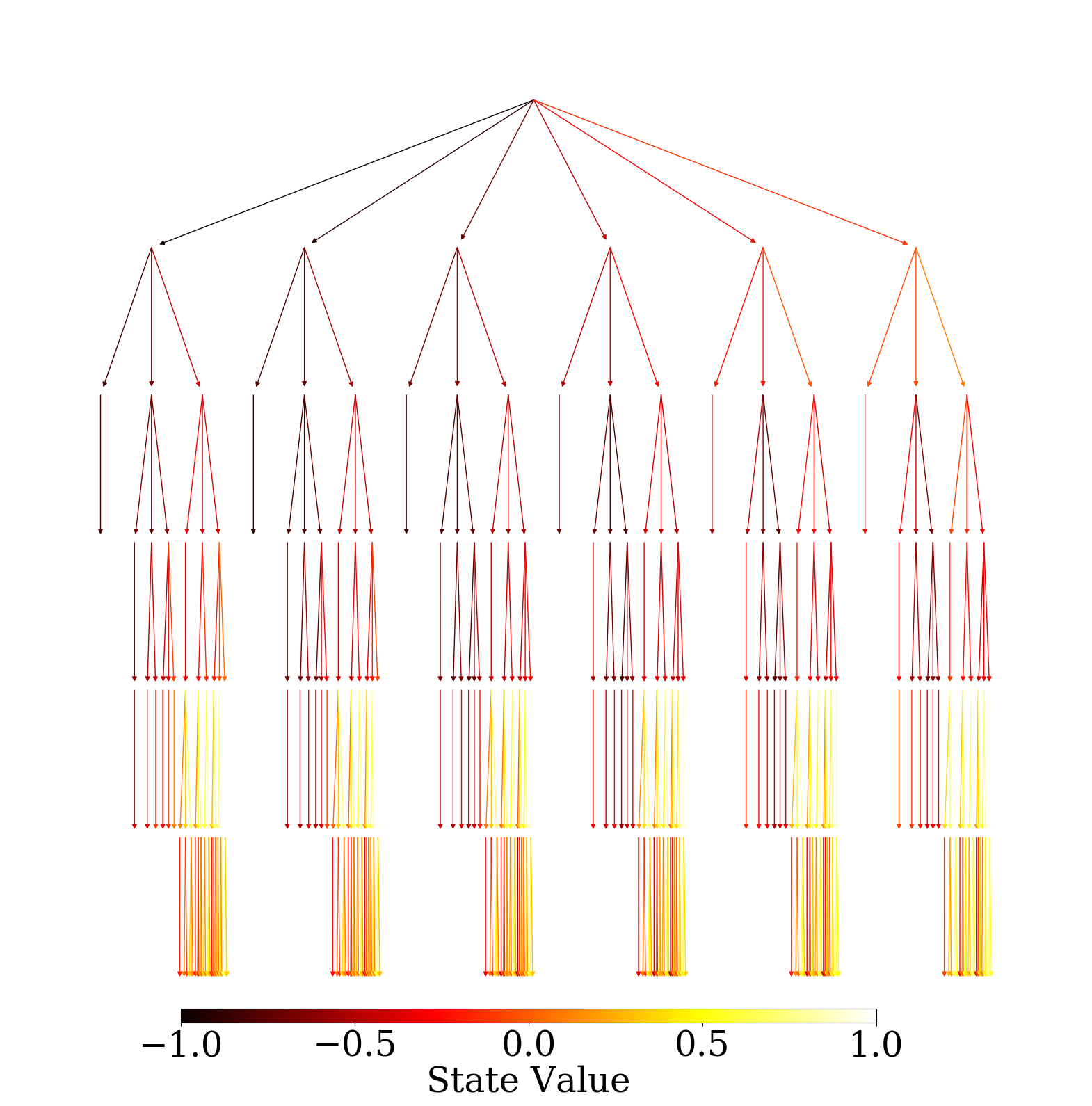}
	}
	\subfigure[Iteration 3]{
		\includegraphics[width=0.235\textwidth]{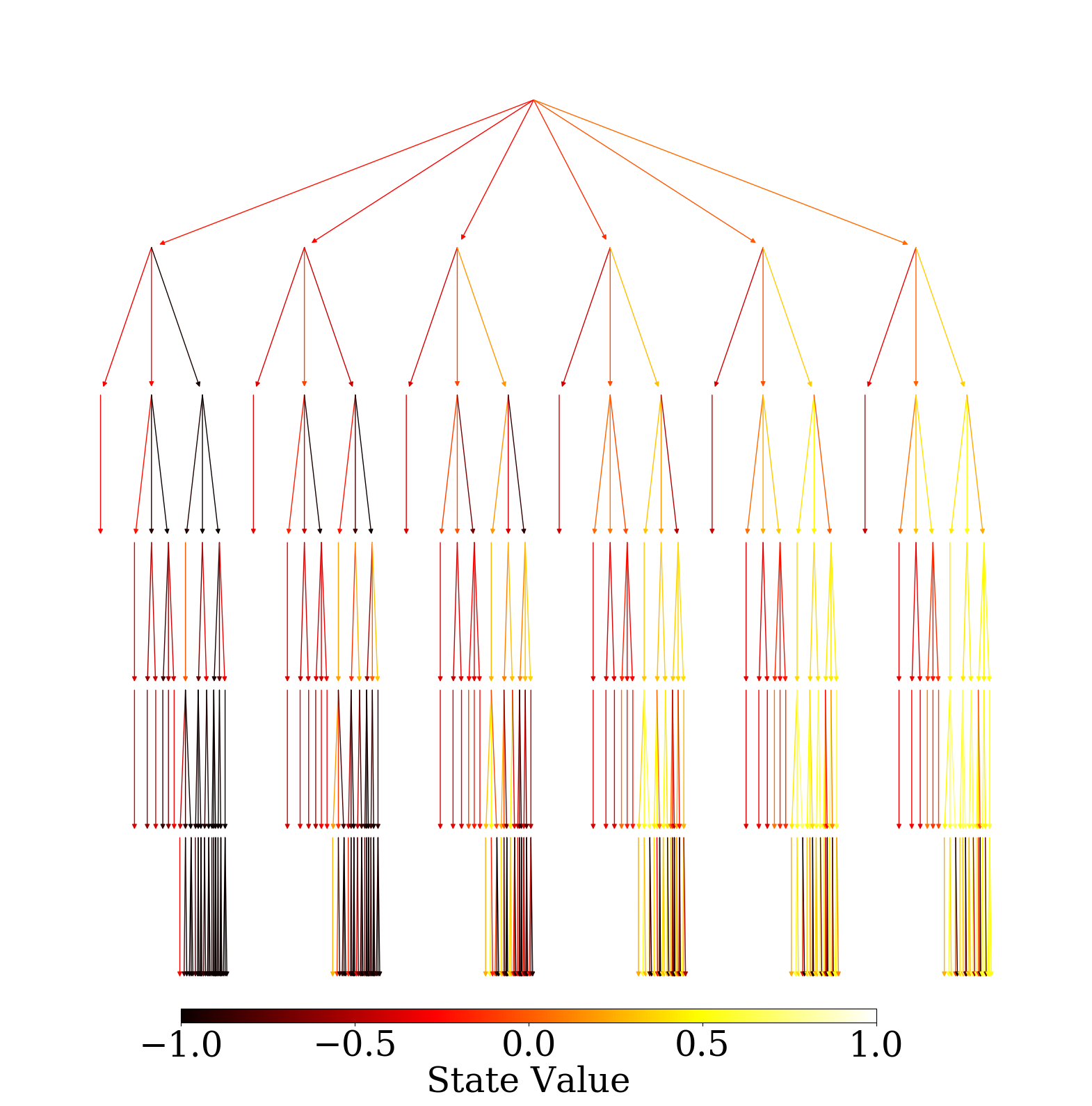}
	}
	\subfigure[Iteration 6]{
		\includegraphics[width=0.235\textwidth]{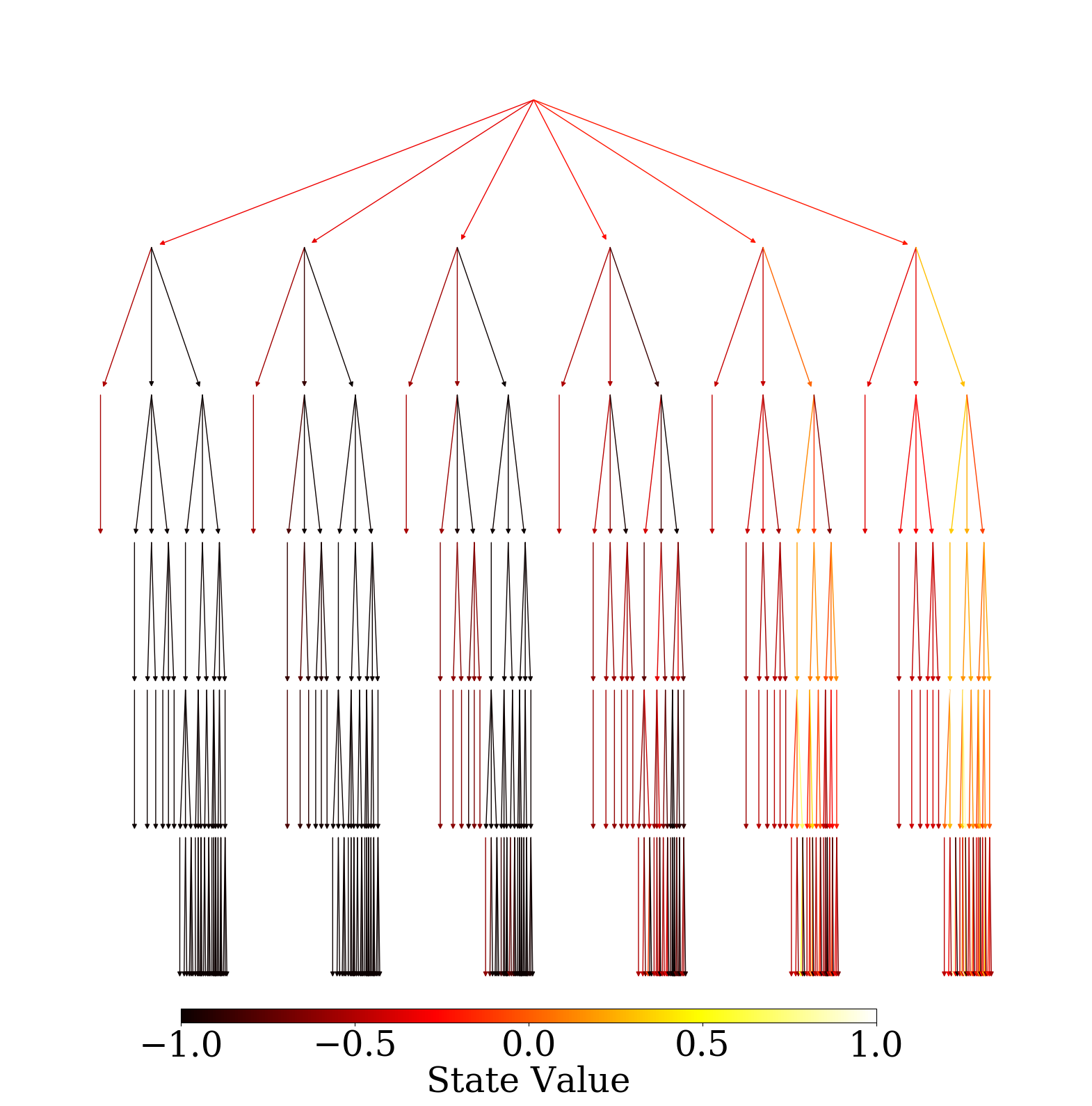}
	}
	\subfigure[Iteration 10]{
		\includegraphics[width=0.235\textwidth]{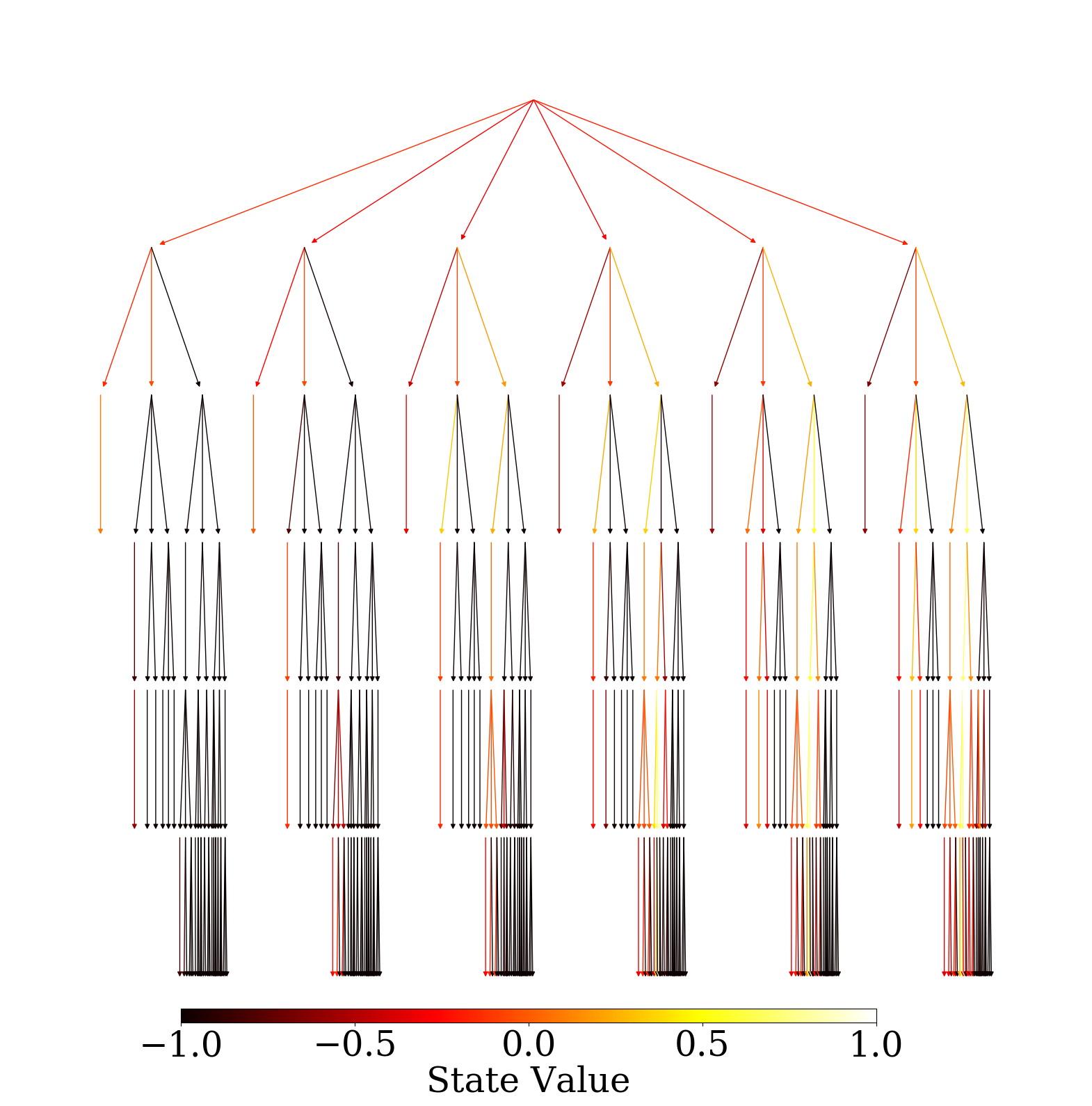}
	}
	\caption{Examples of Q-values of all possible states in the experimental decision tree estimated by the protagonist's policy/value network $f_{\theta}$ during the DRL training iterations for traction-separation model.}
	\label{fig:TS_datagame_decisiontree_statevalues}
\end{figure}

The statistics of the game scores played for the "Calibration/Defense" by the protagonist and the "Falsification/Attack" by the adversary during the DRL iterations are shown in Fig. \ref{fig:TS_learn_violinplot}. 
The improvement of the protagonist's policy is shown by the increase of the median of game scores. 
Fig. \ref{fig:TS_datagame_decisiontree} provides some examples of experiments selected by the protagonist for calibration data. 
The protagonist progressively develops the intelligence to select experiments from multiple displacement jump angles and multiple loading cycles, instead of concentrating on selections only cover very few angles and monotonic loading. 
This is consistent with intuitions from human experts, but is automatically discovered by the AI. 
We further include some examples of estimated Q-values of the experimental decision tree (Fig. \ref{fig:TS_datagame_decisiontree_statevalues}) by the protagonist to illustrate how the agent is learning during the DRL. 
We record the agent's policy/value network $(\vec{p}, v) = f_{\theta}(s)$ trained after each iteration of the DRL. 
Each checkpoint is used to predict the Q-value of each possible state in the experimental decision tree. 
The figure presents the expectations from the protagonist, before choosing any experiments, on how beneficial if an experiment is included in the calibration data. The evolution of the colors illustrates the progressively improved Q-value estimations learned from the game episodes and their rewards collected during the DRL. 

\begin{figure}[h!]\center
	\subfigure[Iteration 0, Episode 11, \newline \hspace{\linewidth} Attack Game Score: -0.085]{
		\includegraphics[width=0.235\textwidth]{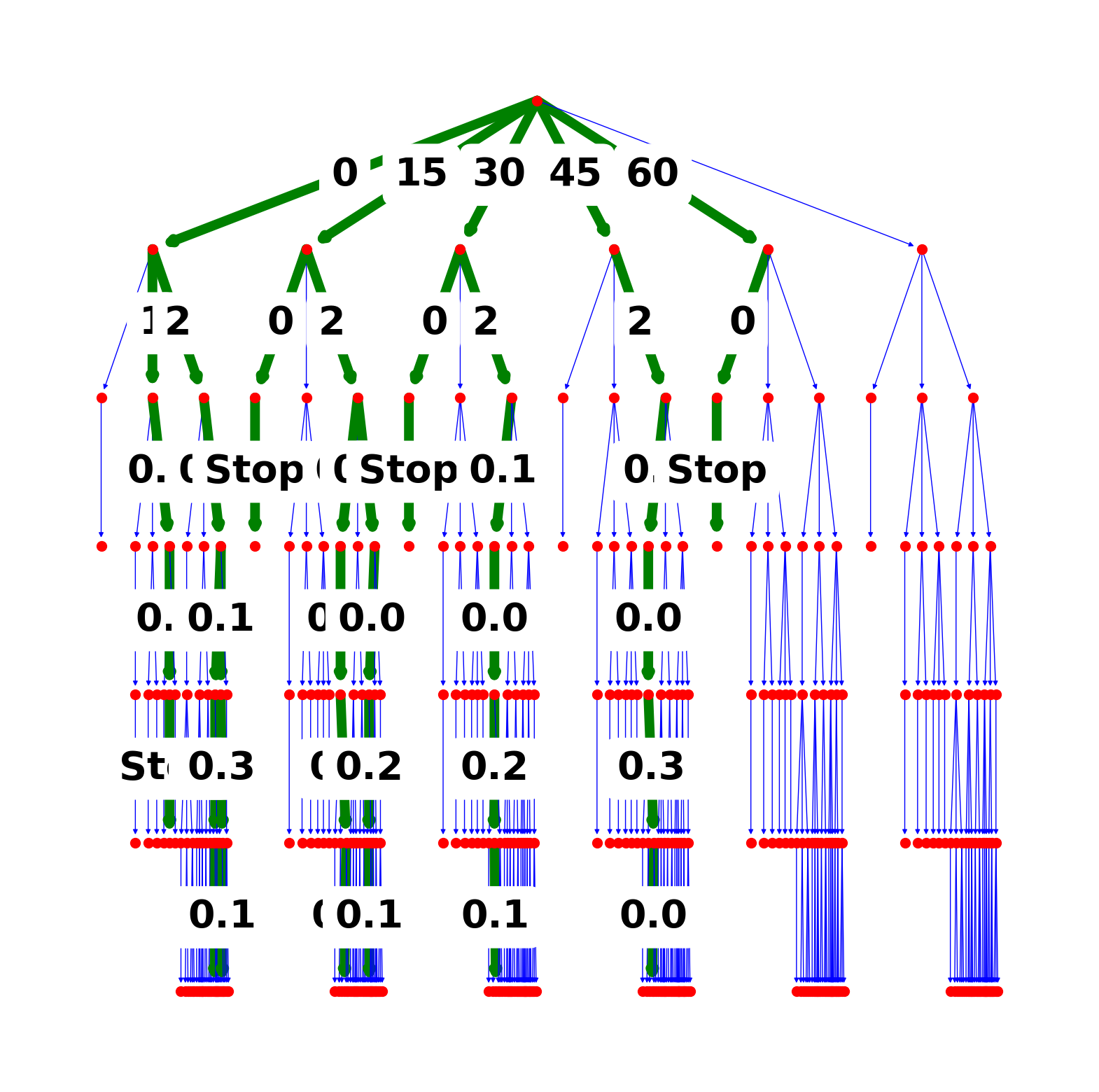}
	}
	\subfigure[Iteration 3, Episode 7, \newline \hspace{\linewidth} Attack Game Score: -0.127]{
		\includegraphics[width=0.235\textwidth]{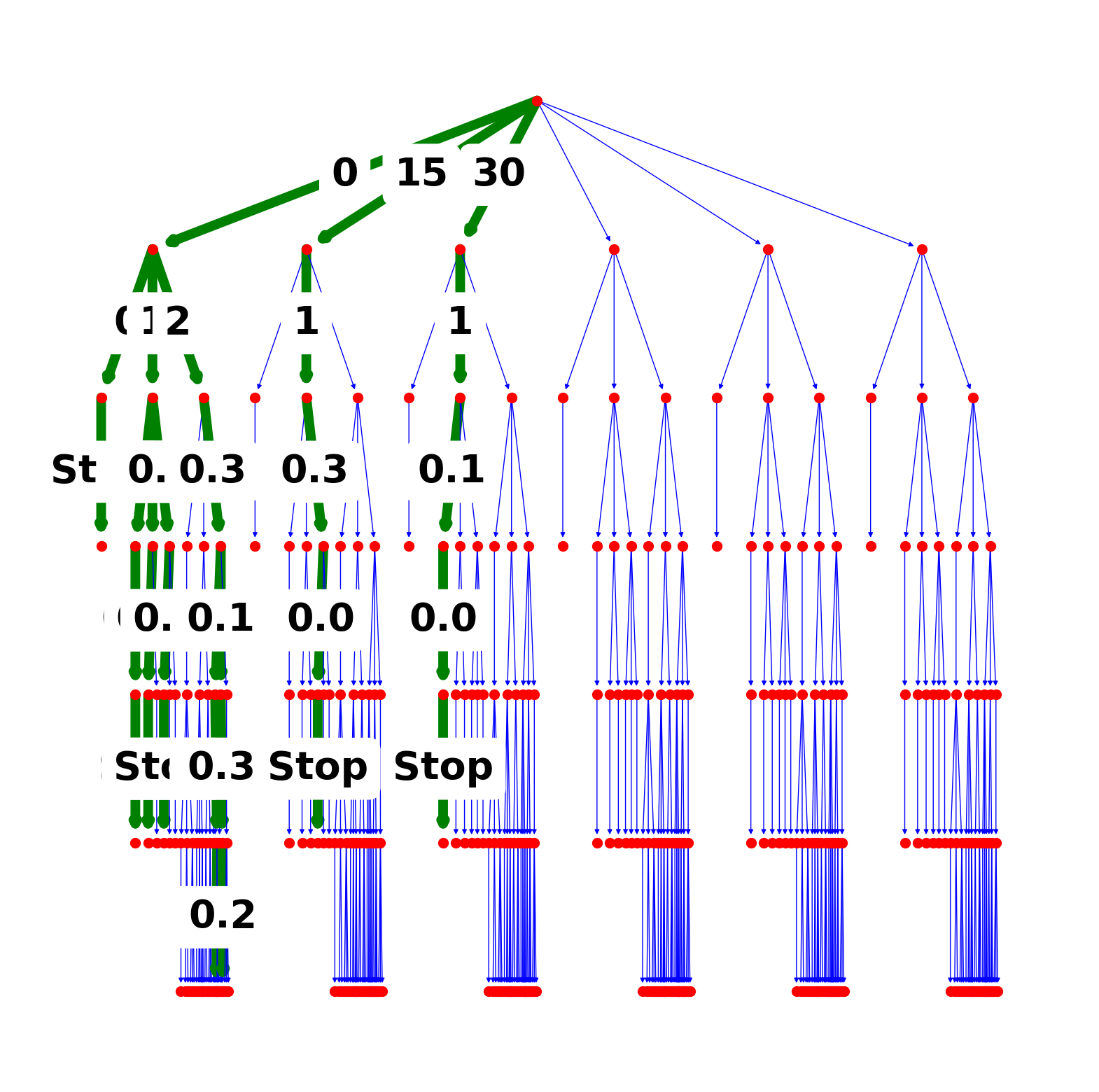}
	}
	\subfigure[Iteration 6, Episode 31, \newline \hspace{\linewidth} Attack Game Score: 0.540]{
		\includegraphics[width=0.235\textwidth]{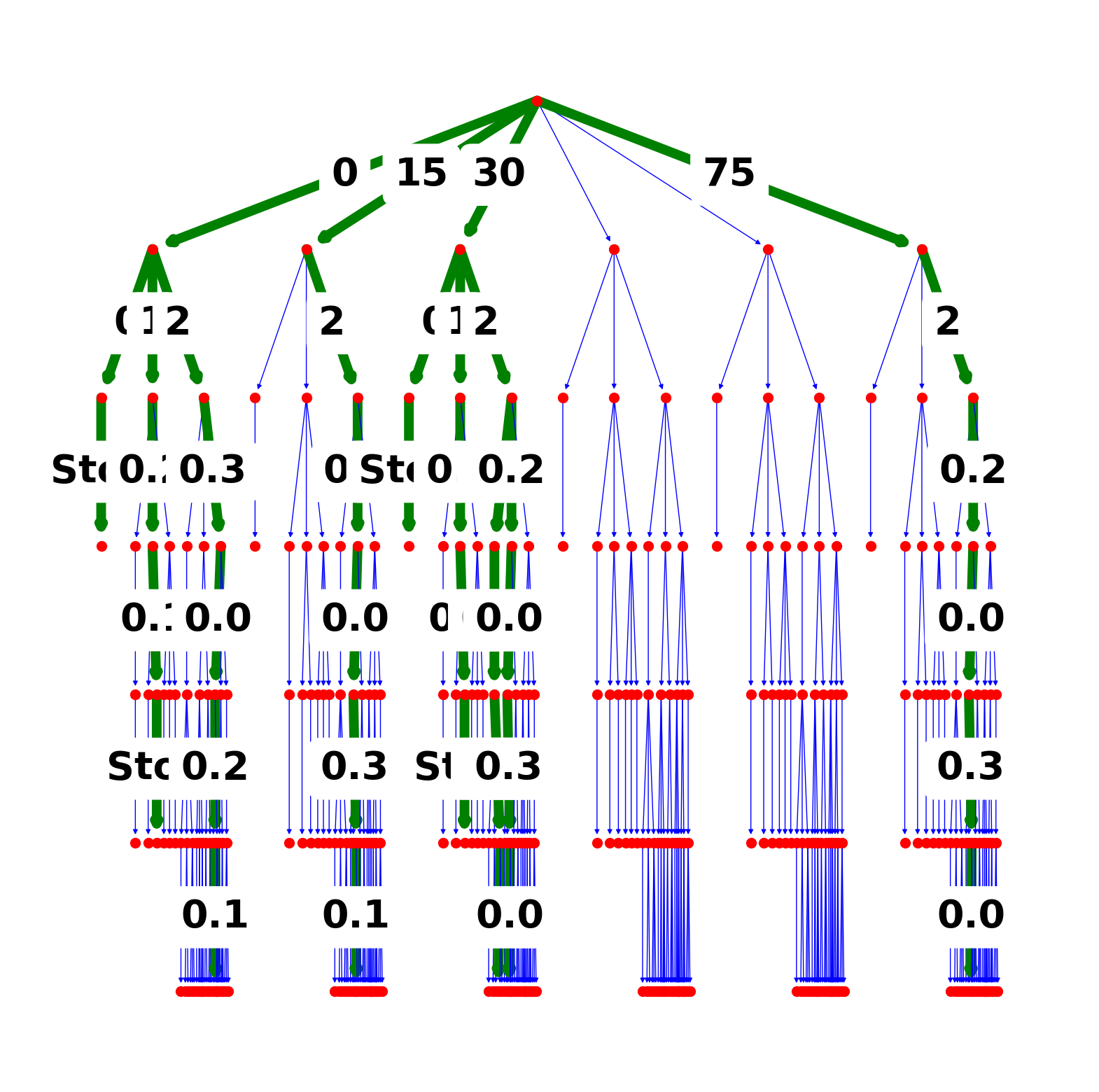}
	}
	\subfigure[Iteration 10, Episode 20, \newline \hspace{\linewidth} Attack Game Score: 0.443]{
		\includegraphics[width=0.235\textwidth]{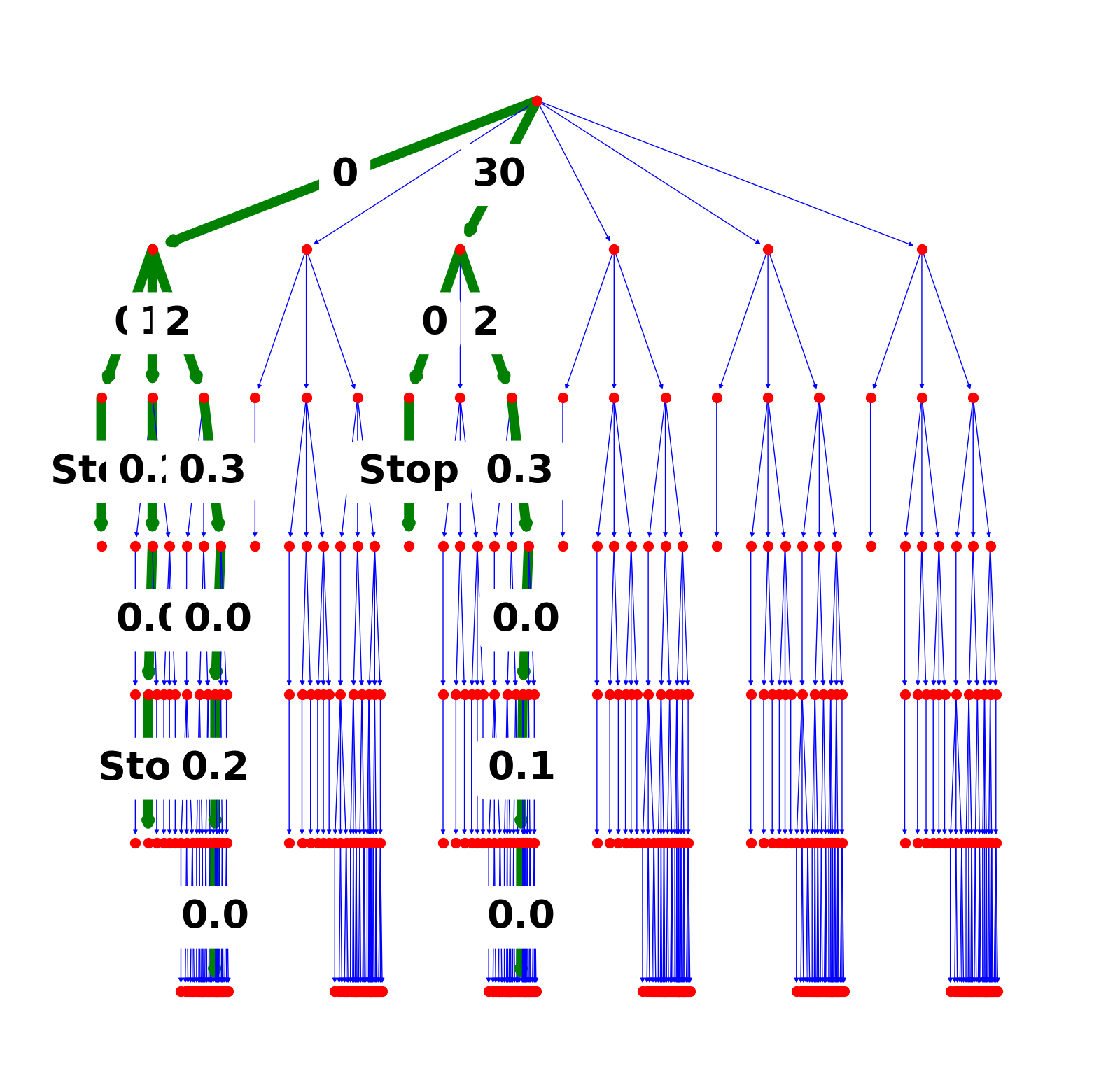}
	}
	\caption{Examples of paths (experiments) in the decision trees selected by the adversary during the DRL training iterations for traction-separation model.}
	\label{fig:TS_attackgame_decisiontree}
\end{figure}

Meanwhile, the adversary's scores also tend to increase as opposed to the previous examples. 
This could be attributed to the fact that the increasingly well-trained model by the protagonist using more effective calibration data, hence the corresponding prediction accuracy on unseen testing data also increases. 
Nevertheless, the adversarial game objective keeps driving the adversary to 
explore the model's weakest performance. 
\textit{This feedback loop plays an important role in forcing the protagonist to find more adequate calibration data to make 
the model more resilient to attacks orchestrated by the 
adversary agent,  based on improved skills learned from previous walks on the decision tree.}
Some example experiments selected by the adversary for testing data along the DRL are provided in Fig. \ref{fig:TS_attackgame_decisiontree}. 

In this example, we observe a slow convergence of the game policies as shown by the score distributions in Fig. \ref{fig:TS_learn_violinplot}. 
We attribute these difficulties to the following factors. 

\begin{figure}[h!]\center
	\subfigure[Iteration 0, Episode 11, \newline \hspace{\linewidth} Defense Game Score: -0.992]{
		\includegraphics[width=0.235\textwidth]{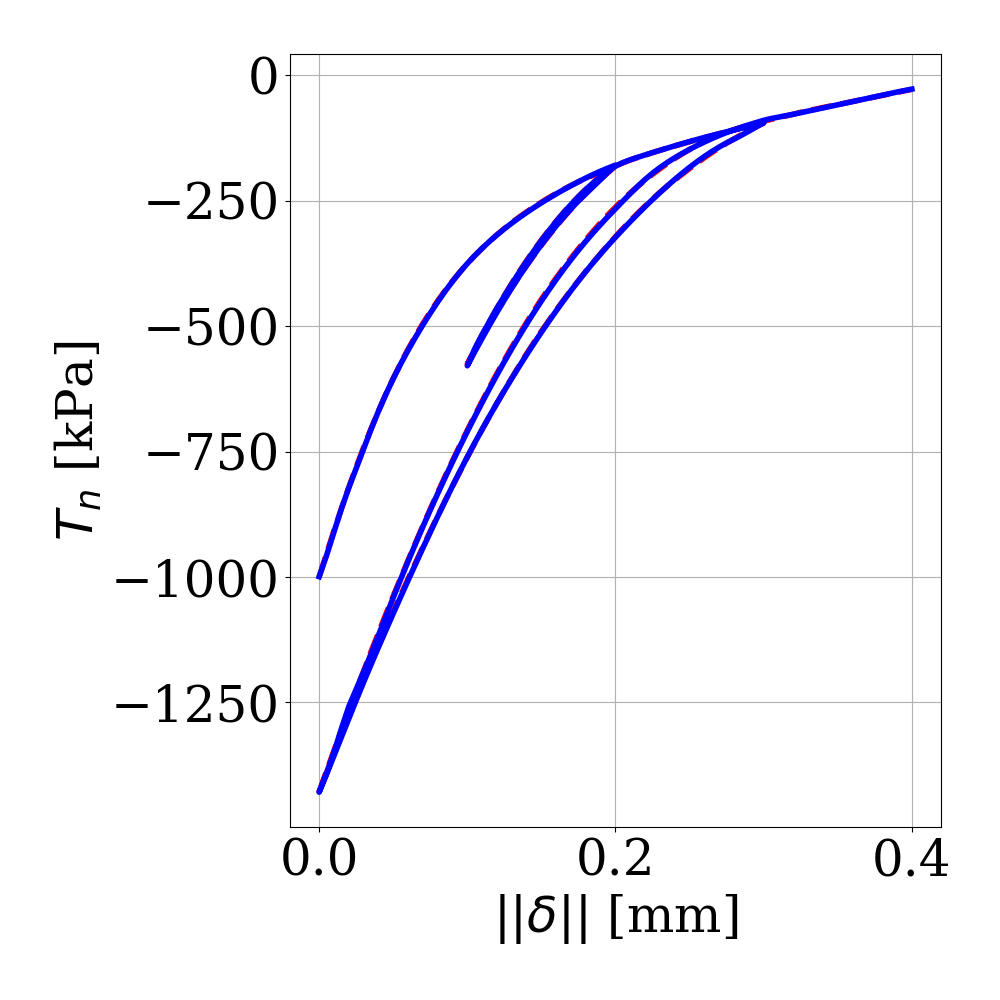}
	}
	\subfigure[Iteration 3, Episode 7, \newline \hspace{\linewidth} Defense Game Score: -0.331]{
		\includegraphics[width=0.235\textwidth]{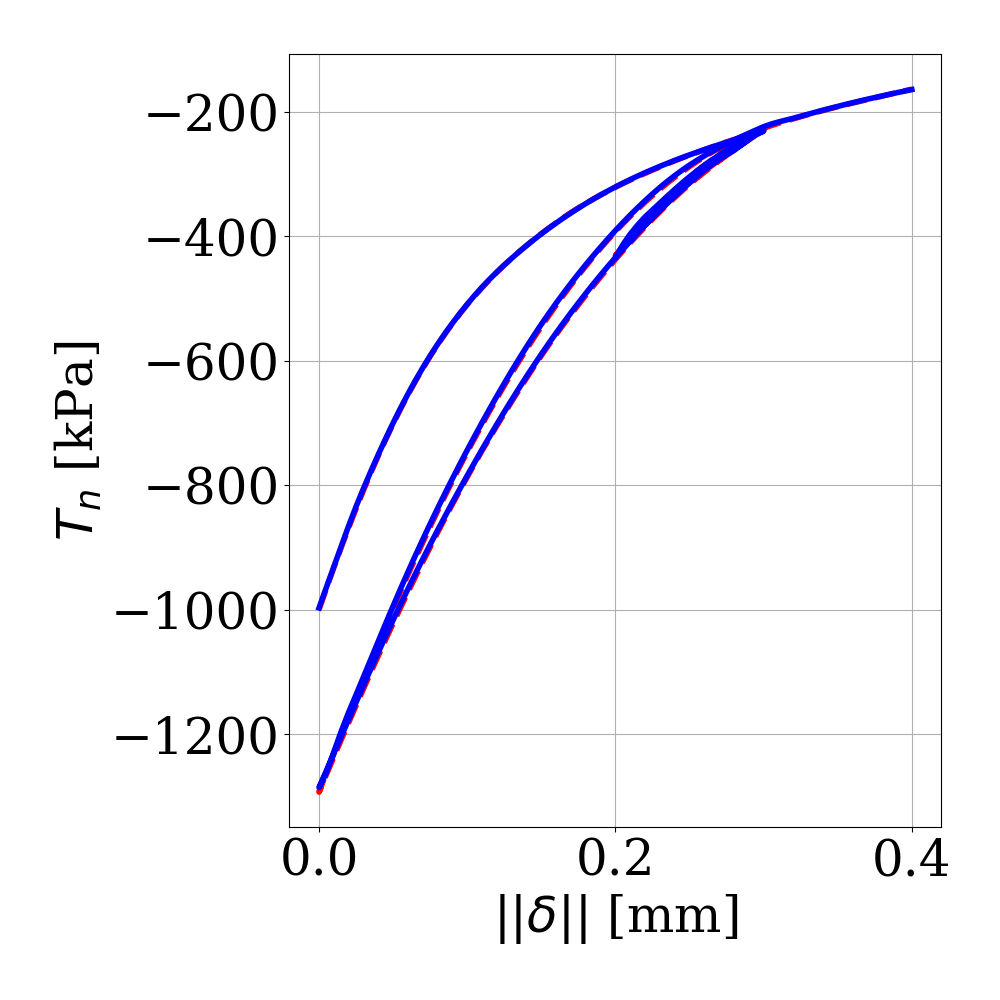}
	}
	\subfigure[Iteration 6, Episode 31, \newline \hspace{\linewidth} Defense Game Score: 0.269]{
		\includegraphics[width=0.235\textwidth]{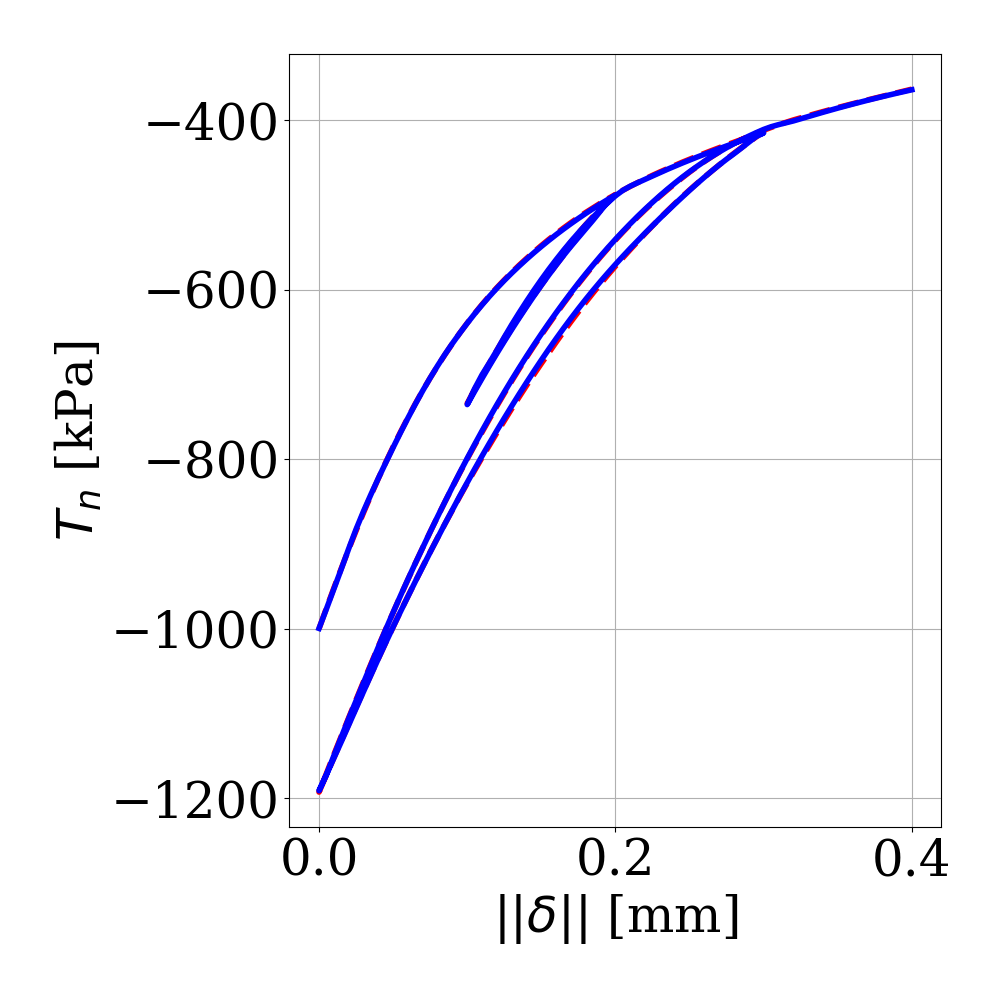}
	}
	\subfigure[Iteration 10, Episode 20, \newline \hspace{\linewidth} Defense Game Score: 0.879]{
		\includegraphics[width=0.235\textwidth]{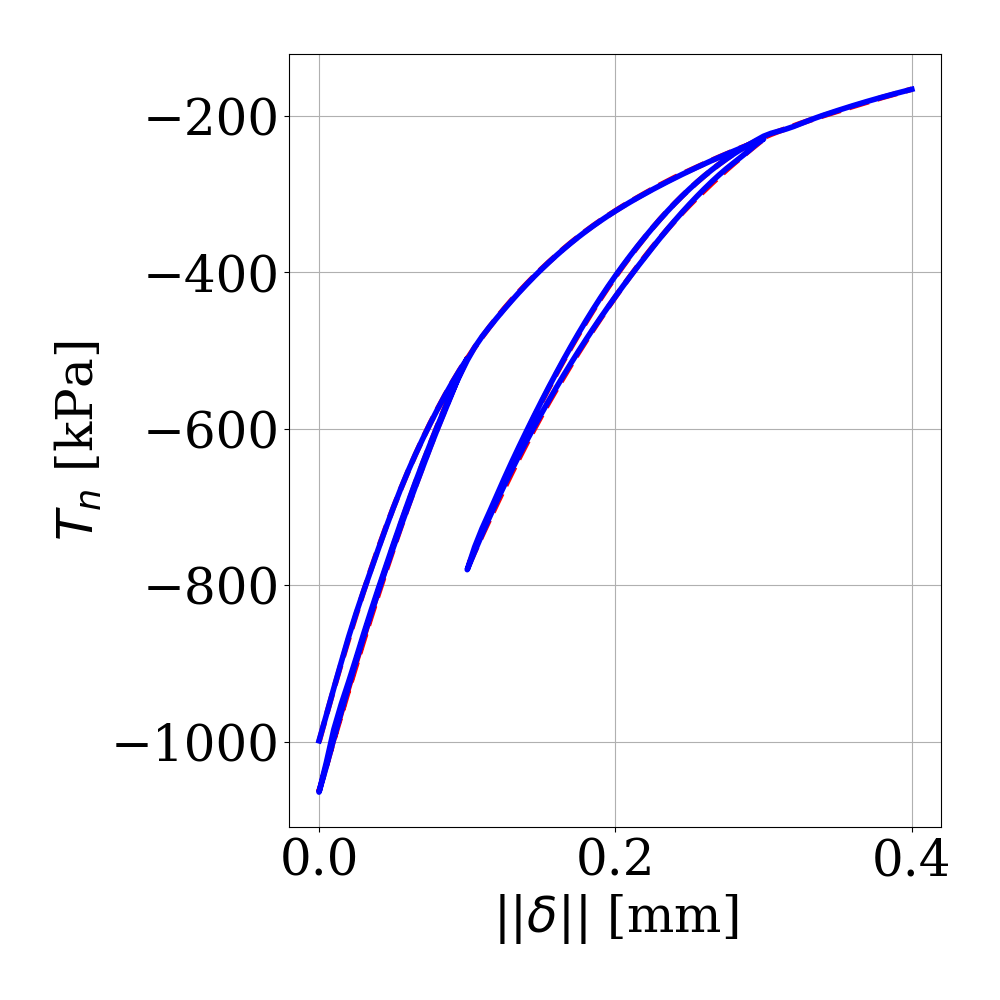}
	}
	\subfigure[Iteration 0, Episode 11, \newline \hspace{\linewidth} Defense Game Score: -0.992]{
		\includegraphics[width=0.235\textwidth]{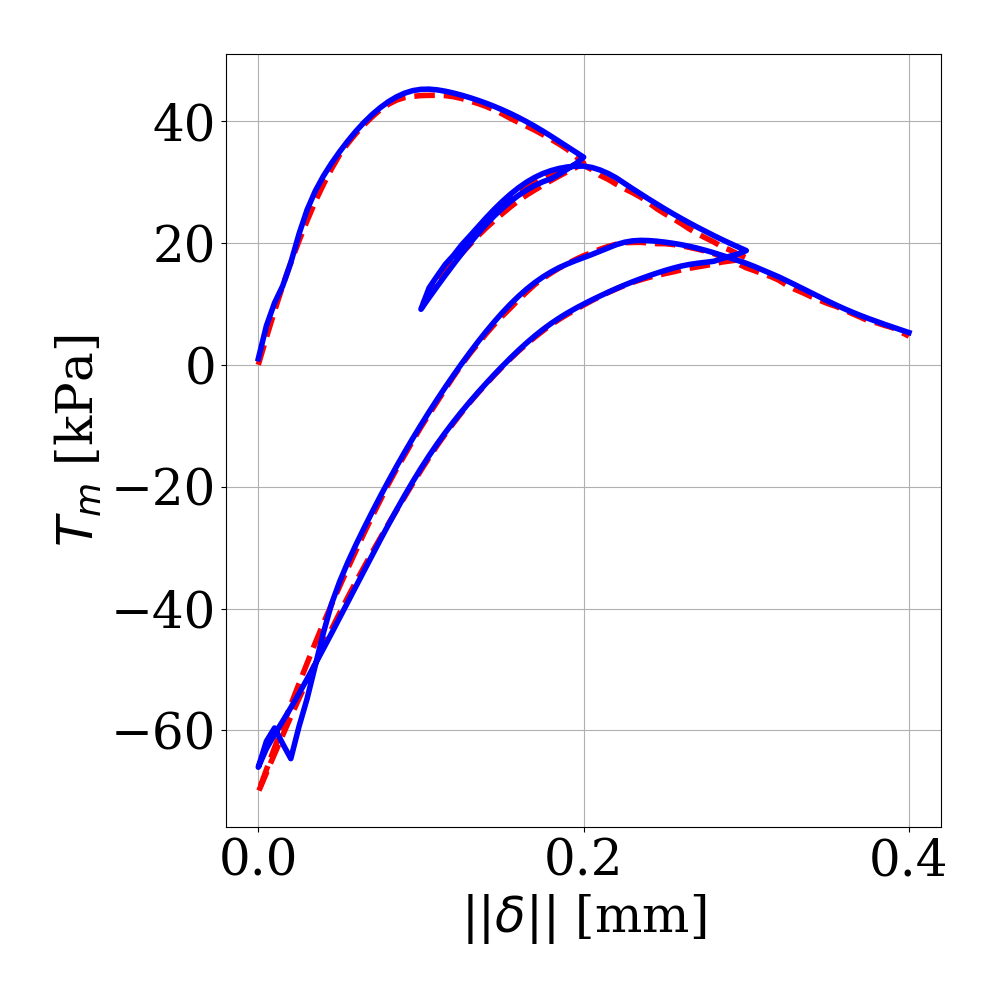}
	}
	\subfigure[Iteration 3, Episode 7, \newline \hspace{\linewidth} Defense Game Score: -0.331]{
		\includegraphics[width=0.235\textwidth]{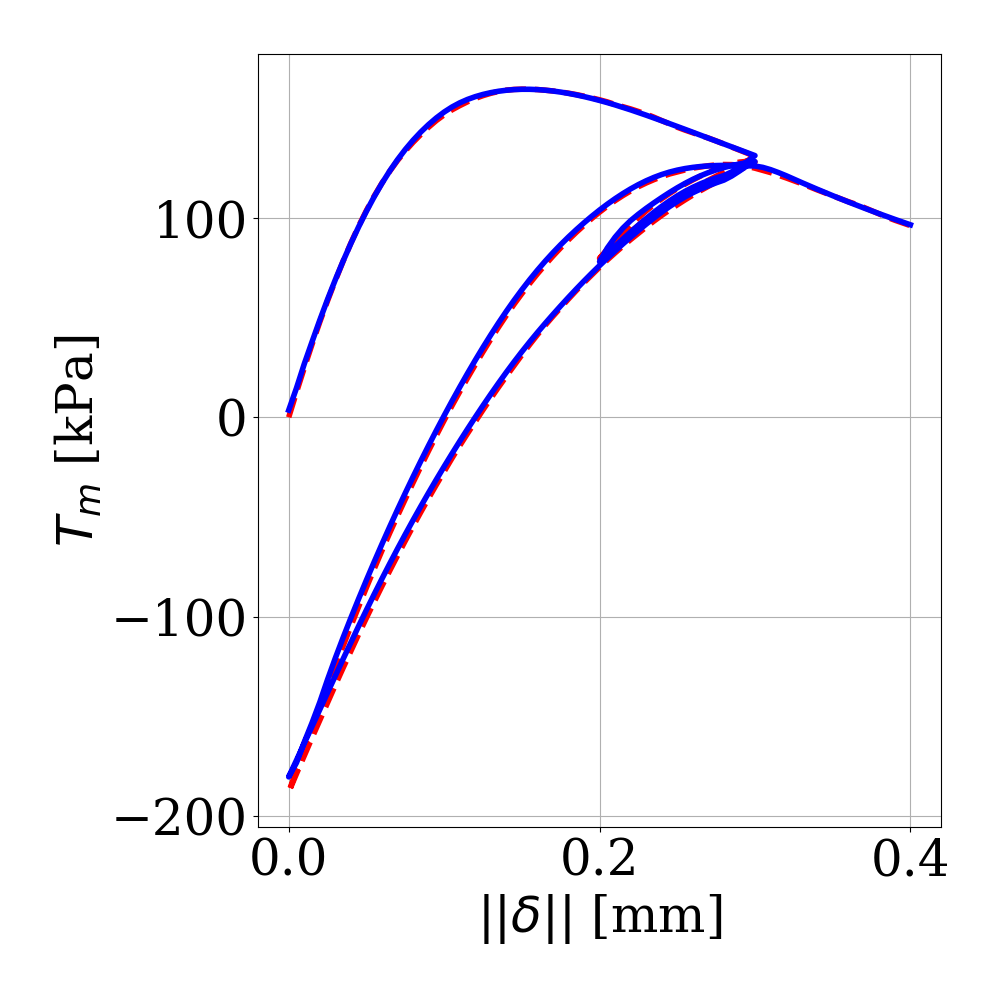}
	}
	\subfigure[Iteration 6, Episode 31, \newline \hspace{\linewidth} Defense Game Score: 0.269]{
		\includegraphics[width=0.235\textwidth]{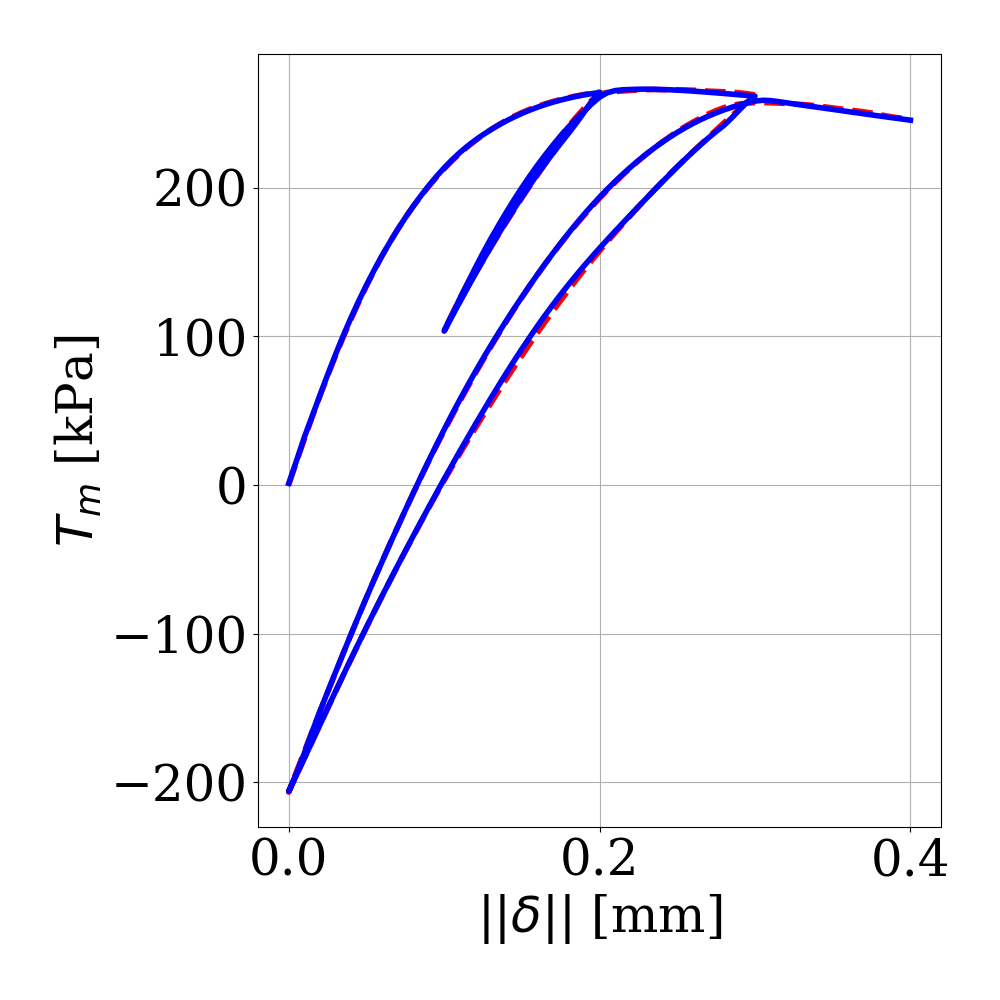}
	}
	\subfigure[Iteration 10, Episode 20, \newline \hspace{\linewidth} Defense Game Score: 0.879]{
		\includegraphics[width=0.235\textwidth]{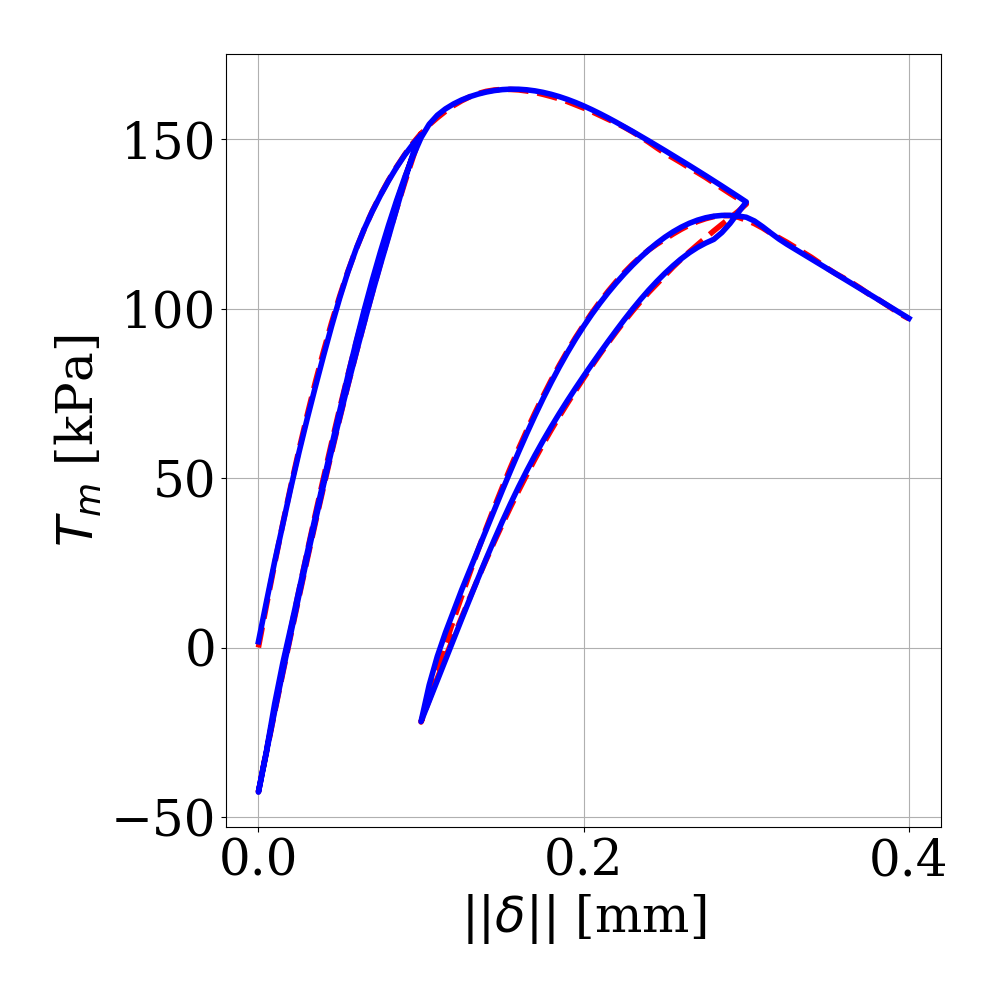}
	}
	\caption{Examples of response curves of the games played by the protagonist during the DRL training iterations for traction-separation model. Experimental data are plotted in red dashed curves, model predictions are plotted in blue solid curves. }
	\label{fig:TS_datagame_curves}
\end{figure}

\begin{enumerate}
\item The game dimension of this example is $228!/(10!(228-10)!) \approx 1e23$ where 228 is the total number of leaves in the decision tree (cf. Section \ref{subsec:decision_tree_interface}) and 10 is the maximum number of paths chosen by the agent ($N^{max}_{path}$), even when the agents try to follow the best policy learned from previous gameplay experiences, a slight deviation from this policy due to the freedom of "exploration" may lead to significant deterioration on the game performance. 
\item The ANN-based traction-separation model is highly adaptive to the calibration data. Neural networks can be trained very accurately to a broad range of calibration data, without changing its architecture. 
The handcrafted models from experts, however, are developed from fixed theory and have fixed mathematical expressions. They are not flexible to significant changes in calibration data. 
This can be seen from the example response curves in Fig. \ref{fig:TS_datagame_curves}. Because the calibration scores are uniformly high, the performance of the protagonist can not be judged solely based on the calibration accuracy. 
\item The performance of the protagonist mainly depends on the prediction accuracy on unseen testing data. 
This accuracy and hence the score the protagonist received is nevertheless depends on the data generated from experiments designed by the adversary. 
\end{enumerate}

At the early stage of the game, the adversary is learning from scratch without any human knowledge provided. As a result, the Q table of the adversary may not be accurate enough to find the best way to attack the trained model properly in the first DRL iterations. 
Some gameplays from the protagonist may yield an apparently high reward, because the adversary may design  experiments very similar to that of the protagonist used for calibration. 
For example, the protagonist may only train the model with monotonic loading and few loading angles, whereas the adversary also test the model on monotonic loadings. 
The reward in this case is apparently to be high. However, since there has not been enough exploration done, 
the Q tables for both agents are not sufficiently accurate to yield a score that carries credibility to judge 
the performance of the unexplored loading paths characterized by a subset of possible walks in the decision tree. 

\begin{figure}[h!]\center
	\subfigure[Iteration 0, Episode 11, \newline \hspace{\linewidth} Attack Game Score: -0.085]{
		\includegraphics[width=0.235\textwidth]{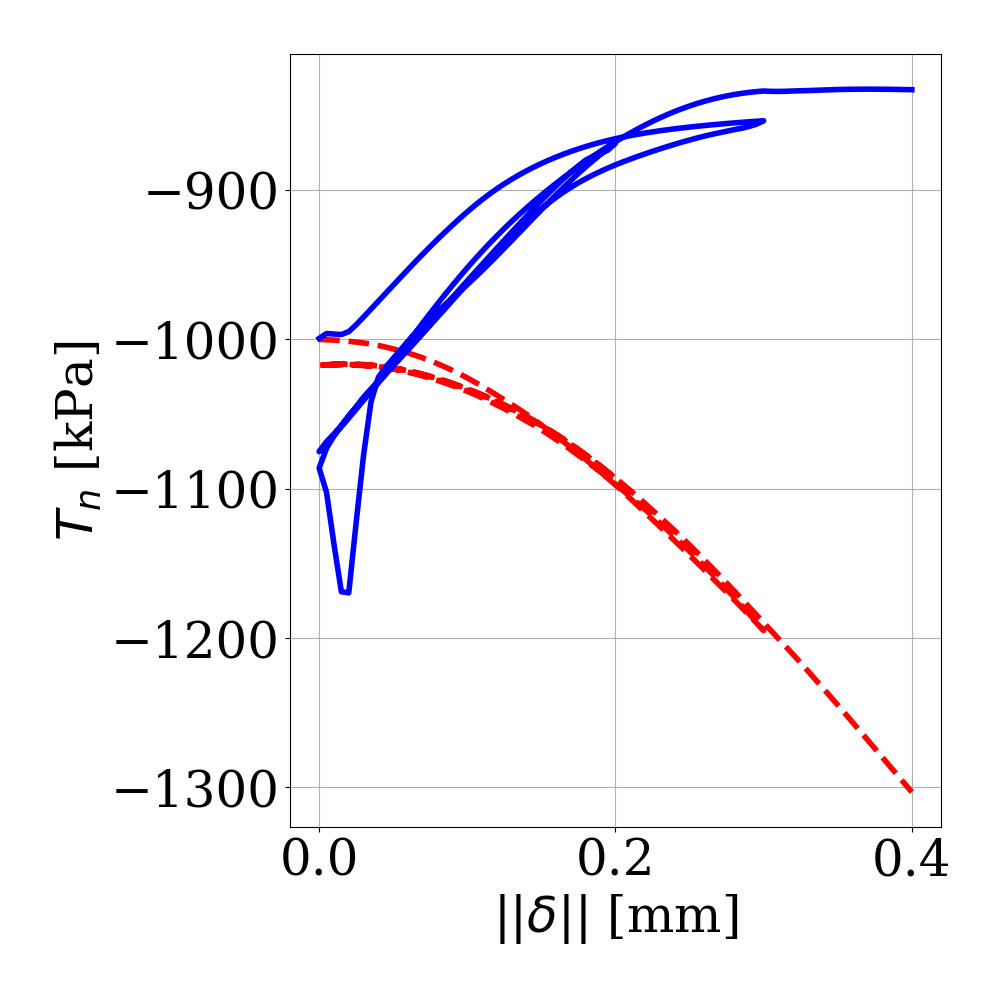}
	}
	\subfigure[Iteration 3, Episode 7, \newline \hspace{\linewidth} Attack Game Score: -0.127]{
		\includegraphics[width=0.235\textwidth]{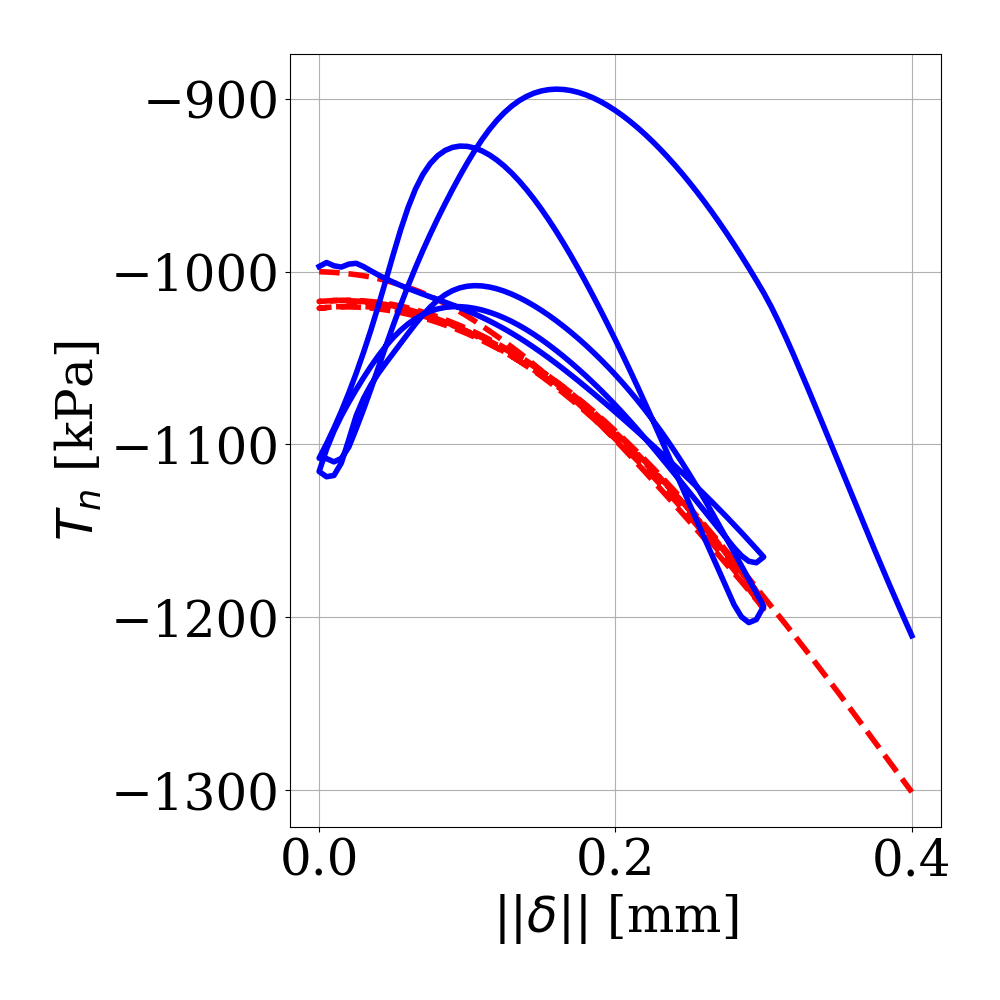}
	}
	\subfigure[Iteration 6, Episode 31, \newline \hspace{\linewidth} Attack Game Score: 0.540]{
		\includegraphics[width=0.235\textwidth]{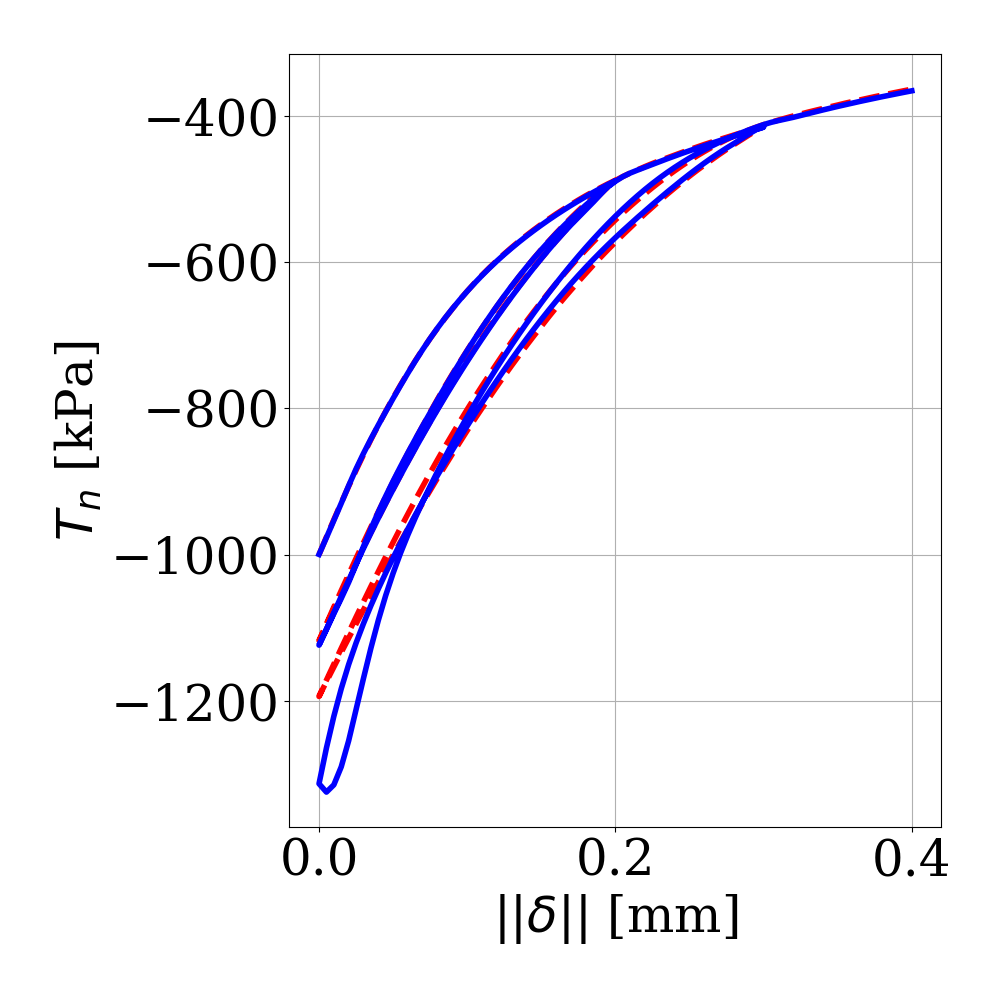}
	}
	\subfigure[Iteration 10, Episode 20, \newline \hspace{\linewidth} Attack Game Score: 0.443]{
		\includegraphics[width=0.235\textwidth]{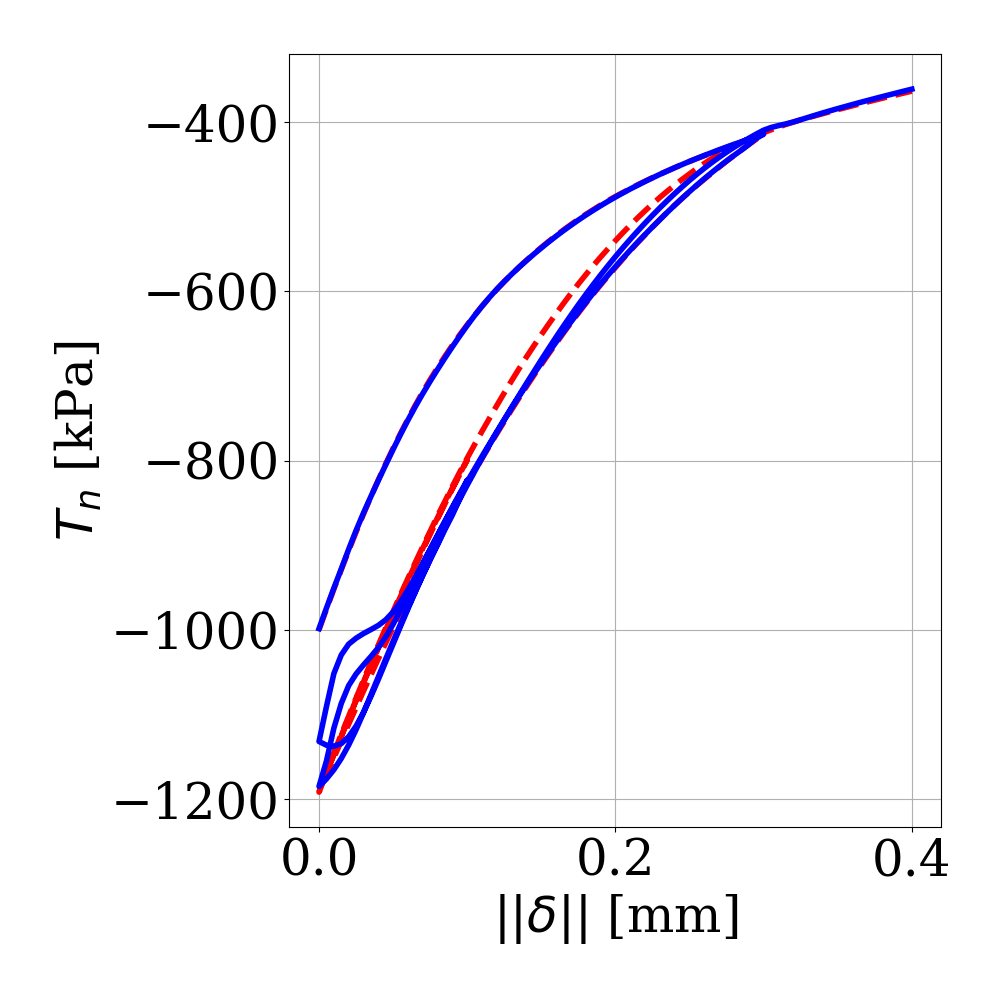}
	}
	\subfigure[Iteration 0, Episode 11, \newline \hspace{\linewidth} Attack Game Score: -0.085]{
		\includegraphics[width=0.235\textwidth]{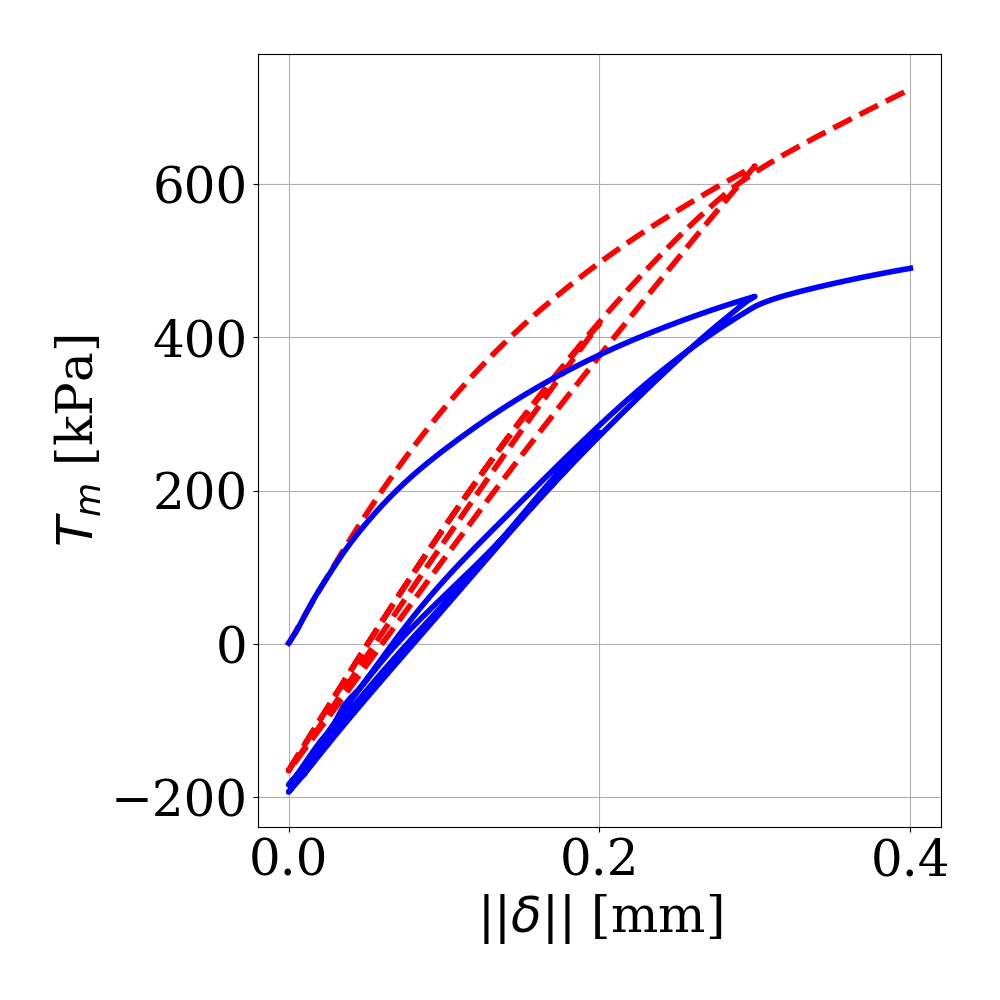}
	}
	\subfigure[Iteration 3, Episode 7, \newline \hspace{\linewidth} Attack Game Score: -0.127]{
		\includegraphics[width=0.235\textwidth]{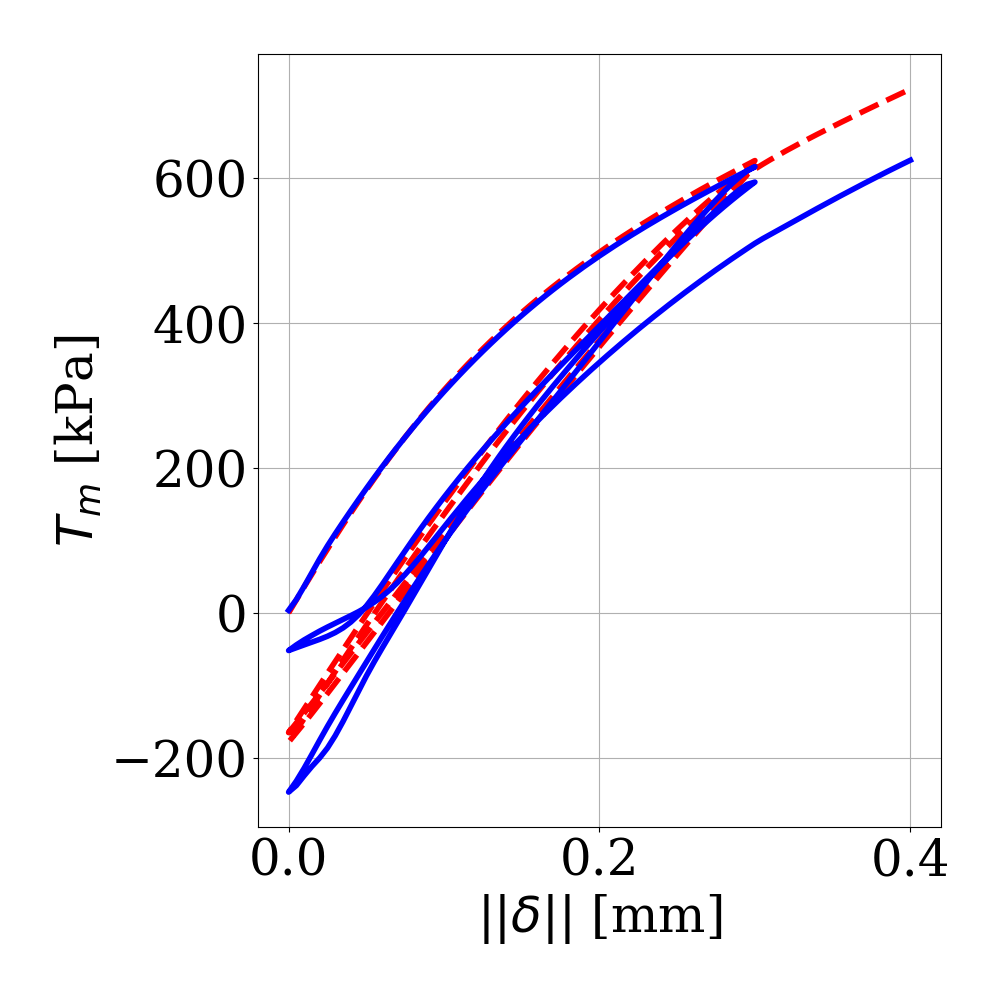}
	}
	\subfigure[Iteration 6, Episode 31, \newline \hspace{\linewidth} Attack Game Score: 0.540]{
		\includegraphics[width=0.235\textwidth]{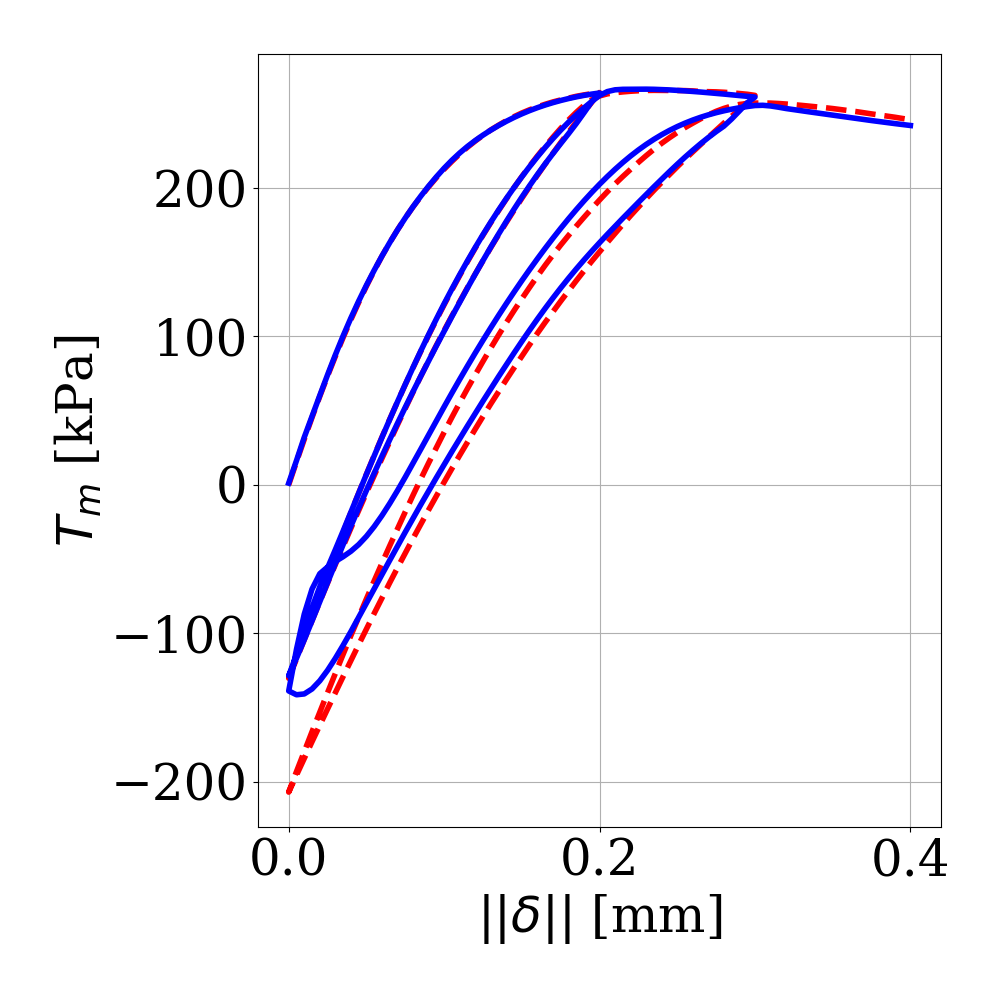}
	}
	\subfigure[Iteration 10, Episode 20, \newline \hspace{\linewidth} Attack Game Score: 0.443]{
		\includegraphics[width=0.235\textwidth]{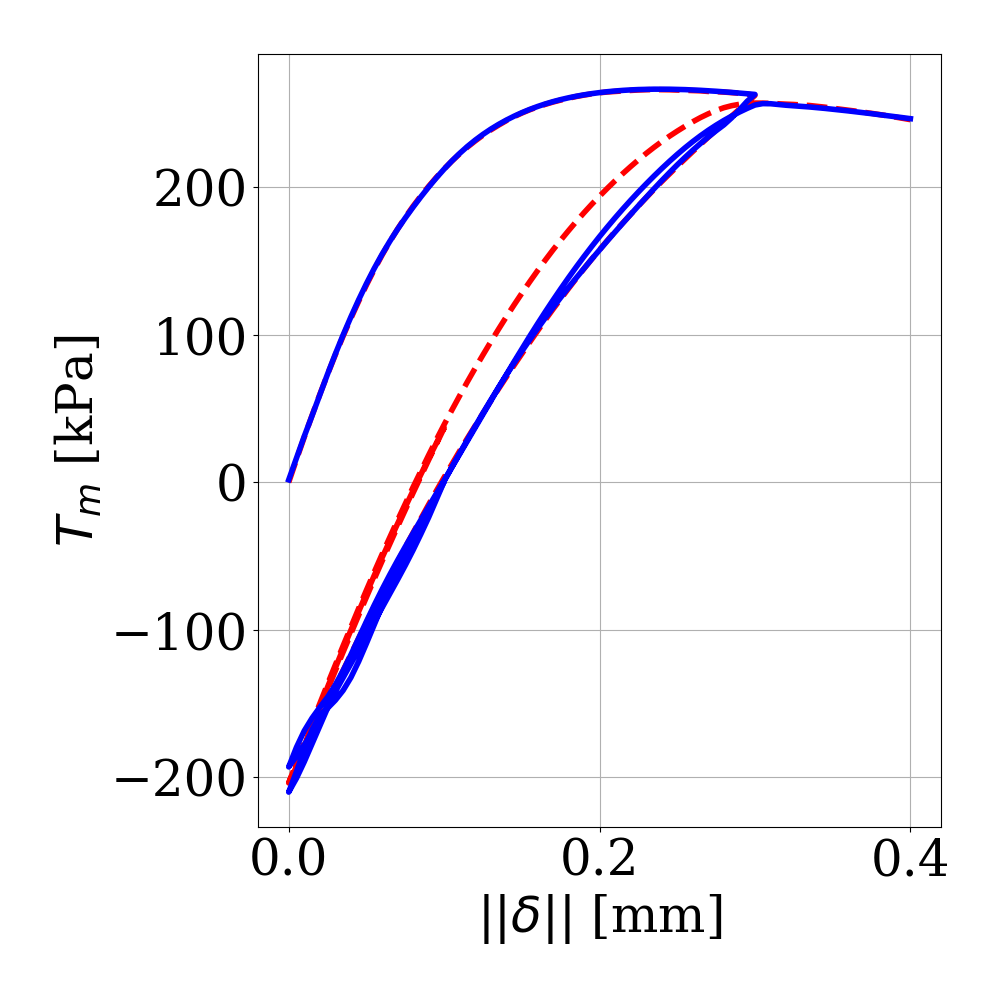}
	}
	\caption{Examples of response curves of the games played by the adversary during the DRL training iterations for traction-separation model. Experimental data are plotted in red dashed curves, model predictions are plotted in blue solid curves. }
	\label{fig:TS_attackgame_curves}
\end{figure}

At the late stage of the game (Iteration 10),  the DRL is capable of improving the blind prediction capability of the data-driven models, as shown in Fig. \ref{fig:TS_attackgame_curves}. This result is attributed to the fact that the Q table of the protagonist has been sufficiently improved such that it is able to design experiments 
that calibrate the model much better. While adversary agent that launch the adversarial attacks are given the same opportunities to improve its own Q table and therefore the policy decision skills, its action no longer exposes any particularly severe weakness of the model. such attributes are encouraging sign that shows the the machine learning model may provide efficiently robust predictions on unseen data. 

More importantly, these three numerical experiments show that the the competition between the two agents is helpful to improve the robustness of the collective performance of both agents. This finding is consistent with previous efforts on the validation and blind predictions of material models such as the Sandia Fracture Challenges \citep{boyce2014sandia, boyce2016second} and the VELAS project \citep{arulanandan1993project, popescu1995comparison}.

% Hence, 
%it might be oversimplifying to assume that the machine-learning model will always perform well 
%at the calibration but more vulnerable to predictions on unseen data.  \alert{not sure why overfitting imply 
% oversimplification of the assumption. Anyway, not sure what the last sentence is really trying to say}

\section{Conclusion and future perspectives}
We introduce a multi-agent non-cooperative meta-modeling game in which the generation of calibration/validation data and the adversary 
testing data aided to falsify the model
are handled by two competing artificial intelligence experimentalist agents. 
Mimicking the competition between a pair of protagonist and adversary in order to calibrate/validate and falsify a constitutive model for a path-dependent process, these two AI agents interact with each other sequentially and exchange information until both agents reach their own objectives and further actions do not gain better individual rewards. 
The winning strategies of the non-cooperative game are efficiently searched by the deep reinforcement learning (DRL) technique. The wide-applicability and efficiency of our approach have been shown through two elasto-plasticity models and a data-driven traction-separation model, with the number of possible game configurations as enormous as $1.5e9$ and $1e23$, respectively. 
To the best knowledge of the authors, this is the first time the strategies of experimentalists who provide data to validate/falsify a history-dependent model 
are formulated in a non-cooperative decision-making game. Both agents are able to continuously improved their knowledge of the constitutive law using  experimental data generated by each others in
 a competitive DRL framework \citep{silver2017mastering, silver2017masteringchess}
and established the Nash equilibria strategies that tell us the quality of the models 
in the applications represented in the decision trees. 
Such innovations are necessary keys to develop powerful AI assistants that take over large amounts of trial-and-error burdens from human researchers for material modeling and knowledge discoveries with decision trees that are too deep to explore manually. 
More importantly, the competitive nature enables us to not only find the optimal setup 
of a constitutive laws, but also find out the weakness of the models in an unbiased third-party manner. 

Further improvements and extensions can be made regarding the following aspects of our current framework. 
(1) Experimental data from real-world granular materials, instead of DEM samples, can be used if the AI agents' decision trees are connected with laboratory instruments. 
(2) The rewards of the game strategies and hence the conclusions drawn from the game are sensitive to the game settings and the score systems (objective functions), e.g., the hyperparameters $N^{max}_{path}$, $E^{max}_{NS}$, $E^{min}_{NS}$, $\alpha_{\text{SCORE}}$. 
These game designs need to be tuned by human experts in order to appropriately investigate the strengths and weaknesses of the model. 
(3) They also depend on the calibration procedures, especially the elasto-plasticity models from human experts in which the material parameters have specific meanings and require initial guesses and bounds. 
The neural network models, on the other hand, can adapt to a wide range of material behaviour and calibrate well, but with material parameters difficult to interpret. 

\section{Acknowledgments}
The work of KW and WCS is supported by National Science Foundation under grants 
contracts CMMI-1846875 and CMMI-1940203, 
the Earth Materials and Processes
program from the US Army Research Office under grant contract 
W911NF-18-2-0306,
the Dynamic Materials and Interactions Program 
from the Air Force Office of Scientific Research under 
grant contract FA9550-17-1-0169 and FA9550-19-1-0318,
as well as the Columbia SEAS Interdisciplinary Research Seed Grant.  
The work of QD is supported in part by
NSF  CCF-1704833, DMS-1719699, DMR-1534910, 
and  ARO  MURI  W911NF-15-1-0562. 
 These supports are gratefully 
acknowledged. 
The views and conclusions contained in this document are those of the authors, 
and should not be interpreted as representing the official policies, either expressed or implied, 
of the sponsors, including the Army Research Laboratory or the U.S. Government. 
The U.S. Government is authorized to reproduce and distribute reprints for 
Government purposes notwithstanding any copyright notation herein.

\appendix \normalsize
\section{Preparation and experiments of DEM samples} \label{sec:dem_samples}
The data for calibration and evaluation of the prediction accuracy of the constitutive models are generated by numerical simulations on representative volume elements (RVEs) of densely-packed spherical DEM particles. 
The open-source discrete element simulation software YADE for DEM is used by the AI experimentalist agents to generate 
data, including the homogenized stress and strain measures and the geometrical attributes such as porosity, coordination number and fabric tensor \citep{vsmilauer2010yade}. 

The RVEs for AI-guided experimentation on bulk granular materials (Section \ref{subsec:decision_tree_bulk}) consist of discrete element particles having radii between $1 \pm 0.3$ mm with a uniform distribution. 
The Cundall's elastic-frictional contact model (\citep{cundall1979discrete}) is used for the inter-particle constitutive law. 
The material parameters are: interparticle elastic modulus $E_{eq}=0.5$ GPa, ratio between shear and normal stiffness $k_s/k_n=0.3$, frictional angle $\varphi=$ \ang{30}, density $\rho=2600$ $kg/m^3$, Cundall damping coefficient $\alpha_{damp}=0.4$. 
Firstly, a random loose packing enclosed in a parallelepiped of edge size $50$ mm is generated. 
The sample is then isotropically compressed to $p_0$ = -300 kPa. 
$e_0$ = 0.60 is approximated using a fictitious frictional angle of 0.05 rad, whereas $e_0$ = 0.55 is approximated using a frictional angle of 0.01 rad. 
These samples are then isotropically compressed to higher confinements $p_0$ = -400 kPa and $p_0$ = -500 kPa. 
The generated RVE samples for the AI experimentalists to choose are presented in Table \ref{tab:dem_samples}. 
The 'DTC', 'DTE', 'TTC' tests are conducted quasi-statically using the YADE engine "PeriTriaxController", and data are recorded at every strain increment of 1e-4. 

\begin{table}[h!]
	\centering
	\begin{tabular}{| p{2cm} | p{2cm} | p{2cm} | p{2cm} | p{2cm} |}
		\hline
		Sample No. & 'Sample $p_0$' & 'Sample $e_0$' & $p_0$ & $e_0$\\ \hline
		1 & '300kPa' & '0.60' & -300 kPa & 0.5955 \\ \hline
		2 & '300kPa' & '0.55' & -300 kPa & 0.5554 \\ \hline
		3 & '400kPa' & '0.60' & -400 kPa & 0.5936 \\ \hline
		4 & '400kPa' & '0.55' & -400 kPa & 0.5538 \\ \hline
		5 & '500kPa' & '0.60' & -500 kPa & 0.5917 \\ \hline
		6 & '500kPa' & '0.55' & -500 kPa & 0.5521 \\ \hline
	\end{tabular}
	\caption{Initial DEM samples for AI-guided experimentation on bulk granular materials.}
	\label{tab:dem_samples}
\end{table}

The RVEs for AI-guided experimentation on granular interfaces (Section \ref{subsec:decision_tree_interface}) consist of discrete element particles having radii between $1 \pm 0.3$ mm with a uniform distribution. 
The Cundall's elastic-frictional contact model is used. 
The material parameters are identical to those of the bulk RVEs. 
The sample with initially random loose packing is isotropically compressed to $p_0$ = -1 MPa using a fictitious frictional angle of 0.01 rad. 
Hence the initial traction is -1 MPa in the normal direction and 0 MPa in the tangential direction. 
The width between the upper and lower surfaces of the sample is 20 mm. 
The mixed-mode shear tests with different loading paths are conducted quasi-statically, and data are recorded at every displacement jump increment of 0.005 mm. 

\section{Material Models in Numerical Examples} \label{sec:example_models}
\subsection{Drucker–Prager elasto-plasticity model (Section \ref{subsec:dpmodel_example})}
The model adopts a linear elasticity law with the elastic stiffness tensor 
\begin{equation}
\tensor{C}^e = K \tensor{I}\otimes\tensor{I} + 2 G (\tensor{I}^4_{sym} - \frac{\tensor{I}\otimes\tensor{I}}{3}),
\label{eq:dp_eq_1}
\end{equation}
where $K$ is the elastic bulk modulus and $G$ is the elastic shear modulus. 
\begin{equation}
\left\{
\begin{aligned}
K&=K_0 = \frac{2(1+\nu)}{3(1-2\nu)} G_0\\
G&=G_0
\end{aligned}
\right.,
\label{eq:dp_eq_2}
\end{equation}
where $G_0$ is the reference shear modulus and $\nu$ is the Poisson ratio. 

The yield surface has the form $f = q + \alpha p$, where $p = \frac{1}{3} \text{tr}(\tensor{\sigma})$, $\tensor{s} = \tensor{\sigma} - p\tensor{I}$, $q = \sqrt{3 J_2} = \sqrt{\frac{3}{2}} ||\tensor{s}||$. 
The potential surface has the form $g = q + \beta p - c_g$. 
$\alpha$ and $\beta$ evolve according to
\begin{equation}
\left\{
\begin{aligned}
\alpha &= a_0 + a_1\bar{\epsilon}^p \exp(a_2 p - a_3 \bar{\epsilon}^p)\\
\beta &= \alpha - \beta_0\\
\end{aligned}
\right. ,
\label{eq:dp_eq_3}
\end{equation}
where $a_0$, $a_1$, $a_2$, $a_3$ and $\beta_0$ are material parameters to calibrate. 
$\bar{\epsilon}^p$ is the accumulated plastic strain.

The calibration using the nonlinear least-squares solver "NL2SOL" in Dakota software \citep{adams2014dakota} requires initial guesses, upper and lower bounds of each parameter. 
They are given in Table \ref{tab:dp_parameters_guess}. 
The elasticity parameters $G_0$ and $\nu$ are calibrated using the first three data points of each calibration experiment. 
They are fixed in the later calibration of plasticity parameters $a_0$, $a_1$, $a_2$, $a_3$, $\beta_0$. 
The target features for calibration objectives and accuracy evaluations include data of the pressure $p$, the deviatoric stress $q$, and the volumetric strain $\varepsilon_v = \text{tr}(\tensor{\varepsilon})$.

\begin{table}[h!]
	\centering
	\begin{tabular}{| p{2cm} | p{3cm} | p{3cm} | p{3cm} |}
		\hline
		Parameter & Initial Guess & Lower Bound & Upper Bound\\ \hline
		$G_0$ & 6e4 kPa & 4e4 kPa & 8e4 kPa \\ \hline
		$\nu$ & 0.25 & 0.1 & 0.4 \\ \hline
		$a_0$ & 1.0 & 0.5 & 1.5 \\ \hline
		$a_1$ & 2e4 & 1e2 & 6e4 \\ \hline
		$a_2$ & 1e-5 1/Pa & 5e-6 1/Pa & 5e-5 1/Pa \\ \hline
		$a_3$ & 60.0 & 20.0 & 200.0 \\ \hline
		$\beta_0$ & 0.5 & 0.2 & 0.8 \\ \hline
	\end{tabular}
	\caption{Initial guesses, upper and lower bounds of the material parameters for Drucker–Prager model.}
	\label{tab:dp_parameters_guess}
\end{table}

\subsection{SANISAND  elasto-plasticity model (Section \ref{subsec:sanisandmodel_example})}
The model is expressed in geomechanics sign convention as in the original paper. 
The model adopts the nonlinear elasticity with dependence on the mean pressure $p$ and the void ratio $e$,
\begin{equation}
\left\{
\begin{aligned}
K&= \frac{2(1+\nu)}{3(1-2\nu)} G\\
G&= G_0 p_{at} \frac{(2.97-e)^2}{1+e} (\frac{p}{p_{at}})^{1/2}
\end{aligned}
\right.,
\label{eq:sanisand_1}
\end{equation}
where $G_0$ and $\nu$ are material parameters, $p_{at}$ = 100 kPa is the atmospheric pressure. 

The yield surface has the shape of a small cone
\begin{equation}
f = ||\tensor{s} - p \tensor{\alpha}|| - \sqrt{2/3}pm,
\label{eq:sanisand_2}
\end{equation}
where we fix $m$ to be 0.01. 

The back stress-ratio tensor $\tensor{\alpha}$ evolves according to
\begin{equation}
\dot{\tensor{\alpha}} = \dot{\lambda} (2/3) h (\tensor{\alpha}_{\theta}^b - \tensor{\alpha}),
\label{eq:sanisand_3}
\end{equation}
where $\dot{\lambda}$ is the rate of the plastic multiplier, and
\begin{equation}
\left\{
\begin{aligned}
h &= \frac{b_0}{(\tensor{\alpha}-\tensor{\alpha}_{in}) : \tensor{n}}\\
b_0 &= G_0 h_0 (1-c_h e)(p/p_{at})^{-1/2}\\
\tensor{\alpha}_{\theta}^b &= \sqrt{2/3} [g(\theta,c) M \exp(-n^b \psi)-m] \tensor{n}\\
g(\theta,c) &= \frac{2c}{(1+c)+(1-c)\cos3\theta}\\
\cos3\theta &= \sqrt{6} \text{tr} (\tensor{n}^3)\\
\tensor{n} &= \frac{\frac{\tensor{s}}{p} - \tensor{\alpha}}{\sqrt{2/3}m}\\
\psi &= e - e_{0} + \lambda_c (p/p_{at})^{\xi}
\end{aligned}
\right.,
\label{eq:sanisand_4}
\end{equation}
where $h_0$, $c_h$, $M$, $c$, $n^b$, $e_{0}$, $\lambda_c$, $\xi$ are material parameters.

The plastic flow direction is defined as
\begin{equation}
\left\{
\begin{aligned}
\tensor{m}^{flow} &= B\tensor{n} -C(\tensor{n}^2-\frac{1}{3}\tensor{I}) + \frac{1}{3}D\tensor{I}\\
B &= 1+\frac{3}{2} \frac{1-c}{c} g(\theta,c) \cos3\theta\\
C &= 3 \sqrt{\frac{3}{2}} \frac{1-c}{c} g(\theta,c)\\
D &= A_d (\tensor{\alpha}_{\theta}^d - \tensor{\alpha}) : \tensor{n}\\
A_d &= A_0 (1+<\tensor{z}:\tensor{n}>)\\
\tensor{\alpha}_{\theta}^d &= \sqrt{2/3} [g(\theta,c) M \exp(n^d \psi)-m] \tensor{n}\\
\dot{\tensor{z}} &= -c_z <- \dot{\lambda} D> (z_{max} \tensor{n} + \tensor{z})
\end{aligned}
\right.,
\label{eq:sanisand_5}
\end{equation}
where $A_0$, $n^d$, $c_z$, $z_{max}$ are additional material parameters. 

The initial guesses, upper and lower bounds of the above parameters for Dakota's "NL2SOL" calibration are given in Table \ref{tab:sanisand_parameters_guess}. 
The calibration procedure is identical to that of Drucker–Prager model. 

\begin{table}[h!]
	\centering
	\begin{tabular}{| p{2cm} | p{3cm} | p{3cm} | p{3cm} |}
		\hline
		Parameter & Initial Guess & Lower Bound & Upper Bound\\ \hline
		$G_0$ & 1e4 kPa & 5e3 kPa & 2e4 kPa \\ \hline
		$\nu$ & 0.25 & 0.1 & 0.4 \\ \hline
		$M$ & 0.75 & 0.5 & 1.0 \\ \hline
		$c$ & 0.9 & 0.7 & 1.0 \\ \hline
		$e_0$ & 0.8 & 0.7 & 0.9 \\ \hline
		$\lambda_c$ & 0.0025 & 0.0001 & 0.005 \\ \hline
		$\xi$ & 1.0 & 0.8 & 1.2 \\ \hline
		$n^b$ & 3.0 & 1.0 & 5.0 \\ \hline
		$n^d$ & 0.5 & 0.01 & 1.0 \\ \hline
		$A_0$ & 1.0 & 0.5 & 1.5 \\ \hline
		$h_0$ & 30.0 & 10.0 & 50.0 \\ \hline
		$c_h$ & 1.0 & 0.5 & 1.5 \\ \hline
		$c_z$ & 600.0 & 400.0 & 800.0 \\ \hline
		$z_{max}$ & 2.5 & 1.0 & 5.0 \\ \hline
	\end{tabular}
	\caption{Initial guesses, upper and lower bounds of the material parameters for SANISAND model.}
	\label{tab:sanisand_parameters_guess}
\end{table}

\subsection{Data-driven traction-separation model (Section \ref{subsec:rnntsmodel_example})}
\label{subsec:rnntsmodel}
Data-driven traction-separation models use artificial neural networks (ANNs) as universal function approximators to continuous functions of various complexity on compact subsets of $R^n$ (Universal approximation theorem, \citep{hornik1989multilayer}). 
Moreover, a special type of ANNs, recurrent neural networks (RNN, e.g., long short-term memory (LSTM) \citep{hochreiter1997long}, gated recurrent units (GRU) \citep{cho2014learning, chollet2015keras}), can capture the functions of a time series of inputs, which is appropriate for replicating the path-dependent material behaviors in granular interfaces. 
The data-driven model in the example firstly uses the histories of normal, tangential, accumulated norm, and maximum experienced norm of the displacement jumps through a RNN to predict the current normal and tangential fabrics of the interface. 
Then these displacement jump and fabric features are input together into a second RNN to predict the current normal and tangential traction across the interface. 
The parameters in each RNN are calibrated with training data of the corresponding input and output features using the backpropagation. 
Each RNN consists of two hidden layers with 32 GRU neurons in each layer, and the output layer is a dense layer with linear activation function. 
All input and output data are pre-processed by standard scaling using mean values and standard deviations \citep{scikit-learn}. 
Each input feature contains its current value and 4 history values prior to the current loading step. 
Each RNN is trained for 1000 epochs using the Adam optimization algorithm \citep{kingma2014adam}, with a batch size of 128.

\bibliographystyle{plainnat}
\bibliography{main}

\end{document}